\documentclass[manuscript,acmsmall,screen,nonacm]{acmart}

\AtBeginDocument{%
  \providecommand\BibTeX{{%
    \normalfont B\kern-0.5em{\scshape i\kern-0.25em b}\kern-0.8em\TeX}}}

\usepackage{amsfonts}
\usepackage{graphicx}
\usepackage{epstopdf}
\usepackage{algorithm,algpseudocode}

\usepackage{tikz}
\usetikzlibrary{arrows.meta, shapes.misc, calc}
\usepackage{amssymb}

\usepackage{subcaption}
\usepackage{pgfplots}
\usepackage{pgfplotstable}

\usepackage{cleveref}

\usepackage{listings}
\crefname{lstlisting}{listing}{listings}
\Crefname{lstlisting}{Listing}{Listings}

\definecolor{DarkGreen}{RGB}{34,139,34}
\definecolor{UFL}{RGB}{229,127,229} 
\definecolor{FFC}{RGB}{144,238,144} 
\definecolor{DOLFIN}{RGB}{173,216,230} 

\newcommand{\N}{N}
\newcommand{\disp}[1]{\displaystyle{#1}}
\newcommand{\ds}{\mathop{}\!\mathrm{d}s}
\newcommand{\dx}{\mathop{}\!\mathrm{d}x}

\newtheorem{remark}{Remark}
\newtheorem{example}{Example}

\DeclareMathOperator{\supp}{supp}

\usepackage{wrapfig}
\newcommand{\R}{\mathbb{R}}
\newcommand{\foralls}{\forall \,} 
\newcommand{\code}[1]{\lstinline!#1!}

\usepackage{mathtools}
\newlength{\arrow}
\newcommand*{\myrightarrow}[1]{\xrightarrow{\mathmakebox[\arrow]{#1}}}

\begin{document}

\title{Abstractions and automated algorithms for mixed domain finite element methods}

\author{C{\'e}cile Daversin-Catty}
\affiliation{%
  \institution{Simula Research Laboratory}
  \streetaddress{P.O. Box 134}
  \postcode{1325}
  \city{Lysaker}
  \country{Norway}
}
\email{cecile@simula.no}

\author{Chris N. Richardson}
\affiliation{%
  \institution{BP Institute, University of Cambridge}
  \streetaddress{Madingley Road}
  \city{Cambridge}
  \postcode{CB3 0EZ}
  \country{United Kingdom}
}
\email{cnr12@cam.ac.uk}

\author{Ada J. Ellingsrud}
\affiliation{%
  \institution{Simula Research Laboratory}
  \streetaddress{P.O. Box 134}
  \postcode{1325}
  \city{Lysaker}
  \country{Norway}
}
\email{ada@simula.no}

\author{Marie E. Rognes}
\affiliation{%
  \institution{Simula Research Laboratory}
  \streetaddress{P.O. Box 134}
  \postcode{1325}
  \city{Lysaker}
  \country{Norway}
}
\email{meg@simula.no}

\renewcommand{\shortauthors}{C. Daversin-Catty, C. N. Richardson, A. J. Ellingsrud and M. E. Rognes}

\begin{abstract}
  Mixed dimensional partial differential equations (PDEs) are
  equations coupling unknown fields defined over domains of differing
  topological dimension. Such equations naturally arise in a wide
  range of scientific fields including geology, physiology, biology
  and fracture mechanics. Mixed dimensional PDEs are also commonly
  encountered when imposing non-standard conditions over a subspace of
  lower dimension e.g. through a Lagrange multiplier. In this paper,
  we present general abstractions and algorithms for finite element
  discretizations of mixed domain and mixed dimensional PDEs of
  co-dimension up to one (i.e.~$n$D-$m$D with $|n-m|\leqslant 1$). We
  introduce high level mathematical software abstractions together
  with lower level algorithms for expressing and efficiently solving
  such coupled systems. The concepts introduced here have also been
  implemented in the context of the FEniCS finite element software.
  We illustrate the new features through a range of examples,
  including a constrained Poisson problem, a set of Stokes-type flow
  models and a model for ionic electrodiffusion.
\end{abstract}



\maketitle

\section{Introduction}
\label{sec:introduction}

Mixed dimensional partial differential equations (PDEs) are systems of
differential equations coupling solution fields defined over domains
of different topological dimensions. Problem settings that call for
such equations are in abundance across the natural
sciences~\cite{Koch:857049, EMI2017}, in multi-physics
problems~\cite{Tambaca2019, Quench2018}, and in
mathematics~\cite{PhDThesisBoon2018, Licht2017}.  For instance, in
geology, fluid flow through faults and fractures in rocks can be
modelled via mixed dimensional PDEs posed on a hierarchy of
interacting domains of heterogeneous dimension~\cite{boon2017mixed,
  Schwenck2015}. In physiology, such equations can model blood flow in
a three-dimensional lumen interacting with a topologically
two-dimensional elastic membrane i.e.~the vessel
wall~\cite{Tambaca2019}. Generally, Lagrange multipliers on
lower-dimensional spaces are commonly used to impose non-standard
boundary conditions or continuity properties over interfaces between
subdomains~\cite{bertoluzza:hal-01420651, EMI2017}, see
e.g.~\Cref{fig:reference-mixed-problem} below for an idealized
example.

For the numerical solution of mixed dimensional PDEs, the finite
element method is a natural approach~\cite{brenner2007mathematical,
  ciarlet2002finite, ern2004theory}. However, the efficient
implementation of finite element discretizations for mixed dimensional
PDEs is non-trivial -- for a number of reasons. First, such
discretizations involve manipulations of multiple meshes and submeshes
of heterogeneous topological dimension. Second, the computation of
local (element-wise) finite element tensors involve integrals of
possibly restrictions of basis functions defined on cells of different
dimensions. Third, the global assembly of the finite element matrices
involve local-to-global mappings across different meshes and
submeshes. And finally, the solution of the resulting linear systems
require efficient and appropriate linear algebra structures. As a
result, the widespread application of mixed dimensional PDEs by domain
specialists is hindered by a lack of numerical solution techniques and
easy-to-use yet efficient software tools.
\begin{figure}
  \centering
  \begin{minipage}[c]{0.44\textwidth}
    \centering
    \begin{tikzpicture}[scale=0.3]
      \draw[black!50] (-4.0,-4.0) -- (4.0,-4.0);
      \draw[black!50] (4.0,-4.0) -- (4.0,4.0);
      \draw[black!50] (4.0,4.0) -- (-4.0,4.0);
      \draw[black!50] (-4.0,4.0) -- (-4.0,-4.0);
      \draw[thick, dashed] (4.0,-4.0) -- (4.0,4.0);
      \draw[thick, dashed] (-4.0,4.0) -- (-4.0,-4.0);
      \draw[black!50] (0, 0) node[] {\huge{$\Omega$}};
      \draw[black] (-4.0, -3.75) node[left] {\small{$\partial \Omega_D$}};
      \draw[black,thick] (0.0, -4.0) -- (0.0, 4.0);
      \draw[black] (0.0, 4.0) node[below right] {\small{$\Gamma$}};
    \end{tikzpicture}
\end{minipage}
\begin{minipage}[c]{0.55\textwidth}
  \begin{subequations}
    \begin{alignat}{3}
      - \Delta u &= f \quad &&\text{on}~ \Omega, \\
      u &= c \quad &&\text{on}~ \Gamma, \\
      u &= 0 \quad &&\text{on}~ \partial \Omega_D, \\
      \frac{\partial u}{\partial n} &= 0
      \quad &&\text{on}~ \partial \Omega \backslash \partial \Omega_D.
    \end{alignat}
    \label{eq:ref-poisson}
  \end{subequations}
\end{minipage}
\caption{Find the solution $u : \Omega \rightarrow \R$ to the Poisson
  equation~\eqref{eq:ref-poisson} with mixed homogeneous boundary
  conditions on a two-dimensional unit square $\Omega = [0,1] \times
  [0,1] \subset \R^2$ such that $u$ is a constant $c$ along the
  topologically one-dimensional interior surface $\Gamma = \{ (x, y)
  \, | \, x = 0.5, y \in (0, 1) \}$. To satisfy the latter constraint,
  introduce a Lagrange multiplier $\lambda : \Gamma \rightarrow \R$.}
  \label{fig:reference-mixed-problem}
\end{figure}

In view of the wide range of applications for mixed dimensional PDEs,
a number of finite element software packages implement some mixed
domain and mixed dimensional finite element features, including
\code{FreeFem++} \cite{2012FreeFem}, \code{Feel++} \cite{2012Feelpp},
\code{deal.II} \cite{2007DealII}, or \code{PorePy}
\cite{2017PorePy}. In particular, \code{FreeFem++} \cite{2012FreeFem}
handles Lagrange multipliers in mixed and mortar methods defining
finite element spaces on boundary meshes. \code{deal.II}
\cite{2007DealII} supports Lagrange multipliers on embedded, possibly
non-matching, meshes, and in particular implements immersed finite
elements methods \cite{HELTAI2012110}. \code{Feel++} \cite{2012Feelpp}
also handles mixed-dimensional problems defining trace meshes, used
for example when implementing domain decomposition and mortar methods
\cite{samake:tel-01092968}. Finally, \code{PorePy} \cite{2017PorePy}
implements mixed dimensional geometrical features providing an
explicit representation of fractures with both finite volumes and
virtual finite element discretizations. Still, the combination of a
generic, automated and high-level software interface would allow for
more rapid development of mixed dimensional discretizations and more
widespread use. 

Over the last 15 years, there has been a significant and growing
interest in generic, high-performance finite element frameworks, as
demonstrated by e.g.~the FEniCS Project~\cite{AlnaesBlechta2015a,
  LoggMardalEtAl2012, FarrellEtAl2013}, the Firedrake
Project~\cite{Firedrake2016}, Feel++~\cite{2012Feelpp},
FreeFEM~\cite{2012FreeFem} and NGSolve~\cite{NGSolve}. A shared design
pattern is the combination of a high-level specification of the
problem discretization, lower-level algorithms for problem solution,
and automated code generation to bridge the gap between. This approach
has been extremely successful, allowing for rapid development of
advanced efficient numerical solvers for non-trivial PDEs and
deployment by application scientists. In particular, FEniCS is
organized as an open source collection of software components
including the high-level domain-specific Unified Form Language
(UFL)~\cite{Alnaes2014}, the FEniCS Form Compiler
(FFC)~\cite{LoggOlgaardEtAl2012a}, and the problem solving environment
DOLFIN~\cite{LoggWells2010a, LoggWellsEtAl2012a}. We refer to the
above references for a more in-depth description of the FEniCS
approach and components.

While FEniCS has offered native support for immersed manifolds since
2012~\cite{FenicsManifolds2013}, support for discretizations of mixed
domain- and mixed dimensional PDEs has been lacking in the core
library. In response and driven by extensive user demand, several
FEniCS extensions have been developed to remedy the situation. For
instance, \code{fenics_ii} \cite{2017FenicsII} implements the concept
of trace spaces, while the \code{multiphenics} Python library
\cite{multiphenics} provides tools aiming to ease the prototyping of
multiphysics problems. However, we argue that native support for mixed
dimensional finite element methods within the core FEniCS framework is
advantageous as it allows for e.g.  increased robustness in part due
to more extensive testing and wider distribution, and easier
development of auxiliary packages and techniques such as e.g.~the
automated derivation of adjoint
models~\cite{FarrellEtAl2013}. Moreover, a formal description of the
abstractions and algorithms involved in finite element methods for
mixed dimensional PDEs is needed.

This work addresses and resolves the gap in available abstractions and
algorithms, and importantly the formal description of such, for the
automated numerical solution of mixed dimensional PDEs via finite
element methods. In particular, we propose and advocate a light-weight
design pattern for mixed dimensional finite element abstractions. We
revise and introduce new abstractions in the Unified Form Language for
mixed function spaces, basis functions and integration domains
allowing for coupled variational formulations defined over mixed
domains. We also introduce a generalized assembly algorithm together
with associated features such as submesh generation and block
matrices. For the automated generation of local element tensor code
from the symbolic representation, i.e. the form compilation, we
present form component extraction algorithms and revised form
compilation strategies.

The concepts and algorithms presented here are implemented in
UFL~\cite{Alnaes2014}, FFC~\cite{LoggOlgaardEtAl2012a}
and DOLFIN~\cite{LoggWells2010a},
and are openly and freely available (see~\cite{DockerContainer,ZenodoExamples}).
The scope of this paper is limited to
mixed domain and mixed dimensional problems of co-dimension one at
most, i.e.~$n$D-$m$D problems with $|n-m|\leqslant 1$ and to
conforming meshes. Various techniques for handling non-matching meshes
are discussed in the literature, such as e.g~\cite{BurmanEtAl2015, 2018MultiMesh},
but not considered further here.

This paper is organized as follows. In \Cref{sec:math-scope} we
describe the mathematical scope of our mixed domain and mixed
dimensional framework. We then address different aspects of the finite
element method applied to mixed domain problems including key
challenges in the subsequent sections. \Cref{sec:meshview} is
dedicated to meshes, nested submeshes and mappings between such.  The
key features in UFL for defining and manipulating mixed domain
function spaces and variational forms are introduced in
\Cref{sec:ufl}. The local-to-global degree of freedom mapping is
introduced in \Cref{sec:dofs} as a key ingredient for the assembly of
mixed domain and mixed dimensional variational forms. Abstract
assembly algorithms, building on the construction of local element
tensors and subsequent insertion using the local-to-global degree of
freedom mappings, are detailed in \Cref{sec:assembly}. An overview of
the revised FEniCS user interface and pipeline is given in
\Cref{sec:interface-overview} with emphasis on automated code
generation of mixed domain and mixed dimensional local tensors and
assembly features.  Importantly, we present numerical results for
various applications in \Cref{sec:numerical-results} ranging from the
an idealized reference example introduced as
\Cref{ex:reference-mixed-problem} to more advanced models highlighting
the relevance of our framework in biomedical
applications. \Cref{sec:conclusions} provides some concluding remarks
while discussing current limitations and future extensions.

\section{Mathematical scope and concepts}
\label{sec:math-scope}

\subsection{Notation}
For convenience, we here provide an overview of the main notation used
in this manuscript. In general, superscripts are used to indicate
subdomain or block indices. In the text, all indices start at $1$. In the code,
the corresponding indices start at $0$. The terms element and
element-wise are used equivalently with cell or cell-wise,
respectively.
\begin{itemize}
\item[$i$, $j$:]
  Indices associated with the number of subdomains.
\item[$n$, $m$:]
  Indices associated with number of basis functions.
\item[$r$, $s$:]
  Indices associated with form arity.
\item[$|S|$:]
  Dimension of a finite set $S$.
\item[$\Omega$, $\Omega^i$:]
  A domain, domain $i$ for $i = 1, \dots, I$.
\item[$d^i$:]
  Topological dimension of $\Omega^i$.
\item[$\mathbb{V}^i$:]
  Vector space relative to $\Omega^i$.
\item[$\mathcal{T}$, $\mathcal{T}^i$:]
  A simplicial mesh, simplicial mesh of $\Omega^i$.
\item[$\mathcal{S}(\mathcal{T})$:]
  The simplicial complex induced by the simplicial mesh $\mathcal{T}$.
\item[$U^i$:]
  Finite element function space defined with respect to $\mathcal{T}^i$.
\item[$\N^i$:]
  Dimension of the finite element space $U^i$: $\N^i = \dim(U^i)$.
\item[$\N^i_K$:]
  Dimension of the finite element space $U^i$ restricted to $K$:
  $\N^i_K = \dim(U^i|_K)$. If $K$ is a cell in $\mathcal{T}^i$, this
  is the local dimension of the finite element space $U^i$.
\item[$\phi^i_n$:]
  Basis function for $U^i$ for $n = 1, \dots, \N^i$.
\item[$\mathcal{N}_K^i$:]
  A set of indices of basis functions in $U^i$ with $K$ in their
  support: \\ $\mathcal{N}_K^i = \{ n \in \{1, \dots, N^i \} \, | \, K
  \subseteq \supp(\phi_n^i)\}$.
\item[$\mathcal{V}$, $\mathcal{V}^i$:]
  Set of $n_v$, $n^i_v$ vertex indices in $\mathcal{T}$, $\mathcal{T}^i$.
\item[$\mathcal{F}$, $\mathcal{F}^i$:]
  Set of $n_f$, $n^i_f$ facet indices in $\mathcal{T}$, $\mathcal{T}^i$.
\item[$\mathcal{C}$, $\mathcal{C}^i$:]
  Set of $n_c$, $n^i_c$ cell indices in $\mathcal{T}$, $\mathcal{T}^i$.
\item[$\mathcal{M}_v^i$, $\mathcal{M}^i$:]
  Child-to-parent vertex and cell index maps.
\item[$\iota_K^i$:]
  Local-to-global degree of freedom map, $K \in \mathcal{T}^i$, for finite element space $U^i$.
\item[$S_K$:]
  Star of $K$, defined as the set of cells in $\mathcal{T}$ containing $K$.
\end{itemize}

\subsection{Mixed domains and meshes}
\label{sec:domains}

We define a mixed domain PDE as a system of PDEs coupling fields $u^i
: \Omega^i \rightarrow \mathbb{V}^i$ where $\Omega^i \subset \R^{d}$
is a bounded domain of geometrical dimension $d$ and topological
dimension $d^i$, and $\mathbb{V}^i$ is a vector space for $i = 1,
\dots, I$.  We assume that there exists an $\Omega \subset
\mathbb{R}^d$, a $d$-dimensional domain that embeds all the subdomains
$\Omega^i \subseteq \Omega$, with $d \geqslant \max_i d^i$. We refer to
$\Omega$ as the parent domain. We assume that $\Omega$ is polyhedral
such that it admits a conforming discretization. The subdomains are
assumed to be of codimension at most one relative to $\Omega$,
i.e.~$|d^{j} - d^{i}| \leqslant 1$ for all $i, j=1, \dots, I$. We will
use the term mixed dimensional PDE for a mixed domain PDE if there are
$i, j$ such that $d^i \not = d^j$.

We assume that the parent domain $\Omega$ is partitioned by a mesh
$\mathcal{T}$ consisting of a finite set of cells $\mathcal{T} =
\{K\}$. For simplicity in terminology, we here consider the case of
simplicial cells (intervals, triangles, tetrahedra). Moreover, we
assume that we can define a conforming mesh $\mathcal{T}^i$ of each
subdomain $\Omega^i$, for $i = 1, \dots, I$, consisting of mesh
entities (vertices, edges, faces, cells) from $\mathcal{T}$. More
precisely, we assume that $\Omega^i = \cup_{k} \{ K^i_k \}$ where the
submesh $\mathcal{T}^i = \{ K^i_k \}_k$ consists of mesh entities
$K^i_k$ from $\mathcal{T}$. In the language of complexes, let
$S(\mathcal{T})$ be the simplicial complex defined by
$\mathcal{T}$. By definition, $\Omega$ is then the underlying space of
$S$. We assume that $\mathcal{T}$ and $\Omega^i$ for $i = 1, \dots, I$
are such that we can define simplicial meshes $\mathcal{T}^i$ with
induced simplicial complexes $\mathcal{S}^i =
\mathcal{S}(\mathcal{T}^i)$ such that $\Omega^i$ is the underlying
space of $\mathcal{S}^i$ and such that $\mathcal{S}^i$ is a subcomplex
of $\mathcal{S}$ for $i = 1, \dots, I$.

\subsection{Finite element function spaces}
We introduce function spaces $U^i$ for $i = 1, \dots, I$, each defined
over $\Omega^i$, such that
\begin{equation}
  U^i = \{ v^i : \Omega^i \rightarrow \mathbb{V}^i \},
\end{equation}
and assume that each unknown $u^i \in U^i$. The solution $u$ of a
mixed domain PDE is hence an $I$-tuple $u = (u^1, \dots, u^I)$ in the
Cartesian product space $U$: 
\begin{equation}
  \label{equ:mixed-fs-def}
  u \in U \equiv U^1 \times U^2 \times \cdots \times U^I.
\end{equation}
We refer to $U$ as a mixed function space with $U^i$ as subspaces.

We are here mainly concerned with finite element spaces $U^i$ defined
relative to the submeshes $\mathcal{T}^i$ for $i = 1, \dots, I$. We
assume that these discrete function spaces are indeed finite element
spaces in the sense that the basis functions have localized support
and can be defined element-wise. We write $\N^i$ for the global
dimension of the finite element space $U^{i}$, and $\N^i_K$ for its
local (element-wise) dimension i.e.~$\dim(U^i |_K)$ for $K \in
\mathcal{T}^i$. We denote by $\{ \phi^i_{n}\}_{n=1}^{N^i}$ the sets of
basis functions spanning the discrete spaces $U^{i}$. Discrete
solutions $u^i \in U^i$, for $i = 1, \dots, I$, can thus be expressed
as a linear combination of these basis functions:
\begin{equation}
  \label{eq:basis-function-def}
  u^i = \sum_{n=1}^{\N^i} \bar{u}^i_n \phi^{i}_{n},
\end{equation}
with expansion coefficients (or, colloquially, degrees of freedom)
$\bar{u}_n^i$ for $n = 1, \dots, N^i$. We emphasize the possibility of
having different kinds of finite element spaces for the different
function spaces. This is especially relevant for multiphysics problems
for which the suitable function space properties can differ from one
field to the other.

\subsection{Variational forms and formulations}
We consider discrete variational formulations of systems of linear or
non-linear PDEs and associated variational forms of arity $r \geq
0$. For time-dependent problems, we presuppose a time-stepping
procedure yielding systems of PDEs at each time step. In general, we
consider systems of PDEs that may be expressed in operator form with
$I \in \mathbb{N}$ equations, each defined over $\Omega^i$ for $i = 1,
\dots, I$. As our main emphasis is on finite element discretizations,
we assume that a discrete variational formulation of the system is
prescribed. 

\subsubsection{Linear variational problems}

We first consider a general system of discrete linear variational
equations: find $u \in U = U^1 \times U^2 \times \dots \times U^I$
such that
\begin{equation}
  \label{equ:vfs-linear}
  a^i(u, v^i) = L^i(v^i) \quad \foralls v^i \in U^i \quad i = 1, \dots, I,
\end{equation}
where $a^i : U \times U^i \rightarrow \R$ is a bilinear form, $L^i:
U^i \rightarrow \R$ is a linear form, and $U^i$ are appropriate finite
element spaces defined over $\Omega^i$ and mapping into
$\mathbb{V}^i$, for $i = 1, \dots, I$. To enhance readability, note
that we present the case of coinciding trial and test subspaces here,
however we include numerical examples with differing test and trial
spaces in Section~\ref{sec:numerical-results}.

By the linearity of $a^i$ and as the approximation space $U$ is defined
as a Cartesian product of function spaces
cf.~\eqref{equ:mixed-fs-def}, each bilinear form $a^i$ can be written
as the sum of bilinear forms $a^{i, j}: U^j \times U^i \rightarrow \mathbb{R}$:
\begin{equation}
  \label{equ:vfs-block-decomposition}
  \exists \, a^{i, j} \mid a^i(u, v^i)
  = \sum \limits_{j=1}^I a^{i, j}(u^j, v^i) \quad \foralls i=1, \dots, I.
\end{equation}
The discrete weak form of the whole coupled system for linear mixed
problems thus consists in finding $u \in U$ such that
\begin{equation}
  \label{eq:linear:var:discrete}
  a(u, v) = L(v) \quad \quad \foralls v \in U, 
\end{equation}
with $v = (v^1, \dots, v^I)$ and 
\begin{equation}
  \label{eq:decomposition-mixed-form-linear}
  a(u, v) = \sum_{i=1}^I \sum_{j=1}^I a^{i, j}(u^{j}, v^{i}) \quad
  \text{ and } \quad L(v) = \sum_{i=1}^I L^i(v^{i}) .
\end{equation}
In general, a variational form $a: U \times U \times \dots \times U $
of arity $r$ can be decomposed into $r$ sums of arity-$r$ forms:
$a^{i_1, i_2, \dots, i_r}: U^{i_r} \times U^{i_{r-1}} \times \dots \times
U^{i_1}: \rightarrow \R$:
\begin{equation}
  \label{eq:block:decomposition:general}
  a(u_r, u_{r-1}, \dots, u_1)
  = \sum_{i_1=1}^I \dots \sum_{i_r=1}^{I} a^{i_1, i_2, \dots, i_r}(u_r^{i_r}, u_{r-1}^{i_{r-1}}, \dots, u_1^{i_1}) .
\end{equation}
for $u_s = (u_s^1, u_s^2, \dots, u_s^I)$ for $s = 1, \dots, r$. We
will refer to $a^{i_1, i_2, \dots, i_r}$ and specifically $a^{i, j}$
and $L^i$ as block forms. We will refer to a (block) form $a^{i_s,
  i_s, \dots, i_s}$ for some $s \in \{ 1, \dots, r \}$ as a diagonal
(block) form. 

The finite element solution of~\eqref{eq:linear:var:discrete}
typically involves the assembly of the bilinear form $a$ and linear
form $L$, i.e.~the construction of a matrix $A$ and a vector $L$ such
that $u$ solves
\begin{equation}
  A \bar{u} = b ,
\end{equation}
where $\bar{u}$ denotes the vector of expansion coefficients for the
discrete field $u$ i.e.
\begin{equation}
  \bar{u} = \{ \bar{u}^i \}_{i=1}^{I}, \quad
  \bar{u}^i = \{ \bar{u}^i_n\}_{n=1}^{\N^i} \quad \foralls i = 1, \dots, I.
\end{equation}
By construction, cf.~\eqref{eq:decomposition-mixed-form-linear}, $A$
is a block matrix and $b$ is a block vector with entries
\begin{equation}
  \label{equ:block-shaped-system-linear}
  \renewcommand{\arraystretch}{1.75}
  \left[
    \begin{array}{c|c|c}
      A^{1,1} & \dots & A^{1,I} \\
      \hline
      \vdots & \ddots & \vdots \\
      \hline
      A^{I,1} & \dots & A^{I,I} \\
    \end{array}
    \right]
  \left[
    \begin{array}{ccc}
      \bar{u}^1 \\
      \hline
      \vdots \\
      \hline
      \bar{u}^I \\
    \end{array}
  \right]
  =
  \left[
    \begin{array}{ccc}
      b^1 \\
      \hline
      \vdots \\
      \hline
      b^I \\
    \end{array}
    \right] .
\end{equation}
The diagonal blocks $A^{i,i}$ represent the uncoupled parts of the
problem while the off-diagonal blocks $A^{i,j}$ for $i \neq j$
represent the interaction between fields living on any two subdomains
$\Omega^i$ and $\Omega^j$. The elements of $A$ and $b$ are defined for
$i, j = 1, \dots, I$ by
\begin{equation}
  \label{eq:decomposition-mixed-form-blocks-linear}
  A^{i,j}_{m,n} = a^{i, j}(\phi^{j}_n, \phi^{i}_m) ~~\text{and}~~ b^i_{m} = L^i(\phi^{i}_m), ~n=1,\dots,\N^{j}, ~m=1,\dots,\N^{i} .
\end{equation}

\begin{example}
\label{ex:reference-mixed-problem}
To illustrate, we detail a variational formulation and the block
structure of the mixed dimensional Poisson example introduced in
\Cref{fig:reference-mixed-problem}. As detailed in Section
\ref{sec:domains}, we assume a mesh $\mathcal{T}$ of the parent domain
$\Omega$ such that a subset of its facets induce a conforming submesh
$\mathcal{T}^2$ of $\Gamma$. We identify $\Omega^1 = \Omega$,
$\mathcal{T}^1 = \mathcal{T}$ and $\Omega^2 = \Gamma$. Further, we let
$U^1 \subset H^1_0(\Omega^1)$ be a finite element space with zero
trace on the Dirichlet boundary $\partial \Omega_D$ only, and let
$U^2$ be a conforming finite element space of $L^2(\Omega^2)$. A
discrete variational formulation describing \eqref{eq:ref-poisson}
then reads: find $(u, \lambda) \in U \equiv U^1 \times U^2$ such that
\begin{equation}
  \label{eq:Poisson-LM-weakform}
  \int_{\Omega} \nabla u \cdot \nabla v \dx + \int_{\Gamma} \lambda v \ds
  + \int_{\Gamma} \eta u \ds
  = \int_{\Omega} f v \dx +  \int_{\Gamma} c \eta \ds ,
\end{equation}
for all $(v, \eta) \in U^1 \times U^2$.

The block decomposition \eqref{eq:decomposition-mixed-form-linear} of
the bilinear form $a(u,v)$ (resp. linear form $L(v)$) gives the
subforms
\begin{align*}
  &a^{1,1}(u,v) = \int_{\Omega} \nabla u \cdot \nabla v \dx, \quad
  a^{1,2}(\lambda,v) = \int_{\Gamma} \lambda v \ds, \quad
  a^{2,1}(u, \eta) = \int_{\Gamma} \eta u \ds, 
\end{align*}
with $a^{2,2}(\lambda, \eta) = 0$, and
\begin{align*}
  L^{1}(v) = \int_{\Omega} f v \dx, \quad
  L^{2}(\eta) = \int_{\Gamma} c \eta \ds.
\end{align*}
The block system corresponding to \eqref{eq:Poisson-LM-weakform}
then reads as follows (with $u = u^1$ and $u^2 = \lambda$):
\begin{equation}
  \label{eq:Poisson-LM-system}
  \renewcommand{\arraystretch}{1.5}
  \left[
    \begin{array}{c|c}
      ~ \\\
      \quad A^{1,1} \quad ~& A^{1,2} \\
      ~ \\
      \hline
      A^{2,1} & A^{2,2}\\
    \end{array}
    \right]
  \left[
    \begin{array}{c}
      ~ \\
      \bar{u}^1\\
      ~ \\
      \hline
      \bar{u}^2\\
    \end{array}
    \right]
  =
  \left[
    \begin{array}{c}
      ~ \\
      b^1\\
      ~ \\
      \hline
      b^2\\
    \end{array}
    \right] ,
\end{equation}
where the blocks $A^{i,j}$ and $b^i$, $i,j = 1,2$ are obtained from
\eqref{eq:decomposition-mixed-form-blocks-linear}.
\end{example}

\subsubsection{Nonlinear variational problems}

Nonlinear mixed domain problems lead to discrete variational formulations of
the form: find $u \in U$ such that
\begin{equation}
  \label{equ:vfs-nonlinear}
  F^i(u; v^i) = 0 \quad \foralls v^i \in U^i,
\end{equation}
where the forms $F^i : U \times U^i \rightarrow \mathbb{R}$ may be
nonlinear in $u \in U$ but are linear in the test functions $v^i \in
U^i$ for $i = 1, \dots, I$. Combining the $I$ equations, the canonical
nonlinear mixed formulation reads as: find $u \in U$ such that
\begin{equation}
  \label{eq:nonlinear:var:discrete}
  F(u; v) = 0 \quad \foralls v \in U,
\end{equation}
with
\begin{equation}
  \label{eq:decomposition-mixed-form-nonlinear}
  F(u; v) = \sum_{i=1}^I F^i(u; v^i) .
\end{equation}

Newton's method or variations are commonly used to solve such
problems. Starting from an initial solution $u_0 = (u^1_0, \dots,
u^I_0)$, each iteration solves the system
\begin{equation}
J(u_k; \cdot, v)(u_{k+1} - u_k) = -F(u_k; v),
\end{equation}
where $J(u_k; \cdot, v)$ is the Jacobian of $F(u_k; v)$ at the $k$-th
iteration. The discrete system at each iterate again has a
block-shaped pattern
\begin{equation}
  \label{equ:block-shaped-system-nonlinear}
  \renewcommand{\arraystretch}{1.75}
  \left[
    \begin{array}{c|c|c}
      J^{1,1} & \dots & J^{1,I} \\
      \hline
      \vdots & \ddots & \vdots \\
      \hline
      J^{I,1} & \dots & J^{I,I} \\
    \end{array}
    \right]
  \left[
    \begin{array}{ccc}
      \bar{\delta u}^0 \\
      \hline
      \vdots \\
      \hline
      \bar{\delta u}^I
    \end{array}
  \right]
  =
  - \left[
    \begin{array}{ccc}
      F^{1} \\
      \hline
      \vdots \\
      \hline
      F^{I} \\
    \end{array}
    \right] ,
\end{equation}
and $u_{k+1} = u_{k} + \delta u$, with blocks defined by:
\begin{equation}
  \label{eq:decomposition-mixed-form-blocks-nonlinear}
  J^{i,j}_{m,n} = \frac{\partial F^{i}(u_k; \phi^i_m)}{\partial u^j} \left( \phi^j_n \right) ~~\text{and}~~
  F^{i}_{m} = F^{i}(u_k; \phi^{i}_m),
\end{equation}
for $n=1, \dots, \N^j$, $m=1,\dots,\N^i$ and $i,j = 1, \dots, I$.

\subsection{Integration domains}
\label{sec:integration:domains}

Typically in finite element applications, the mixed variational forms,
e.g.~$a$ and $L$ in \eqref{eq:linear:var:discrete} and $F$ in
\eqref{eq:nonlinear:var:discrete}, are given as sums of integrals over
different subdomains $\Omega^i$. We assume that all variational forms can
be represented by sums over mesh entities, for instance as sums of
integrals over cells in a domain, see
e.g.~\cite{LoggMardalEtAl2012}. For mixed dimensional problems, the
subdomains $\Omega^i$ will have different topological dimensions $d^i \leqslant d$,
where $d$ is the topological dimension of the parent domain $\Omega$.
We introduce the notation $x = (x_1, \dots,
x_{d})$ for the coordinates of a point $x \in \Omega$.  In the
following, the notation $\dx = \dx_1 \times \dots \times \dx_{d}$ is
used in integrals over a $d$-dimensional domain. We use the notation
$\ds$ to integrate over a co-dimension 1 subdomain of $\Omega$.

\medskip

The implementation of finite element discretizations of mixed domain
problems within a high level framework such as e.g. the FEniCS Project
involves a number of new concepts and algorithmic extensions in
comparison with single domain problems. We dive into these aspects in
the subsequent sections.

\section{Data structures for nested submeshes}
\label{sec:meshview}

In this section, we discuss and suggest data structures for
representing submeshes, in particular for nested submeshes. As
detailed in Section~\ref{sec:domains}, we consider the case where all
subdomains $\Omega^i$ share a parent domain $\Omega$ and where each
subdomain is covered by a conforming submesh $\mathcal{T}^i$ of the
parent complex generated by $\mathcal{T}$. Two mappings between meshes
are essential for mixed domain finite element assembly: first,
mappings between submesh entities and parent mesh entities, and
second, mappings between mesh entities in different submeshes. In the
below, we formalize these concepts and describe their implementation
in the FEniCS context.

\subsection{Mappings for nested submeshes}
\label{subsec:meshview-mappings}

We assume that each mesh is represented by the combination of its
topology (defining the mesh entities and connections between these)
and geometry (defining the spatial vertex
coordinates)~\cite{Logg2009}. We denote by $n_v$, $n_f$ and $n_c$ the
number of vertices, facets and cells in $\mathcal{T}$, respectively,
and let $\mathcal{V}$, $\mathcal{F}$ and $\mathcal{C}$ be the
corresponding sets of vertex, facet and cell indices:
\begin{equation}
  \label{equ:entities-set-parent}
  \mathcal{V} = \{v_k\}_{k=1}^{n_{v}}, ~\mathcal{F} = \{f_k\}_{k=1}^{n_{f}} ~\text{and}~ \mathcal{C} = \{c_k\}_{k=1}^{n_{c}} .
\end{equation}
We assume that any vertex, facet and cell can be identified by the
respective index $v_k$, $f_k$ and $c_k$ and its (mesh entity)
type. For each submesh, we denote by $\mathcal{V}^i$ and
$\mathcal{C}^i$ the sets of vertex and cell indices of
$\mathcal{T}^{i}$, whose indices $v^i_k$ and $c^i_k$ are independent
from the parent mesh numbering cf.~\eqref{equ:entities-set-parent}:
\begin{equation}
  \label{equ:entities-set-child}
  \mathcal{V}^i = \{v^i_k\}_{k=1}^{n_v^i}, ~\mathcal{C}^i = \{c^i_k\}_{k=1}^{n_{c}^i} .
\end{equation}

We now introduce two maps that link the vertex indices in the submesh
with the corresponding vertex index in the parent mesh, and the cell
indices in the submesh with the corresponding mesh entity index in the
parent mesh. In particular, for each submesh $\mathcal{T}^i$, we
define its (child-to-parent) vertex map $\mathcal{M}_v^i$ and cell map $\mathcal{M}^i$:
\begin{equation}
  \label{eq:meshview-mappings}
  \mathcal{M}_v^i : \mathcal{V}^i \rightarrow \mathcal{V} 
  \quad \text{and} \quad 
  \mathcal{M}^i : \mathcal{C}^i \rightarrow  \mathcal{E} \in \{\mathcal{C}, \mathcal{F}\} .
\end{equation}
We note that if the submesh $\mathcal{T}^{i}$ has the same topological
dimension as its parent $\mathcal{T}$, each cell in the submesh is a
cell in the parent mesh and so $\mathcal{E} = \mathcal{C}$. However,
the cells of a submesh of codimension $e \geq 1$ are mesh entities of
codimension $e$ in the parent mesh $\mathcal{T}$, and $\mathcal{E} =
\mathcal{F}$ when $e = 1$. These concepts applied to the reference
example \eqref{eq:ref-poisson} are illustrated in
\Cref{fig:build-submeshes}.
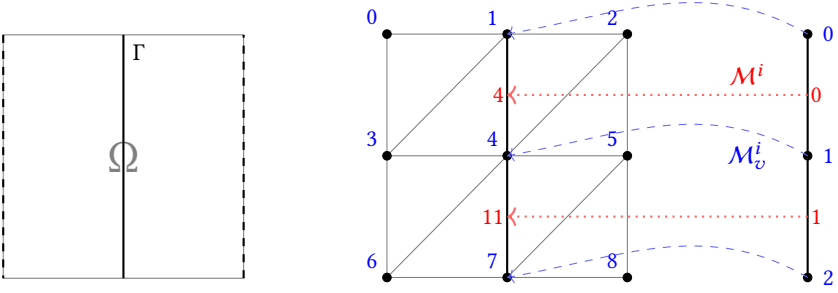
\begin{figure}
  \centering
  \begin{subfigure}[c]{0.4\textwidth}
    \centering
    \begin{tikzpicture}[scale=0.4]
      \draw[black!50] (-4.0,-4.0) -- (4.0,-4.0);
      \draw[black!50] (4.0,-4.0) -- (4.0,4.0);
      \draw[black!50] (4.0,4.0) -- (-4.0,4.0);
      \draw[black!50] (-4.0,4.0) -- (-4.0,-4.0);
      \draw[thick, dashed] (4.0,-4.0) -- (4.0,4.0);
      \draw[thick, dashed] (-4.0,4.0) -- (-4.0,-4.0);
      \draw[black!50] (0, 0) node[] {\huge{$\Omega$}};
      \draw[black,thick] (0.0, -4.0) -- (0.0, 4.0);
      \draw[black] (0.0, 4.0) node[below right] {\small{$\Gamma$}};
    \end{tikzpicture}
  \end{subfigure}
  \begin{subfigure}[c]{0.5\textwidth}
    \centering
    \begin{tikzpicture}[scale=0.4]
      \draw[black!50] (-4.0,2.0) -- (4.0,2.0);
      \draw[black!50] (4.0,2.0) -- (4.0,10.0);
      \draw[black!50] (4.0,10.0) -- (-4.0,10.0);
      \draw[black!50] (-4.0,10.0) -- (-4.0,2.0);
      \draw[black,thick] (0.0, 2.0) -- (0.0, 10.0);
      \draw[black!50] (-4.0,6.0) -- (4.0,6.0);
      \draw[black!50] (-4.0,6.0) -- (0.0, 10.0);
      \draw[black!50] (0.0,6.0) -- (4.0, 10.0);
      \draw[black!50] (-4.0,2.0) -- (0.0, 6.0);
      \draw[black!50] (0.0,2.0) -- (4.0, 6.0);
      \draw (0,0) [fill] (-4,2) circle [radius=4pt];
      \draw (0,0) [fill] (-4,6) circle [radius=4pt];
      \draw (0,0) [fill] (-4,10) circle [radius=4pt];
      \draw (0,0) [fill] (0,2) circle [radius=4pt];
      \draw (0,0) [fill] (0,6) circle [radius=4pt];
      \draw (0,0) [fill] (0,10) circle [radius=4pt];
      \draw (0,0) [fill] (4,2) circle [radius=4pt];
      \draw (0,0) [fill] (4,6) circle [radius=4pt];
      \draw (0,0) [fill] (4,10) circle [radius=4pt];
      \draw[black,thick] (10.0,2.0) -- (10.0, 10.0);
      \draw (0,0) [fill] (10,2) circle [radius=4pt];
      \draw (0,0) [fill] (10,6) circle [radius=4pt];
      \draw (0,0) [fill] (10,10) circle [radius=4pt];
      \draw [->, dotted, thick, red!60] (10,8) -- (0,8);
      \draw [->, dotted, thick, red!60] (10,4) -- (0,4);
      \draw [->, dashed, blue!60] (10,10) to [out=150,in=10] (0,10);
      \draw [->, dashed, blue!60] (10,6) to [out=150,in=10] (0,6);
      \draw [->, dashed, blue!60] (10,2) to [out=150,in=10] (0,2);
      \draw (9.8, 4) node[right] {\textcolor{red}{\small{$1$}}};
      \draw (9.8, 8) node[right] {\textcolor{red}{\small{$0$}}};
      \draw (8, 8) node[above] {\textcolor{red}{\small{$\mathcal{M}^i$}}};
      \draw (10.2, 2) node[right] {\textcolor{blue}{\small{$2$}}};
      \draw (10.2, 6) node[right] {\textcolor{blue}{\small{$1$}}};
      \draw (10.2, 10) node[right] {\textcolor{blue}{\small{$0$}}};
      \draw (8, 6) node[] {\textcolor{blue}{\small{$\mathcal{M}^i_{v}$}}};
      \draw (-4, 10) node[above left] {\textcolor{blue}{\small{$0$}}};
      \draw (0, 10) node[above left] {\textcolor{blue}{\small{$1$}}};
      \draw (4, 10) node[above left] {\textcolor{blue}{\small{$2$}}};
      \draw (-4, 6) node[above left] {\textcolor{blue}{\small{$3$}}};
      \draw (0, 6) node[above left] {\textcolor{blue}{\small{$4$}}};
      \draw (4, 6) node[above left] {\textcolor{blue}{\small{$5$}}};
      \draw (-4, 2) node[above left] {\textcolor{blue}{\small{$6$}}};
      \draw (0, 2) node[above left] {\textcolor{blue}{\small{$7$}}};
      \draw (4, 2) node[above left] {\textcolor{blue}{\small{$8$}}};
      \draw (0.2, 4) node[left] {\textcolor{red}{\small{$11$}}};
      \draw (0.2, 8) node[left] {\textcolor{red}{\small{$4$}}};
    \end{tikzpicture}
  \end{subfigure}
  \caption{Mapping between a submesh and its parent. The parent mesh
    $\mathcal{T}$ is the 2D mesh representing $\Omega$, while the
    submesh $\mathcal{T}^{1}$ represents the interface $\Gamma$ and
    can be constructed as a subset of $\mathcal{T}$ facets. The
    parent mesh has $n_{v} = 9$ vertices ($\mathcal{V} = \{0, \dots,
    8\}$) and $n_{f} = 16$ facets ($\mathcal{F} = \{0, \dots, 15\}$),
    while the submesh $\mathcal{T}^{1}$
    has $n_v^1 = 3$ vertices ($\mathcal{V}^1 = \{0,1, 2\}$)
    and $n_{c}^1 = 2$ cells ($\mathcal{C}^1 = \{0,1\}$). The
    mappings $\mathcal{M}_v^1 : \mathcal{V}^1 \rightarrow \mathcal{V}
    $ and $\mathcal{M}^1: \mathcal{C}^1 \rightarrow \mathcal{F}$
    cf.~\eqref{eq:meshview-mappings} give $\mathcal{M}_v^1(0) = 1$,
    $\mathcal{M}_v^1(1) = 4$, $\mathcal{M}_v^1(2) = 7$ and
    $\mathcal{M}^1(0) = 4$, $\mathcal{M}^1(1) = 11$.  }
  \label{fig:build-submeshes}
\end{figure}
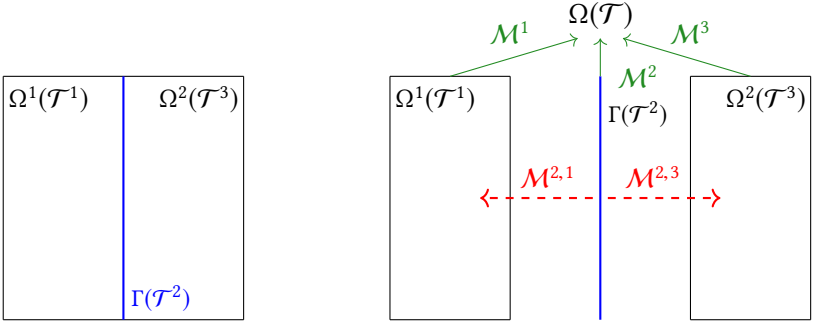
\begin{figure}
  \centering
  \begin{subfigure}[b]{0.4\textwidth}
    \centering
    \begin{tikzpicture}[scale=0.4]
      \draw[black] (-4.0,2.0) -- (4.0,2.0);
      \draw[black] (4.0,2.0) -- (4.0,10.0);
      \draw[black] (4.0,10.0) -- (-4.0,10.0);
      \draw[black] (-4.0,10.0) -- (-4.0,2.0);
      \draw[blue, thick] (0.0, 2.0) -- (0.0, 10.0);
      \draw[blue] (1.25, 2) node[above] {\small{$\Gamma(\mathcal{T}^{2})$}};
      \draw (-2.5, 10) node[below] {$\Omega^1(\mathcal{T}^{1})$};
      \draw (2.5, 10) node[below] {$\Omega^2(\mathcal{T}^{3})$};
    \end{tikzpicture}
  \end{subfigure}
  \begin{subfigure}[b]{0.5\textwidth}
    \centering
    \begin{tikzpicture}[scale=0.4]
      \draw (0, 0) node[] {\large{$\Omega(\mathcal{T})$}};
      \draw[black] (-7.0,-10.0) -- (-3.0,-10.0);
      \draw[black] (-3.0,-10.0) -- (-3.0,-2.0);
      \draw[black] (-3.0,-2.0) -- (-7.0,-2.0);
      \draw[black] (-7.0,-2.0) -- (-7.0,-10.0);
      \draw (-5.5, -2) node[below] {$\Omega^1(\mathcal{T}^{1})$};
      \draw[black] (3.0,-10.0) -- (7.0,-10.0);
      \draw[black] (7.0,-10.0) -- (7.0,-2.0);
      \draw[black] (7.0,-2.0) -- (3.0,-2.0);
      \draw[black] (3.0,-2.0) -- (3.0,-10.0);
      \draw (5.5, -2) node[below] {$\Omega^2(\mathcal{T}^{3})$};
      \draw[blue, thick] (0.0, -10.0) -- (0.0, -2.0);
      \draw (1.25, -2.5) node[below] {\small{$\Gamma(\mathcal{T}^{2})$}};
      \draw[DarkGreen, <-] (-0.75,-0.75) -- (-5, -2);
      \draw[DarkGreen] (-3, -1.25) node[above] {$\mathcal{M}^1$};
      \draw[DarkGreen, <-] (0.75,-0.75) -- (5, -2);
      \draw[DarkGreen] (3, -1.25) node[above] {$\mathcal{M}^3$};
      \draw[DarkGreen, <-] (0.0,-0.75) -- (0,-2);
      \draw[DarkGreen] (1.25, -2) node[] {$\mathcal{M}^2$};

      \draw[thick, red, dashed, ->] (-0.25,-6) -- (-4,-6);
      \draw[thick, red, dashed, ->] (0.25,-6) -- (4,-6);
      \draw[red] (-1.75, -6) node[above] {$\mathcal{M}^{2, 1}$};
      \draw[red] (1.75, -6) node[above] {$\mathcal{M}^{2, 3}$};
    \end{tikzpicture}
  \end{subfigure}
  \caption{The assembly of the systems can require additional mappings
    $\mathcal{M}^{i, j}$ to relate the submeshes $\mathcal{T}^{i}$ and
    $\mathcal{T}^{j}$ with $i \neq j$, assuming they are built from
    the same parent $\mathcal{T}$.  For example, the additional
    mapping $\mathcal{M}^{2, 1}: \mathcal{C}^2 \rightarrow
    \mathcal{F}^1$ is needed and only built if $a^{1,2}(\phi^{2}_n,
    \phi^{1}_m)$ or $a^{2,1}(\phi^{1}_m, \phi^{2}_n)$ is non-zero,
    ~$n=1,\dots,\N^{2}, ~m=1,\dots,\N^{1}$.}
  \label{fig:extra-mappings}
\end{figure}

A mixed domain problem can couple an arbitrary but finite number $I$
of fields $u^i \in \Omega^i$, $i=1, \dots, I$.  The assembly of the
systems \eqref{equ:block-shaped-system-linear},
\eqref{equ:block-shaped-system-nonlinear} can then require additional
mappings $\mathcal{M}^{i, j}$ to relate the submeshes
$\mathcal{T}^{i}$ and $\mathcal{T}^{j}$ involved in e.g.~$a^{i,j}$
\eqref{eq:decomposition-mixed-form-linear} for $i \neq j$ as
illustrated by \Cref{fig:extra-mappings}, assuming their intersection
$\mathcal{T}^i \cap \mathcal{T}^j$ is non-empty.
Assume that $d^j \geq d^i$ without further loss of generality.
We can then express the map from
cell indices of $\mathcal{T}^i \in \mathcal{T}^i \cap \mathcal{T}^j$
to corresponding mesh entity indices in $\mathcal{T}^j$ as
\begin{equation}
  \mathcal{M}^{i, j}: \mathcal{C}^i \rightarrow \mathcal{E}^j,
  \quad \mathcal{E}^j \in \{ \mathcal{F}^j, \mathcal{C}^j \}.
\end{equation}
As all submeshes $\mathcal{T}^{i}$, $i=1, \dots, I$ share the same
parent mesh $\mathcal{T}$ by assumption, we can use the mappings
$\mathcal{M}^i$ and $\mathcal{M}^j$ \eqref{eq:meshview-mappings}
to establish the relation between $\mathcal{C}^i$ and
$\mathcal{C}^j$. If $\mathcal{T}^i$ and $\mathcal{T}^j$ have the same
topological dimension ($d^i = d^j$),
then the cell map between the two submeshes can be
expressed directly as
\begin{equation}
  \label{eq:extra-mappings}
  \mathcal{M}^{i, j} = \left ( \mathcal{M}^j \right )^{-1} \circ \mathcal{M}^i
  : \mathcal{C}^i \longrightarrow \mathcal{C}^j .
\end{equation}

On the other hand, if $d^i \not = d^j$ i.e.~$d^i = d^j - 1$, then the
computation of $\mathcal{M}^{i, j}$ requires additional intermediate
steps. The mapping $\mathcal{M}^i : \mathcal{C}^i \rightarrow
\mathcal{F}$ \eqref{eq:meshview-mappings} gives the facet index $f_k
\in \mathcal{T}$ associated with the lower dimensional cell index $c_k
\in \mathcal{T}^i$. The mesh connectivity, relating entities of
various dimension within the same mesh, denoted as $(d-1) \rightarrow
d$ in \cite{Logg2009}, gives the indices of the (two) cells adjacent
to $f_k$ in $\mathcal{T}$. The inverse mapping $\left( \mathcal{M}^j
\right)^{-1}: \mathcal{C} \rightarrow \mathcal{C}^j$ gives their
equivalent indices in the submesh $\mathcal{T}^{j}$. Finally, the
facet $f \in \mathcal{F}^j$ shared by these cells can be found via the
mesh connectivity $d^j \rightarrow (d^j-1)$ in $\mathcal{T}^j$.  The
mapping $\mathcal{M}^{i, j} : \mathcal{C}^i \longrightarrow
\mathcal{F}^j$ \eqref{equ:extra-mapping-codim1} is then obtained:
\begin{equation}
  \label{equ:extra-mapping-codim1}
  \mathcal{C}^i \myrightarrow{\mathcal{M}^i} \mathcal{F} \myrightarrow{(d-1) \rightarrow d} \{ \mathcal{C}, \mathcal{C} \}
  \myrightarrow{\left ( \mathcal{M}^{j} \right )^{-1} \times 2} \{ \mathcal{C}^j, \mathcal{C}^j \} \myrightarrow{d^j \rightarrow (d^j-1)} \mathcal{F}^j
\end{equation}

\subsection{Nested submesh algorithms in FEniCS}
\label{subsec:nested-submeshes-algo}

In this section, we discuss algorithms for nested submeshes and
associated parent-child relationships in the context of the
FEniCS/DOLFIN finite element library.

The DOLFIN \code{Mesh} class provides data structures and algorithms
for computational meshes holding the underlying geometry and topology
through dedicated objects \code{MeshGeometry} and \code{MeshTopology}
\cite{Logg2009, LoggWells2010a}. The \code{MeshGeometry} stores the
coordinates of the mesh vertices, while the \code{MeshTopology}
defines the mesh entities (vertices, edges, facets and cells) and
their connections.
The mesh entities are labeled by pairs $e = (d, e_j)$ defining each entity $e$
from its index $e_j$ within the set of entities of topological dimension $d$.
To represent discrete functions defined over mesh entities, for instance
a map from cell indices to specific integer values, DOLFIN provides
the class(es) \code{MeshFunction}.

To efficiently represent meshes for mixed domain discretizations, we
introduce a new lightweight \code{MeshView} class. This class is
designed to allow for representing and building submeshes
$\mathcal{T}^{i}$ as new \code{Mesh} objects while storing their
relationship with the parent mesh $\mathcal{T}$. A \code{MeshView}
object links two meshes (for instance a submesh and its parent mesh,
or two submeshes) by holding pointers to the parent (or associated)
mesh $\mathcal{T}$ together with the vertex and cells maps
$\mathcal{M}_v^i$ and $\mathcal{M}^i$. Further, we let the
\code{MeshTopology} of a (sub)mesh hold a map of \code{MeshView}s with
the identifier of the parent (or associated) meshes as keys. To reduce
complexity, we consider one generation of meshes: i.e. we only support
parent-child and sibling meshes.

To construct a submesh, we assume that a \code{MeshFunction} defined
over the parent mesh encodes the selected subset of mesh entities by
an integer, referred to as a tag. The \code{MeshView} class implements
a \code{create} function which builds the submesh $\mathcal{T}^{i}$
and its child-to-parent maps $\mathcal{M}_v^i$ and $\mathcal{M}^i$
  \eqref{eq:meshview-mappings} from this \code{MeshFunction} and the
  corresponding tag. This function returns a new \code{Mesh} object,
  with a pointer to the \code{MeshView} object in its
  \code{MeshTopology}, and its use is illustrated in
  \Cref{code:meshview}.
\begin{lstlisting}[mathescape=true, language=Python,
    caption={{[Python]} Creation of the submesh $\mathcal{T}^{2}$ from the parent mesh $\mathcal{T}$
      (see \Cref{fig:build-submeshes} and \Cref{fig:extra-mappings})
      for \Cref{ex:reference-mixed-problem}
      using the \code{MeshView} class.},
    label=code:meshview]
# Define function over facets in parent mesh to represent tags
marker = MeshFunction("size_t", mesh, 1, 0)
for f in facets(mesh):
  marker[f] = 0.5 - eps < f.midpoint().x() < 0.5 + eps
# Build the submesh $\mathcal{T}^{2}$ from the facets marked as 1
submesh = MeshView.create(marker, 1)
\end{lstlisting}

In addition to mapping between mesh entities of a submesh and its
parent, mixed domain form assembly typically requires knowledge of
mesh entity mappings between different submeshes as
illustrated by \Cref{fig:extra-mappings}. We also represent
these mappings via the \code{MeshView} class, and provide a new
\code{build_mapping} function to construct these maps. In particular,
for submeshes $\mathcal{T}^i$ and $\mathcal{T}^j$ with a shared parent
mesh $\mathcal{T}$, \code{build_mapping} creates a new \code{MeshView}
object pointing to the associated mesh $\mathcal{T}^j$, a cell map
defined by~\eqref{eq:extra-mappings} and an empty vertex map by
default. This \code{MeshView} is then added to the map of
\code{MeshView}s associated with the \code{MeshTopology} of submesh
$\mathcal{T}^i$, in addition to its initial parent mesh view. Mappings
between submeshes are built (and stored) on-the-fly during mixed
domain form assembly as illustrated in \Cref{code:build-mapping}.
In particular, we do not build unnecessary mappings
$\mathcal{M}^{i, j}: \mathcal{C}^i \rightarrow \mathcal{E}^j$.
\begin{lstlisting}[mathescape=true, language=C++,
    caption={{[C++]} Illustration of additional mappings between submeshes being built on-the-fly during form assembly.
      For each subform $a^{i,j}$ (\code{_a[i][j]}), \code{build_mapping}
      is called to build the mapping between the integration mesh (\code{_a[i][j]->mesh()})
      and each basis function mesh (\code{mesh0}) unless it exists i.e.~unless
      the mapping is already listed in the \code{MeshTopology} of the integration mesh.},
    label=code:build-mapping]
// List of $a^{i,j}$ integration mesh mappings (MeshView map)
auto mesh_mapping = _a[i][j]->mesh()->topology().mapping();
// Add mapping with TestFunction mesh if needed
auto mesh0 = _a[i][j]->function_space(0)->mesh();
if(_a[i][j]->mesh() != mesh0 && !mesh_mapping.count(mesh0->id()))
  _a[i][j]->mesh()->build_mapping(mesh0);
\end{lstlisting}

\subsection{Algorithmic complexity of submesh algorithms}
\pretolerance=1000

The algorithmic complexity of the submesh construction can be
estimated as follows. The construction of each submesh
$\mathcal{T}^{i}$ requires iterating over the $n_{c}^i$ marked
entities in the parent $\mathcal{T}$.  The mapping $\mathcal{M}^i$
\eqref{eq:meshview-mappings} is then obtained and can be stored
directly. The storage of the $n_v^i$ vertices of $\mathcal{T}^{i}$ to
build the underlying \code{MeshGeometry} and the mapping
$\mathcal{M}_v^i$ \eqref{eq:meshview-mappings} requires iterating over
the local vertices of each cell $c$ in the submesh $\mathcal{T}^i$,
representing a complexity of $\mathcal{O}(n_{v}^i)$.
The \code{MeshTopology} holds the previously built mappings
$\mathcal{M}^i$ and $\mathcal{M}_v^i$ and the numbering of
$\mathcal{T}^i$ entities obtained directly by an iterative loop
over both mappings whose complexity is $\mathcal{O}(n_v^i +
n_{c}^i)$. In parallel, the \code{MeshTopology} also requires the
global numbering within the scope of the parallel computation
communicator.  Since $\mathcal{M}^i$ and $\mathcal{M}_v^i$ are
locally built on each processor, the global numbering needs additional
loops and parallel communications to establish the ownership of the
shared entities. Each additional mapping \eqref{eq:extra-mappings}
requires iterating over the $n_{c}^i$ cells of $\mathcal{T}^{i}$ to
find their counterparts in $\mathcal{T}^{j}$, $i \neq j$, hence
representing a $\mathcal{O}(n_{c}^i)$ complexity.
Taking all into account, the submesh construction scales
with the size of the mesh, and is a scalable parallel algorithm.

\subsection{MPI-parallelism of nested submeshes and mappings}
DOLFIN \cite{LoggMardalEtAl2012} is designed to be seamlessly
parallel, meaning that the same code can be used to perform both
serial and parallel simulations.  On distributed memory architectures,
the parallel support relies on the Message Passing Interface (MPI).
DOLFIN automatically performs mesh partitioning in parallel using the
libraries ParMETIS \cite{ParMetis2011} or SCOTCH \cite{Scotch1996}. Each processor
holds only a portion of the global mesh, stored as a standard
\code{Mesh} object, for which it is responsible. Data exchange between
processors then requires the computation of
local-to-global\footnote{In this section local-to-global refers to
  process-to-communicator (local to process, global to communicator).}
maps on each process.  The nested submeshes $\mathcal{T}^{i}$ are
assumed to be built from a common parent mesh $\mathcal{T}$.  When
running a mixed-dimensional simulation in parallel, the partitioning
of the submeshes stems from the partitioning of the parent mesh
$\mathcal{T}$, i.e.~no auxiliary partitioning is
performed. Thus, it is possible for a submesh to be distributed over
only some of the available processors. And vice versa, it may be that
a processor does not own any entities of a given submesh.

As described in Section~\ref{subsec:nested-submeshes-algo}, the
submeshes $\mathcal{T}^{i}$ are represented as standard \code{Mesh}
objects storing a \code{MeshView} in their \code{MeshTopology}.  As
for the global mesh $\mathcal{T}$, the submeshes $\mathcal{T}^{i}$
require a local-to-global mapping to communicate data between
processors, which implies establishing the ownership of each submesh
entity among the processors. We assume that the cells can belong to
only one partition, i.e.~we do not introduce ghost cells.
Each cell $c^i_k$ in $\mathcal{C}^{i}$ is owned by the processor
owning the corresponding entity $\mathcal{M}^i(c^i_k)$ in
$\mathcal{T}$. However, the vertices located at the interface between
partitions are shared by a set of processors. Among these, the
vertices are assumed to be owned by the processor with the lowest
rank. This processor holds the underlying local-to-global mapping and
sends it to the other processors involved. Thus, a cell owned by a
processor of rank $i$ may have vertices owned by a processor of rank
$j$, where $j<i$.

\section{Form language abstractions and algorithms for mixed domains}
\label{sec:ufl}

The Unified Form Language (UFL) \cite{Alnaes2014, AlnaesBlechta2015a}
is a domain-specific language for finite element spaces, tensor
algebra and variational forms. It provides a flexible interface for
defining variational formulations of differential equations, through
abstractions closely mimicking the mathematical syntax. UFL includes a
set of predefined base finite element families, including but not
limited to Lagrange~\cite{brenner2007mathematical}, Discontinuous
Galerkin~\cite{Arnold2000}, Raviart--Thomas~\cite{RaviartThomas1977},
Brezzi-Douglas-Marini~\cite{BrezziEtAl1985},
N\'edelec~\cite{Nedelec1980, Nedelec1986}, of arbitrary polynomial
dimension. The UFL finite element definition mimics that of
Ciarlet~\cite{Ciarlet1976}, and in particular, a finite element is
defined relative to a reference element (and not to a mesh). Mixed
finite elements can be defined as Cartesian products of the base
element families, assuming that all subelements share a common
reference cell. A UFL function space is defined by a pairing of a
(mixed) finite element and a domain (representing e.g.~the
mesh). However, for mixed domain and dimensional problems, these
abstractions are not sufficient.

To extend UFL with abstractions for mixed domain variational problems,
we advocate a lightweight approach, essentially representing mixed
function spaces (in contrast to function spaces over mixed elements)
as tuples of function spaces. This design choice mirrors our design
choice for finite element assembly of mixed domain variational forms
using block tensors. We detail the new UFL abstractions for mixed
(domain) function spaces and integration in
Sections~\ref{sec:ufl:abstractions} and~\ref{subsec:measures} below.
To facilitate mixed domain assembly, we have also extended UFL with
new algorithms for splitting mixed domain variational forms into sums
of subforms, described in Section~\ref{sec:ufl:algorithms}. Finally,
we describe UFL form validation and typical non-admissible operations
in Section~\ref{subsec:UFL-form-verification}.

\subsection{Mixed function spaces and functions}
\label{sec:ufl:abstractions}

To represent a mixed domain discrete function space $U = U^1 \times
\dots \times U^I$ composed of a finite number $I$ of finite element
function spaces $U^i$ for $i = 1, \dots, I$, we introduce a
new\footnote{The keyword \code{MixedFunctionSpace} existed in previous versions of
  UFL, but was deprecated in version 2016.1.0. It has
  now been reintroduced in a more generic context
  handling mixed domain and mixed dimensional function spaces.} UFL class
\code{MixedFunctionSpace}.  This lightweight class simply holds a
tuple of the component spaces $(U^1, \dots, U^I)$, and sample usage is
provided in \Cref{code:ufl-function-space-product}. The key
operational aspect of the \code{MixedFunctionSpace} abstraction is the
identification of the relative position of a subfunction space and
argument within the product space.  
\begin{lstlisting}[mathescape=true, language=Python,
    caption={{[Python]} Example of \code{MixedFunctionSpace} usage in UFL/DOLFIN to define
      a mixed function space with two component spaces defined relative
      to different meshes.},
    label=code:ufl-function-space-product]
# U = U^1 x U^2
U1 = FunctionSpace(mesh, "CG", 2)
U2 = FunctionSpace(submesh, "DG", 1)
U = MixedFunctionSpace(U1, U2)
\end{lstlisting}

UFL distinguishes between two types of functions appearing in
variational forms: (i) \code{Argument}s representing the basis functions
for a function space and (ii) \code{Coefficient}s representing any function
in a function space, i.e.~a weighted linear combination of basis
functions. A \code{TrialFunction} and \code{TestFunction} represent
pre-indexed \code{Argument}s with index corresponding to the order of
the argument in the form(s). UFL assumes that a variational form is
always linear in its \code{Argument}s but possibly nonlinear in its
\code{Coefficient}s. To define test and trial functions on a mixed
function space, and arguments in general, the syntax
\code{TestFunctions}, \code{TrialFunctions} and \code{Arguments} have
been adopted. This syntax is illustrated in \Cref{code:arguments}
below. These operators, when acting on a \code{MixedFunctionSpace}
with $I$ subspaces, return a tuple of basis functions $u^i$ for $i =
1, \dots, I$, as a list of
\code{Argument} objects embedding the block index $i$ of the
underlying function spaces $U^i$.
\begin{lstlisting}[mathescape=true, language=Python,
    caption={{[Python]} Basis functions (\code{Arguments}, and more specifically \code{TestFunctions} and \code{TrialFunctions}),
    or any functions (\code{Coefficients}) of a \code{MixedFunctionSpace} are defined as a list of functions embedding the block
    index of the underlying function space.},
    label=code:arguments]
(u0, [...], uI) = TrialFunctions(U) # $u = (u^1, \dots, u^I)$
(v0, [...], vI) = TestFunctions(U)  # $v = (v^1, \dots, v^I)$
(w0, [...], wI) = Arguments(U)      # $w = (w^1, \dots, w^I)$
(f0, [...], fI) = Coefficients(U)   # $f = (f^1, \dots, f^I)$
\end{lstlisting}

\subsection{Measures for mixed domain variational forms}
\label{subsec:measures}

In UFL \cite{Alnaes2014}, integrals are expressed through
multiplication (\code{*}) by a measure representing the integral type.
The main integral types are: \code{dx} for integrals over the interior
of the domain, \code{ds} for the exterior facets i.e.~the integrals
over the boundary, and \code{dS} for the set of interior
facets. Integrals over different parts of the domain can be expressed
using markers given as an optional parameter \code{subdomain_data} and
specifying the corresponding tag in the form expression.  The terminal
operands involved in the form integrals are \code{Argument}s and
\code{Coefficient}s, carrying their associated function space(s) and
thereby the associated mesh(es) (see \Cref{code:arguments}). When the
form arguments belong to the same function space i.e.~for
monodomain problems or for diagonal block forms, only one mesh is
involved and the integration domain can thus be deduced without
explicit definition by the measure.

For off-diagonal block forms, involving function spaces defined over
different submeshes, the integration domain must be explicitly
specified by measure (re)definition. Typical usage is illustrated in
\Cref{code:measure-definition}. 
\begin{lstlisting}[mathescape=true, language=Python,
    caption={{[Python]} Example of redefining measures to define the integration
      domain of a form integral, giving explicitly the integral type,
      the integration mesh and possibly a marker to handle integrals
      over different parts of the domain.},
    label=code:measure-definition]
dx1 = Measure("dx", domain=U1.mesh()) # Integral over $\Omega^1$
dx2 = Measure("dx", domain=U2.mesh(), subdomain_data=...) # $\Omega^2$
ds1 = Measure("ds", domain=U1.mesh()) # $\partial \Omega^1$
\end{lstlisting}
When a measure is used in a form (integral), we assume that the
measure's domain and the function space mesh of at least one of the
form arguments coincide. Further, the redefined measures should define
integration over cells for a lower dimensional mesh rather than
integration over facets for a higher dimensional mesh.

With the measures defined in \Cref{code:measure-definition}, we can
express the variational formulation of the Poisson problem introduced
in \Cref{ex:reference-mixed-problem} as follows
(\Cref{code:poisson-vf-def}).
\begin{lstlisting}[mathescape=true, language=Python,
    caption = {{[Python]} UFL/DOLFIN implementation of the variational form
      \eqref{eq:Poisson-LM-weakform} from
      \Cref{ex:reference-mixed-problem} with explicit specification of
      the integration domains.}, label=code:poisson-vf-def]
a = inner(grad(u),grad(v))*dx1 + v*l*dx2(1) + u*e*dx2(1)
\end{lstlisting}

\subsection{Mixed domain variational form algorithms}
\label{sec:ufl:algorithms}

A key advantage of UFL and similar domain specific languages is the
ability to manipulate e.g.~variational forms at the symbolic level. In
the context of mixed domain variational forms, a key operation is to
extract subform blocks i.e.~to compute a decomposition into subforms
(such as e.g.~\eqref{eq:decomposition-mixed-form-linear}) of a mixed
domain form. For instance, to assemble mixed domain variational forms,
we advocate a block-by-block approach for the sake of flexibility,
efficiency and reuse. This is also the approach considered by
\cite{multiphenics, 2017FenicsII}. For automated block-by-block
assembly, the automated extraction of subforms from a variational form
defined over a mixed function space is convenient.

To extract the subforms $a^{i,j}$ from a bilinear form $a$ (and the
analogous for linear forms), we have introduced a UFL function
\code{extract_blocks}. Its underlying algorithm relies on the directed
acyclic graph (DAG) representation used by UFL \cite{Alnaes2014,
  LoggMardalEtAl2012} to represent the form integrands. For any given
bilinear form $a$, its DAG expression tree allows the identification
and extraction of the terms involving the pair $(U^j, U^i)$ of
subspaces as the subform $a^{i, j}$, given the corresponding indexing
$i, j = 1, \dots, I$. This algorithm relies on the embedding of the
block index with the \code{MixedFunctionSpace}. The function
\code{extract_blocks} can either return the whole list of subforms of
a given form, or a specific subform $a^{i,j}$ given the indices
$(i,j)$. A code example demonstrating the usage of this function is
presented in \Cref{code:extract-blocks} below.

\begin{lstlisting}[mathescape=true,language=Python,
    caption={{[Python]} The \code{extract_blocks} function is used to extract the
      subforms $a^ {i,j}$ of the mixed domains form $a(u,v) = \sum_i
      \sum_j a^{i,j}(u^{j}, v^{i})$ from the arguments indexing.  The
      subform $a^{1,2}$ from \Cref{ex:reference-mixed-problem}
      implemented as in \Cref{code:poisson-vf-def} can be obtained
      using \code{a_12 = extract_blocks(a,0,1)}. Note again that the
      indices start at $0$ in the code.}, label=code:extract-blocks]
# $a(u,v) = \sum_i \sum_j a^{i,j}(u^{j}, v^{i})$
a = u0*v0*dx0 + ... ui*vj*dxi + ... + uI*vI*dxI
# as $\equiv [a^{0,0}, \dots, a^{i,j}, a^{i,j+1}, \dots, a^{i+1,j} \dots, a^{I,I}]$
as = extract_blocks(a)
# a_ij $\equiv a^{i,j}(u^{j}, v^{i})$
a_ij = extract_blocks(a,i,j)
\end{lstlisting}

\subsection{UFL mixed domain form verification}
\label{subsec:UFL-form-verification}
The following code checks have been introduced to prevent confusion or
misuse of the mixed domain features. All UFL verification assertions
for single domain variational forms have been extended to mixed domain
forms by application to each block subform.

Regarding the mixed function spaces definition, a
\code{MixedFunctionSpace} is not a \code{FunctionSpace}, but rather a
list of \code{FunctionSpace} objects. Arguments
(resp. coefficients/functions) defined from a
\code{MixedFunctionSpace} form a list of \code{Argument}s
(resp. \code{Coefficient}s) corresponding to each block.  Thus, only
the plural version of the related keywords are allowed:
\code{TrialFunction(V)} is not allowed when \code{V} is a
\code{MixedFunctionSpace}.
Instead, \code{TrialFunctions(V)} should be used.

The coupling of arguments and/or functions from different function
spaces in a form requires the underlying objects to be defined from a
\code{MixedFunctionSpace}. In other words, combining
\code{TrialFunction(V1)} and \code{TestFunction(V2)} with \code{V1}
and \code{V2} defined as different \code{FunctionSpace}s is not
supported. One should instead introduce \code{V} as a
\code{MixedFunctionSpace(V1,V2)} and define the arguments
(\code{TestFunctions(V)} and \code{TrialFunctions(V)}) from the
latter.

The assembly of off-diagonal blocks combining arguments from different
function spaces requires a mapping between the (sub)meshes involved,
as discussed in \Cref{sec:meshview}. To define this mapping, we assume
that the submeshes share a common parent mesh. This assertion is
checked at the mesh data structure level, e.g.~when building mappings
between the submeshes within the \code{build_mapping}
function. Finally, a form integral is not valid if the integration
mesh defined through the integral's measure does not coincide with one
of the meshes associated with the form arguments.

\section{Mixed domain function spaces and degrees of freedom}
\label{sec:dofs}

In this section, we discuss the transfer of local-to-global
degree-of-freedom maps between submeshes. These maps are used in mixed
domain assembly algorithms, described in detail in
Section~\ref{sec:assembly}.

Consider a mesh $\mathcal{T}^i = \{K \}$ and a finite element space
$U^i$ defined relative to $\mathcal{T}^i$. The standard
local-to-global mapping $\iota_{c^i}^{i}$ for the finite element space
$U^i$ and a cell $K \in \mathcal{T}^i$ with cell index $c^i \in
\mathcal{C}^i$ maps the set of local basis function indices to the
corresponding global indices:
\begin{equation}
  \label{equ:local-to-global-dofs}
  \iota_{c^i}^{i} : \{1, \dots, \N_{K}^{i} \} \rightarrow \{1, \dots \N^i \} ,
\end{equation}
where $\N^i$ denotes the (global) dimension of $U^i$ and $\N_{K}^i$
denotes the (local) dimension of $U^i|_{K}$ for each $K$, see
e.g.~\cite{LoggMardalEtAl2012} for more details.  We assume that the
local-to-global map $\iota^i_{c^i}$ is available for each submesh
$\mathcal{T}^i$.

\begin{figure}[!ht]
  \centering
  \begin{subfigure}[b]{0.3\textwidth}
    \centering
    \begin{tikzpicture}[scale=0.4]
      \draw[blue!75] (-4.0,2.0) -- (4.0,2.0);
      \draw[blue!75] (4.0,2.0) -- (4.0,6.0);
      \draw[blue!75] (-4.0,6.0) -- (-4.0,2.0);
      \draw[blue!75] (-4.0,6.0) -- (4.0,6.0);
      \draw[blue!75] (0.0,2.0) -- (4.0, 6.0);
      \draw[blue!75] (0.0, 2.0) -- (0.0, 6.0);
      \draw[blue!75] (-4.0,2.0) -- (0.0, 6.0);
      \draw[dashed,black!50] (-4.0,10.0) -- (-4.0,6.0);
      \draw[dashed,black!50] (4.0,10.0) -- (4.0,6.0);
      \draw[dashed,black!50] (4.0,10.0) -- (-4.0,10.0);
      \draw[dashed,black!50] (-4.0,6.0) -- (0.0, 10.0);
      \draw[dashed,black!50] (0.0,6.0) -- (4.0, 10.0);
      \draw[dashed,black!50] (0.0,10.0) -- (0.0,6.0);
      \draw[blue] (0,0) [fill] (-4,2) circle [radius=4pt];
      \draw[blue] (0,0) [fill] (-4,6) circle [radius=4pt];
      \draw[black!50] (0,0) [fill] (-4,10) circle [radius=4pt];
      \draw[blue] (0,0) [fill] (0,2) circle [radius=4pt];
      \draw[blue] (0,0) [fill] (0,6) circle [radius=4pt];
      \draw[black!50] (0,0) [fill] (0,10) circle [radius=4pt];
      \draw[blue] (0,0) [fill] (4,2) circle [radius=4pt];
      \draw[blue] (0,0) [fill] (4,6) circle [radius=4pt];
      \draw[black!50] (0,0) [fill] (4,10) circle [radius=4pt];
      \draw (-4, 6) node[above left] {\textcolor{blue}{\small{$0$}}};
      \draw (0, 6) node[above left] {\textcolor{blue}{\small{$1$}}};
      \draw (4, 6) node[above left] {\textcolor{blue}{\small{$2$}}};
      \draw (-4, 2) node[below left] {\textcolor{blue}{\small{$3$}}};
      \draw (0, 2) node[below left] {\textcolor{blue}{\small{$4$}}};
      \draw (4, 2) node[below left] {\textcolor{blue}{\small{$5$}}};
      \draw[loosely dashed, black,line width=0.5mm] (-4.0, 2.0) -- (0.0, 2.0);
      \draw[loosely dashed, black,line width=0.5mm] (0.0, 2.0) -- (0.0, 6.0);
      \draw[loosely dashed, black,line width=0.5mm] (0.0, 6.0) -- (-4.0, 2.0);
      \draw (-4,2) node[above, xshift = 0.4cm, yshift = -0.05cm] {\footnotesize $0$};
      \draw (0,6) node[below, xshift = -0.1cm, yshift = -0.1cm] {\footnotesize $1$};
      \draw (0,2) node[above, xshift = -0.2cm, yshift = -0.05cm] {\footnotesize $2$};
      \draw (-1.25,3.5) node[] {\small $c^1$};
    \end{tikzpicture}
    \caption{$K$ in $\mathcal{T}^1$}
    \label{fig:dof-local-to-global-1}
  \end{subfigure}
  \begin{subfigure}[b]{0.3\textwidth}
    \centering
    \begin{tikzpicture}[scale=0.4]
      \draw[DarkGreen!75] (-4.0,2.0) -- (0.0,2.0);
      \draw[DarkGreen!75] (0.0,10.0) -- (-4.0,10.0);
      \draw[DarkGreen!75] (-4.0,10.0) -- (-4.0,2.0);
      \draw[DarkGreen!75] (-4.0,6.0) -- (0.0,6.0);
      \draw[DarkGreen!75] (-4.0,6.0) -- (0.0, 10.0);
      \draw[DarkGreen!75] (-4.0,2.0) -- (0.0, 6.0);
      \draw[DarkGreen!75] (0.0, 2.0) -- (0.0, 6.0);
      \draw[DarkGreen!75] (0.0,6.0) -- (0.0, 10.0);
      \draw[dashed,black!50] (0.0,6.0) -- (4.0, 10.0);
      \draw[dashed,black!50] (0.0,2.0) -- (4.0, 6.0);
      \draw[dashed,black!50] (0.0,2.0) -- (4.0,2.0);
      \draw[dashed,black!50] (0.0,6.0) -- (4.0,6.0);
      \draw[dashed,black!50] (4.0,10.0) -- (0.0,10.0);
      \draw[dashed,black!50] (4.0,2.0) -- (4.0,10.0);
      \draw[DarkGreen] (0,0) [fill] (-4,2) circle [radius=4pt];
      \draw[DarkGreen] (0,0) [fill] (-4,6) circle [radius=4pt];
      \draw[DarkGreen] (0,0) [fill] (-4,10) circle [radius=4pt];
      \draw[DarkGreen] (0,0) [fill] (0,2) circle [radius=4pt];
      \draw[DarkGreen] (0,0) [fill] (0,6) circle [radius=4pt];
      \draw[DarkGreen] (0,0) [fill] (0,10) circle [radius=4pt];
      \draw[black!50] (0,0) [fill] (4,2) circle [radius=4pt];
      \draw[black!50] (0,0) [fill] (4,6) circle [radius=4pt];
      \draw[black!50] (0,0) [fill] (4,10) circle [radius=4pt];
      \draw (-4, 10) node[above left] {\textcolor{DarkGreen}{\small{$0$}}};
      \draw (0, 10) node[above left] {\textcolor{DarkGreen}{\small{$1$}}};
      \draw (-4, 6) node[above left] {\textcolor{DarkGreen}{\small{$2$}}};
      \draw (0, 6) node[above left] {\textcolor{DarkGreen}{\small{$3$}}};
      \draw (-4, 2) node[below left] {\textcolor{DarkGreen}{\small{$4$}}};
      \draw (0, 2) node[below left] {\textcolor{DarkGreen}{\small{$5$}}};
      \draw[loosely dashed, black,line width=0.5mm] (-4.0, 2.0) -- (0.0, 2.0);
      \draw[loosely dashed, black,line width=0.5mm] (0.0, 2.0) -- (0.0, 6.0);
      \draw[loosely dashed, black,line width=0.5mm] (0.0, 6.0) -- (-4.0, 2.0);
      \draw (-4,2) node[above, xshift = 0.4cm, yshift = -0.05cm] {\footnotesize $0$};
      \draw (0,6) node[below, xshift = -0.1cm, yshift = -0.1cm] {\footnotesize $1$};
      \draw (0,2) node[above, xshift = -0.2cm, yshift = -0.05cm] {\footnotesize $2$};
      \draw (-1.25,3.5) node[] {\small $c^2$};
    \end{tikzpicture}
    \caption{$K$ in $\mathcal{T}^2$}
    \label{fig:dof-local-to-global-2}
  \end{subfigure}
  \begin{subfigure}[b]{0.3\textwidth}
    \centering
    \begin{tikzpicture}[scale=0.4]
      \draw[black!50] (-4.0,2.0) -- (0.0,2.0);
      \draw[black!50] (0.0,6.0) -- (4.0,6.0);
      \draw[black!50] (-4.0,2.0) -- (4.0, 6.0);
      \draw[black!50] (0.0,2.0) -- (4.0, 6.0);
      \draw[red!75] (0.0, 2.0) -- (0.0, 6.0);
      \draw[red!75] (0.0,10.0) -- (0.0,6.0);
      \draw[dashed,black!50] (0.0,2.0) -- (4.0,2.0);
      \draw[dashed,black!50] (4.0,2.0) -- (4.0,6.0);
      \draw[dashed,black!50] (-4.0,6.0) -- (-4.0,2.0);
      \draw[dashed,black!50] (-4.0,6.0) -- (0.0,6.0);
      \draw[dashed,black!50] (-4.0,10.0) -- (-4.0,6.0);
      \draw[dashed,black!50] (4.0,10.0) -- (4.0,6.0);
      \draw[dashed,black!50] (4.0,10.0) -- (-4.0,10.0);
      \draw[dashed,black!50] (-4.0,6.0) -- (0.0, 10.0);
      \draw[dashed,black!50] (0.0,6.0) -- (4.0, 10.0);
      \draw[draw=black!50, fill=red!10]  (-4,2) -- (0,2) -- (0,6) -- cycle;
      \draw[draw=black!50, fill=red!10]  (0,2) -- (0,6) -- (4,6) -- cycle;
      \draw[red!75] (0.0, 2.0) -- (0.0, 6.0);
      \draw[black!50] (-4.0,2.0) -- (0.0, 2.0);
      \draw[black!50] (0.0,6.0) -- (4.0, 6.0);
      \draw[black!50] (0.0,2.0) -- (4.0, 6.0);
      \draw[black!50] (-4.0,2.0) -- (0.0, 6.0);
      \draw[black!50] (0,0) [fill] (-4,2) circle [radius=4pt];
      \draw[black!50] (0,0) [fill] (-4,6) circle [radius=4pt];
      \draw[black!50] (0,0) [fill] (-4,10) circle [radius=4pt];
      \draw[red!75] (0,0) [fill] (0,2) circle [radius=4pt];
      \draw[red!75] (0,0) [fill] (0,6) circle [radius=4pt];
      \draw[red!75] (0,0) [fill] (0,10) circle [radius=4pt];
      \draw[black!50] (0,0) [fill] (4,2) circle [radius=4pt];
      \draw[black!50] (0,0) [fill] (4,6) circle [radius=4pt];
      \draw[black!50] (0,0) [fill] (4,10) circle [radius=4pt];
      \draw (0, 10) node[above left] {\textcolor{red!75}{\small{$0$}}};
      \draw (0, 6) node[above left] {\textcolor{red!75}{\small{$1$}}};
      \draw (0, 2) node[below left] {\textcolor{red!75}{\small{$2$}}};
      \draw[loosely dashed, black,line width=0.5mm] (0.0, 2.0) -- (0.0, 6.0);
      \draw (0,6) node[below, xshift = -0.15cm, yshift = -0.1cm] {\footnotesize $0$};
      \draw (0,2) node[above, xshift = -0.15cm, yshift = -0.05cm] {\footnotesize $1$};
      \draw (-0.6,3.75) node[] {\small $c^3$};
    \end{tikzpicture}
    \caption{$K$ in $\mathcal{T}^3$}
    \label{fig:dof-local-to-global-3}
  \end{subfigure}  
  \caption{Illustration of submesh mappings and local-to-global
    degrees of freedom numbering.
    \subref{fig:dof-local-to-global-1} The submesh $\mathcal{T}^1$
    defined as the lower half of $\mathcal{T}$ with global degrees of
    freedom numbering (in blue) of $U^1(\mathcal{T}^1)$.  The cell $K$
    with index $c^1 \in \mathcal{C}^1$ has $N^1_K = 3$ local degrees
    of freedom. The local-to-global relationship
    \eqref{eq:local-to-global-dofs} gives $\iota^i_{c^1}(0) = 3$,
    $\iota^i_{c^1}(1) = 1$ and $\iota^i_{c^1}(2) = 4$.
    \subref{fig:dof-local-to-global-2} The submesh $\mathcal{T}^2$
    defined as the left half of $\mathcal{T}$ with global degrees of
    freedom numbering (in green) of $U^2(\mathcal{T}^2)$.  The cell
    $K$ with index $c^2 \in \mathcal{C}^2$ has $N^2_K = 3$ local
    degrees of freedom (in black) with $\iota^i_{c^2}(0) = 4$,
    $\iota^i_{c^2}(1) = 3$ and $\iota^i_{c^2}(2) = 5$.
    \subref{fig:dof-local-to-global-3}
    The lower-dimensional submesh $\mathcal{T}^3$ defined
    as the middle vertical line of $\mathcal{T}$ with global degrees of
    freedom numbering (in red) of $U^3(\mathcal{T}^3)$.
    The cell $K$ with index $c^3 \in \mathcal{C}^3$ has $N^3_K = 2$ local
    degrees of freedom (in black) with $\iota^i_{c^3}(0) = 1$
    and $\iota^i_{c^3}(1) = 2$.
    The star $S_{K}$ of $K$ is composed of two cells (in light red).}
  \label{fig:dof-local-to-global}
\end{figure}
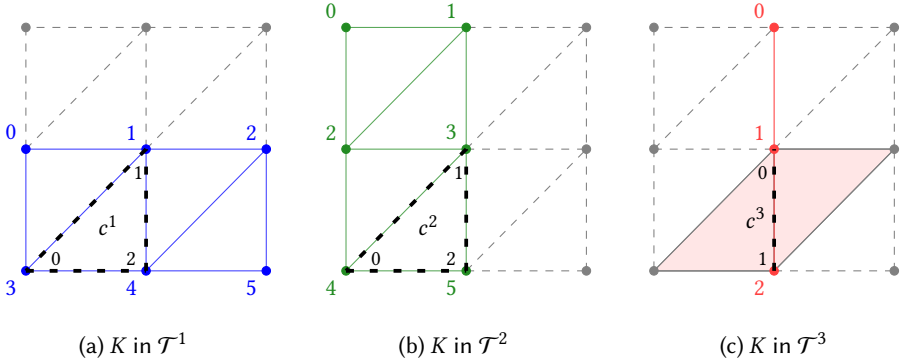

First, consider the case of two submeshes $\mathcal{T}^1$ and
$\mathcal{T}^2$ with a parent mesh $\mathcal{T}$ with $d = d^1 = d^2$
as illustrated in \Cref{fig:dof-local-to-global}.  Consider two
function spaces $U^1 = U^1(\mathcal{T}^1)$ and $U^2 =
U^2(\mathcal{T}^2)$.  For each $K \in \mathcal{T}^1$ (resp. $K \in
\mathcal{T}^2$) with index $c^1 \in \mathcal{C}^1$ (resp. $c^2 \in
\mathcal{C}^2$), we can use the mapping $\mathcal{M}^1$
(resp. $\mathcal{M}^2$) to access the cell index $c$ relative to the
parent mesh $\mathcal{T}$:
\begin{equation}
  c = \mathcal{M}^1(c^1) = \mathcal{M}^2(c^2).
\end{equation}
When there is a coupling between $\mathcal{T}^1$ and $\mathcal{T}^2$,
the mapping $\mathcal{M}^{2, 1}$ can be used to get the cell index 
$c^1 \in \mathcal{C}^1$ of $K$ relative to $\mathcal{T}^1$ from
its index $c^2 \in \mathcal{C}^2$ relative to $\mathcal{T}^2$:
\begin{equation}
  c^1 = \mathcal{M}^{2, 1}(c^2). 
\end{equation}
Subsequently, we can define the global index $n$ for the local
(degree-of-freedom) index $l$ for $K$ in $U^i$ via
\begin{equation}
  \label{eq:local-to-global-dofs}
  n = \iota^i_{c^i}(l) ,
\end{equation}
for $l = 1, \dots, N^i_{K}$.

Next, consider the case of two submeshes $\mathcal{T}^1$ and
$\mathcal{T}^3$ with a parent mesh $\mathcal{T}$ with $d = d^1 > d^3$
and a function space $U^1$ defined relative to $\mathcal{T}^1$. For
each $K \in \mathcal{T}^3$, we define its star $S_{K}$
as the set of cells in $\mathcal{T}$ containing $K$ (see
\Cref{fig:dof-local-to-global-3}).  Take $\tilde K \in S_{K}$ and let
$c \in \mathcal{C}$ be its cell index relative to $\mathcal{T}$. By
stipulation, $\tilde K$ is also a cell in $\mathcal{T}^1$, but its cell index
$c^1$ relative to $\mathcal{T}^1$ is given by
\begin{equation*}
  c^1 = (\mathcal{M}^1)^{-1}(c) .
\end{equation*}
This relation can thus be used to transfer local-to-global maps of
degrees of freedom.

\section{Mixed domain finite element assembly}
\label{sec:assembly}

The finite element assembly of a variational form $a$ of arity $r$ is
the computation of the $r$-tensor $A$ resulting from evaluating the
variational form over its range of basis functions. A variational form
can consist of multiple integrals, each with its own integration
domain. A typical finite element assembly algorithm iterates over the
cells $K$ of each integration domain $\mathcal{K}$ to compute the
global finite element tensor $A$ by (i) evaluating local (cell-wise)
element tensors $A^K$ and (ii) inserting (or adding) these into the
global tensor via a local-to-global degree-of-freedom mapping. For
more details on finite element assembly in general, see
e.g.~\cite{LoggMardalEtAl2012}.

\subsection{Mixed domain assembly challenges and discussion of approach}

In single domain finite element assembly, the basis functions and
coefficients are defined on one mesh $\mathcal{T}$ and the integration
domains are defined relative to this mesh. For the assembly of mixed
domain variational forms, we here consider a block-by-block approach
as illustrated by e.g.~the decomposition
\eqref{eq:decomposition-mixed-form-linear} and the resulting block
linear system \eqref{equ:block-shaped-system-linear} for bilinear
forms and the general \eqref{eq:block:decomposition:general}. In
particular, we assemble each integral of each block form
separately. Diagonal block forms are defined relative to a single
domain, and can thus be assembled using standard single domain
assembly algorithms. We therefore do not discuss these further here,
but rather focus on the off-diagonal blocks. These present a number of
additional challenges:
\begin{itemize}
\item
  Assembly of off-diagonal blocks requires knowledge of the relationships
  between the integration mesh given by the form measure and the
  meshes involved in the trial and test spaces. These relationships
  are obtained through mappings between the parent-child or
  sibling meshes as discussed in~\Cref{sec:meshview}.
\item
  New techniques are required for the evaluation (and form
  compilation) of local element tensors over function spaces defined
  over different domains and dimensions.
\end{itemize}

In single domain finite element assembly, each local element tensor
corresponds to the contribution from a single element. In mixed
dimensional finite element assembly, the finite element tensor can
again be formed by combining local element tensor
contributions. However, the local element tensor concept is more
multifaceted. Below, we introduce two local tensor concepts for mixed
dimensional variational forms: the composite local element tensor and
the local element tensor.

\subsection{Mixed domain assembly of cell integrals}

We analyze the assembly of a mixed domain variational block form of
arity $r$ in further detail, using a bilinear form ($r = 2$) as a
guiding case. The discussion is analogous for general
$r$-forms. Consider an off-diagonal block form $a^{i, j} : U^j \times
U^i \rightarrow \R$ for a fixed $i \not = j$ and assume without loss
of generality that $\Omega^i \cap \Omega^j \not = \emptyset$, but that
$\Omega^i \not = \Omega^j$. To alleviate notation, we just write $a =
a^{i, j}$, and set $i = 1$ and $j = 2$, again without loss of
generality. We further assume that $a$ represents a single integral,
cf.~Section~\ref{sec:integration:domains}, noting that sums of
integrals are easily handled. Thus, we have that
\begin{equation}
  \label{eq:sum:K}
  a: U^2 \times U^1 \rightarrow \R, \quad
  a(\cdot, \cdot) = \sum_{K \in \mathcal{K}} a^K(\cdot, \cdot),
\end{equation}
where $\mathcal{K}$ is the integration domain of dimension $d^\mathcal{K}$,
assumed to be (a subset of) the cells in either $\mathcal{T}^1$ or $\mathcal{T}^2$
(cf.~Section~\ref{subsec:measures}). Specifically, we need to evaluate
\begin{equation}
  \label{eq:mixed-form-blocks-local}
  A^{K}_{m,n} = a^{K}(\phi^{2}_{n}, \phi^{1}_{m}),
\end{equation}
for all $m = 1, \dots, \N^{1}$, $n = 1, \dots, \N^{2}$.

\subsubsection{Case 1 (homogeneous dimension) $d^{\mathcal{K}} = d^1 = d^2$}

If $\Omega^1$ and $\Omega^2$ both have the topological dimension of
the integration mesh $d^{\mathcal{K}} = d^1 = d^2 \leq d$, the
finite element spaces $U^1$ and $U^2$ can be defined over the same
reference cell. In this case, the assembly of the local tensors
$A^{K}$~\eqref{eq:mixed-form-blocks-local} can be handled by standard
techniques. However, the insertion into the global tensor $A$ requires
knowledge of the global degree of freedom numberings $n^1$ and $n^2$
relative to $U^1$ and $U^2$, respectively. These indices are obtained
through the local-to-global mappings \eqref{eq:local-to-global-dofs}
as illustrated in \Cref{alg:samedim-assembly}.
\begin{algorithm}[H]
  \caption{Assembly of bilinear cell integrals over homogeneous dimensions}
  \label{alg:samedim-assembly}
  \begin{algorithmic}[1]
    \For{$K$ in $\mathcal{K}$}
    \State{Compute the cell indices $c^1$ and $c^2$ of $K$ relative to $\mathcal{T}^1$ and $\mathcal{T}^2$, respectively}
    \State{Compute $A_{K}$}
    \For{$l_1 \gets 1, \dots, N^1_K$ and $l_2 \gets 1, \dots, N^2_K$}
    \State{Compute $m = \iota_{c^1}^1(l_1)$ and $n = \iota_{c^2}^2(l_2)$}
    \State{Add entry $(l_1, l_2)$ of $A_{K}$ to $A$ at entry $(m, n)$}
    \EndFor
    \EndFor
  \end{algorithmic}
\end{algorithm}

\subsubsection{Case 2 (codimension one) $d^{\mathcal{K}} = d^2 = d^1 - 1$}

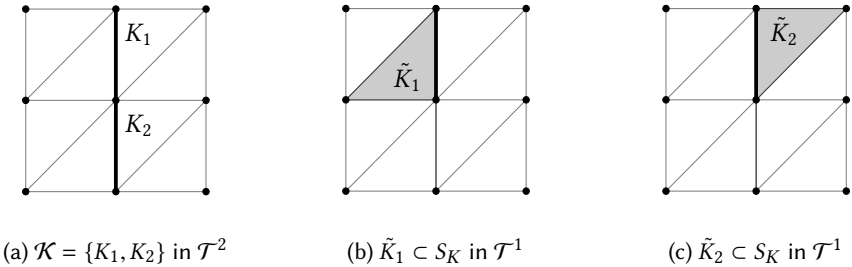
\begin{figure}[!ht]
  \centering
    \begin{subfigure}[c]{0.3\textwidth}
    \centering
    \begin{tikzpicture}[scale=0.3]
      \draw[black!50] (-4.0,2.0) -- (4.0,2.0);
      \draw[black!50] (4.0,2.0) -- (4.0,10.0);
      \draw[black!50] (4.0,10.0) -- (-4.0,10.0);
      \draw[black!50] (-4.0,10.0) -- (-4.0,2.0);
      \draw[black] (0.0, 2.0) -- (0.0, 6.0);
      \draw[black!50] (-4.0,6.0) -- (4.0,6.0);
      \draw[black!50] (-4.0,6.0) -- (0.0, 10.0);
      \draw[black!50] (0.0,6.0) -- (4.0, 10.0);
      \draw[black!50] (-4.0,2.0) -- (0.0, 6.0);
      \draw[black!50] (0.0,2.0) -- (4.0, 6.0);
      \draw (0,0) [fill] (-4,2) circle [radius=4pt];
      \draw (0,0) [fill] (-4,6) circle [radius=4pt];
      \draw (0,0) [fill] (-4,10) circle [radius=4pt];
      \draw (0,0) [fill] (0,2) circle [radius=4pt];
      \draw (0,0) [fill] (0,6) circle [radius=4pt];
      \draw (0,0) [fill] (0,10) circle [radius=4pt];
      \draw (0,0) [fill] (4,2) circle [radius=4pt];
      \draw (0,0) [fill] (4,6) circle [radius=4pt];
      \draw (0,0) [fill] (4,10) circle [radius=4pt];
      \draw[black,line width=0.5mm] (0.0, 2.0) -- (0.0, 10.0);
      \draw (0,8) node[above right] {$K_1$};
      \draw (0,4) node[above right] {$K_2$};
    \end{tikzpicture}
    \caption{$\mathcal{K} = \{ K_1, K_2 \}$ in $\mathcal{T}^2$}
    \label{fig:local-tensors-mixed-dim-1}
  \end{subfigure}
  \begin{subfigure}[c]{0.3\textwidth}
    \centering
    \begin{tikzpicture}[scale=0.3]
      \draw[black!50] (-4.0,2.0) -- (4.0,2.0);
      \draw[black!50] (4.0,2.0) -- (4.0,10.0);
      \draw[black!50] (4.0,10.0) -- (-4.0,10.0);
      \draw[black!50] (-4.0,10.0) -- (-4.0,2.0);
      \draw[black] (0.0, 2.0) -- (0.0, 6.0);
      \draw[black!50] (-4.0,6.0) -- (4.0,6.0);
      \draw[black!50] (-4.0,6.0) -- (0.0, 10.0);
      \draw[black!50] (0.0,6.0) -- (4.0, 10.0);
      \draw[black!50] (-4.0,2.0) -- (0.0, 6.0);
      \draw[black!50] (0.0,2.0) -- (4.0, 6.0);
      \draw[black!75, fill=black!20]  (-4,6) -- (0,6) -- (0,10) -- cycle;
      \draw (-1.25,7) node[] {$\tilde K_{1}$};
      \draw[black,line width=0.5mm] (0.0, 6.0) -- (0.0, 10.0);
      \draw (0,0) [fill] (-4,2) circle [radius=4pt];
      \draw (0,0) [fill] (-4,6) circle [radius=4pt];
      \draw (0,0) [fill] (-4,10) circle [radius=4pt];
      \draw (0,0) [fill] (0,2) circle [radius=4pt];
      \draw (0,0) [fill] (0,6) circle [radius=4pt];
      \draw (0,0) [fill] (0,10) circle [radius=4pt];
      \draw (0,0) [fill] (4,2) circle [radius=4pt];
      \draw (0,0) [fill] (4,6) circle [radius=4pt];
      \draw (0,0) [fill] (4,10) circle [radius=4pt];
    \end{tikzpicture}
    \caption{$\tilde K_{1} \subset S_K$ in $\mathcal{T}^1$}
    \label{fig:local-tensors-mixed-dim-2}
  \end{subfigure}
  \begin{subfigure}[c]{0.3\textwidth}
    \centering
    \begin{tikzpicture}[scale=0.3]
      \draw[black!50] (-4.0,2.0) -- (4.0,2.0);
      \draw[black!50] (4.0,2.0) -- (4.0,10.0);
      \draw[black!50] (4.0,10.0) -- (-4.0,10.0);
      \draw[black!50] (-4.0,10.0) -- (-4.0,2.0);
      \draw[black] (0.0, 2.0) -- (0.0, 6.0);
      \draw[black!50] (-4.0,6.0) -- (4.0,6.0);
      \draw[black!50] (-4.0,6.0) -- (0.0, 10.0);
      \draw[black!50] (0.0,6.0) -- (4.0, 10.0);
      \draw[black!50] (-4.0,2.0) -- (0.0, 6.0);
      \draw[black!50] (0.0,2.0) -- (4.0, 6.0);
      \draw[black!75, fill=black!20]  (0,6) -- (4,10) -- (0,10) -- cycle;
      \draw (1.25,9) node[] {$\tilde K_2$};
      \draw[black,line width=0.5mm] (0.0, 6.0) -- (0.0, 10.0);
      \draw (0,0) [fill] (-4,2) circle [radius=4pt];
      \draw (0,0) [fill] (-4,6) circle [radius=4pt];
      \draw (0,0) [fill] (-4,10) circle [radius=4pt];
      \draw (0,0) [fill] (0,2) circle [radius=4pt];
      \draw (0,0) [fill] (0,6) circle [radius=4pt];
      \draw (0,0) [fill] (0,10) circle [radius=4pt];
      \draw (0,0) [fill] (4,2) circle [radius=4pt];
      \draw (0,0) [fill] (4,6) circle [radius=4pt];
      \draw (0,0) [fill] (4,10) circle [radius=4pt];
    \end{tikzpicture}
    \caption{$\tilde K_{2} \subset S_K$ in $\mathcal{T}^1$}
    \label{fig:local-tensors-mixed-dim-3}
  \end{subfigure}
  \caption{Mesh entities involved in the local element tensors of
    mixed dimensional terms for bilinear forms. To exemplify, we
    consider a parent mesh $\mathcal{T}$ of dimension $d = 2$ and a
    continuous piecewise linear finite element space $U^1$ defined
    over a mesh $\mathcal{T}^1 = \mathcal{T}$ of topological dimension
    $d^1 = 2$ and a continuous piecewise linear finite element space
    $U^2$ defined over a mesh $\mathcal{T}^2$ of topological dimension
    $d^2 = 1$.
    \subref{fig:local-tensors-mixed-dim-1} Let $\mathcal{K} = \{K_1, K_2\} \subset
    \mathcal{T}^2$ be the integration domain.
    \subref{fig:local-tensors-mixed-dim-2}-\subref{fig:local-tensors-mixed-dim-3}
    For each $K \in
    \mathcal{K}$, we define its star $S_K = \{ \tilde K_1, \tilde K_2
    \} \subset \mathcal{T}^1$ as the (two) adjoining cells in
    $\mathcal{T}^1$ each with $K$ as a facet.}
  \label{fig:local-tensors-mixed-dim}
\end{figure}

Assume that $U^1$ and $U^2$ have different topological dimensions with
$d^1 > d^2$, and more specifically that $d^1 = d^2 + 1$. The set
$\mathcal{K}$ in~\eqref{eq:sum:K} must then be (a subset of) the cells
in $\mathcal{T}^2$ and (a subset of) the facets in
$\mathcal{T}^1$. Each cell in $\mathcal{K}$ is either an interior
facet in $\mathcal{T}^1$, in which case it is shared between two
cells, or an exterior facet, in which case it belongs to a single cell
and is located on the boundary of $\mathcal{T}^1$. We focus on the
case of interior facets. The case of exterior facets is analogous but
simpler. For instance, if $\mathcal{T}$ is a two-dimensional mesh of
triangles, $\mathcal{T}^1$ is a (sub)mesh of triangles, and
$\mathcal{T}^2$ is a (topologically one-dimensional) submesh of
intervals, then $\mathcal{K}$ must be a subset of the intervals in
$\mathcal{T}^2$. This setting is illustrated for
\Cref{ex:reference-mixed-problem} in
\Cref{fig:local-tensors-mixed-dim}.

Consider an element $K \in \mathcal{K}$. We note that $a^K(\phi^{2}_n,
\phi^{1}_m)$ will be zero for all $m = 1, \dots, \N^{1}$, $n = 1,
\dots, \N^{2}$ for which $\phi^1_m |_K = 0$ or $\phi^2_n |_K =
0$. Conversely, $a^K(\phi^{2}_n, \phi^{1}_m)$ is potentially non-zero
if $K$ is in the support of both $\phi_m^{1}$ and $\phi_n^{2}$ i.e. $K
\subset \supp(\phi_m^{1}) \cap \supp(\phi_n^{2})$. We denote the set
of $U^i$ basis function indices with $K$ in their support by
$\mathcal{N}_K^i$ i.e. $\mathcal{N}_K^{i} = \{n \in \{1, \dots, \N^i\} |
K \subseteq \supp(\phi_n^i) \}$. Thus, potentially $a_K(\phi^{2}_n,
\phi^{1}_m) \not = 0$ for $m \in \mathcal{N}_K^{1}$ and $n \in
\mathcal{N}_K^{2}$ , and $a_K(\phi^{2}_n, \phi^{1}_m) = 0$
otherwise. Since $K \in \mathcal{T}^2$, the number of $U^2$ basis
functions with $K$ in their support equals the local (cell) dimension
of $U^2$: $\N^{2}_K = \dim(\mathcal{N}_K^2) = \dim(U^2 |_K)$. On the
other hand, for $K$ viewed as an interior facet in $\mathcal{T}^1$, we
define its star $S_K = \{\tilde K_{1},\tilde K_{2}\}$ as the set of the (two)
cells in $\mathcal{T}^1$ with $K$ as a facet (see \Cref{fig:local-tensors-mixed-dim}).
The number of $U^1$ basis functions with $K$ in their support equals the dimension of
$U^1$ restricted to the star: $\N^1_{S_K} = \dim(\mathcal{N}_K^{1}) =
\dim(U^1 |_{S_K})$. For exterior facets, we simply define the star as
the single cell with $K$ as a facet.

To proceed, we introduce two new local tensor concepts for mixed
dimensional variational forms. We define the \emph{composite local
  element tensor} $A^K_{q, l}$ as the (potentially) non-zero
contributions from the cell $K \in \mathcal{K} \subseteq
\mathcal{T}^2$ to the global tensor:
\begin{equation}
  \label{eq:composite-local-tensor-def}
  A^{K}_{\kappa, l} = a^{K}(\phi^2_n, \phi^1_m), \quad m = \iota^1_{\tilde c}(\kappa), n = \iota^2_{c^2}(l), \quad \kappa = 1, \dots, N^1_{S_K}, l = 1, \dots, N^2_K.
\end{equation}
where $\iota^i_{\tilde c}$ is a map from (local) degree-of-freedom indices
of cells $\tilde K \subset S_K$ with index $\tilde c$
to global degree-of-freedom indices for
$U^i$ (here for $i = 1$).
Subsidiary, for each $K \in \mathcal{K}$ and for each
$\tilde{K} \in S_K$, we define the \emph{local element tensor}
$A^{\tilde{K}, K}_{k, l}$ as
\begin{equation}
  \label{eq:local-tensor-def}
  A^{\tilde{K}, K}_{k, l} = a^{K}(\phi^2_n, \phi^1_m), \quad m = \iota^1_{c^1}(k), n = \iota^2_{c^2}(l), \quad k = 1, \dots, N^1_{\tilde{K}}, l = 1, \dots, N^2_K.
\end{equation}
for any $\tilde{K} \in \mathcal{T}^1$ with index $c^1$
and $K \in \mathcal{K} \subseteq \mathcal{T}^2$ with index $c^2$.
We note that $m = \iota^1_{c^1}(k)$ can be
computed via the submesh mapping transfer of the local-to-global
mapping as described in Section~\ref{sec:dofs}. Since
$K \in \mathcal{T}^2$, the local-to-global mapping $\iota_{c^2}^2$ is
immediately available.
The composite local element tensor can be expressed in terms of the
local element tensors as
\begin{equation}
  \label{eq:local-to-composite-tensor}
  A^K_{\kappa, l} =  A^{\tilde{K}, K}_{k, l}, \quad \kappa = \gamma^1(k), k = 1, \dots, N^1_{\tilde{K}}, l = 1, \dots, N^2_K, \quad \tilde{K} \in S_K.
\end{equation}
where $\kappa = \gamma^1(k)$ is an appropriate map of local basis
function indices on $\tilde{K}$ to composite local basis function
indices on $S_K$. However, we note that the composite local element
tensor $A^K$ need not be formed explicitly; rather selected parts of
the local element tensors $A^{K, \tilde K}$ for $\tilde K \in S_K$ can
be added directly to the global tensor.
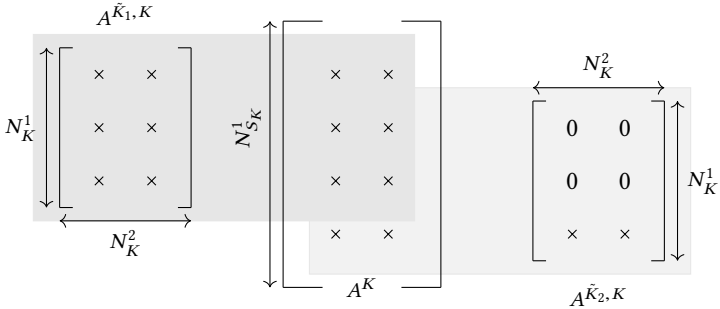
\begin{figure}[!ht]
  \centering
  \begin{tikzpicture}[scale=0.7]
    \tikzset{cross/.style={cross out, draw=black, minimum size=2*(#1-\pgflinewidth), inner sep=0pt, outer sep=0pt},cross/.default={2pt}}
    \draw[gray!20, fill=gray!10] (6.25,2.75) -- (6.25,-0.75) -- (-1,-0.75) -- (-1,2.75) -- cycle;
    \draw[gray!20, fill=gray!20] (-6.25,3.75) -- (-6.25,0.25) -- (1,0.25) -- (1,3.75) -- cycle;
    \draw (0.5,0) node[cross]{};
    \draw (0.5,1) node[cross]{};
    \draw (0.5,2) node[cross]{};
    \draw (0.5,3) node[cross]{};
    \draw (-0.5,0) node[cross]{};
    \draw (-0.5,1) node[cross]{};
    \draw (-0.5,2) node[cross]{};
    \draw (-0.5,3) node[cross]{};
    \draw (-1.5,4) -- (-0.75,4);
    \draw (-1.5,-1) -- (-0.75,-1);
    \draw (-1.5,-1) -- (-1.5,4);
    \draw (1.5,4) -- (0.75,4);
    \draw (1.5,-1) -- (0.75,-1);
    \draw (1.5,-1) -- (1.5,4);
    \draw (0,-1) node[]{\footnotesize{$A^{K}$}};
    \draw[<->] (-1.75,-1) -- (-1.75,4);
    \draw (-2.2,2.6) node[left, rotate=90]{\footnotesize{$N_{S_K}^1$}};
    \draw (-4,1) node[cross]{};
    \draw (-4,2) node[cross]{};
    \draw (-4,3) node[cross]{};
    \draw (-5,1) node[cross]{};
    \draw (-5,2) node[cross]{};
    \draw (-5,3) node[cross]{};
    \draw (-5.75,0.5) -- (-5.75,3.5);
    \draw (-5.5,0.5) -- (-5.75,0.5);
    \draw (-5.5,3.5) -- (-5.75,3.5);
    \draw[<->] (-6,0.5) -- (-6,3.5);
    \draw (-6,2) node[left]{\footnotesize{$\N_{K}^{1}$}};
    \draw (-3.25,0.5) -- (-3.25,3.5);
    \draw (-3.25,0.5) -- (-3.5,0.5);
    \draw (-3.25,3.5) -- (-3.5,3.5);
    \draw[<->] (-5.75,0.25) -- (-3.25,0.25);
    \draw (-4.5, 0.25) node[below]{\footnotesize{$\N_{K}^{2}$}};
    \draw (-4.5, 3.75) node[above]{\footnotesize{$A^{\tilde K_1, K}$}};
    \draw (4,0) node[cross]{};
    \draw (4,1) node[]{0};
    \draw (4,2) node[]{0};
    \draw (5,0) node[cross]{};
    \draw (5,1) node[]{0};
    \draw (5,2) node[]{0};
    \draw (5.75,-0.5) -- (5.75,2.5);
    \draw (5.5,-0.5) -- (5.75,-0.5);
    \draw (5.5,2.5) -- (5.75,2.5);
    \draw[<->] (6,-0.5) -- (6,2.5);
    \draw (6,1) node[right]{\footnotesize{$\N_{K}^{1}$}};
    \draw (3.25,-0.5) -- (3.25,2.5);
    \draw (3.25,-0.5) -- (3.5,-0.5);
    \draw (3.25,2.5) -- (3.5,2.5);
    \draw[<->] (5.75,2.75) -- (3.25,2.75);
    \draw (4.5, 2.75) node[above]{\footnotesize{$\N_{K}^{2}$}};
    \draw (4.5, -0.75) node[below]{\footnotesize{$A^{\tilde K_2, K}$}};
  \end{tikzpicture}
  \caption{Non-zero contributions to the element tensors for mixed
    dimensional (bilinear) forms cf.~\Cref{ex:reference-mixed-problem}
    and \Cref{fig:local-tensors-mixed-dim}. The composite local
    element tensor $A^{K}$, corresponding to the potentially non-zero
    entries of $A^K$ for $K \in \mathcal{K} \subseteq \mathcal{T}^2$
    can be defined by selecting contributions from $\tilde K_1$ and
    $\tilde K_2$. $\N_{K}^{i}$ is the local (element-wise) dimension
    of $U^i$ for $i = 1, 2$. The entries of the local tensor
    $A^{\tilde K_2, K}$ that have already been accounted for though
    $A^{\tilde K_1, K}$ can be zeroed in order to avoid adding a
    contribution twice.}
  \label{fig:contributions-local-tensor}
\end{figure}

To avoid counting the same contribution twice when directly adding the
local local tensors $A^{\tilde K, K}$ to the global tensor $A$, the
following approach may be used. If the star is composed of two cells
$S_K = \{ \tilde{K}_1, \tilde{K}_2\}$ as shown in
\Cref{fig:local-tensors-mixed-dim}, the local tensor $A^{\tilde{K}_2,
  K}$ contains entries that have already been added to the global
tensor when adding $A^{\tilde{K}_1, K}$. In particular, the entries
$A^{\tilde{K}_2, K}_{k_2,l}$, $l = 1, \dots, \N_K^2$ may be replaced
by zero when it exists a $k_1 \in \{ 1, \dots \N_{\tilde{K}_1}^{1}\}$
such that $\gamma^1(k_2) = \gamma^1(k_1)$ (see
\Cref{fig:contributions-local-tensor}).

\begin{example}
To illustrate the above ideas and concepts, we consider the
computation of the finite element tensors for
\Cref{ex:reference-mixed-problem}. We let $U^1$ and $U^2$ be finite
element spaces of continuous piecewise linears defined relative to
$\mathcal{T}^1$ and $\mathcal{T}^2$, respectively, and consider the
off-diagonal matrix block $A^{1, 2}$ corresponding to the form $a^{1,
  2}$ for each cell $K \in \mathcal{T}^2$ (corresponding to interior
facets in $\mathcal{T}^1$):
\begin{equation}
  \label{eq:mixed-form-blocks-local-poisson}
  a^{K, 1, 2}(\phi^2_n, \phi^1_m) = \int_{K} \phi^{2}_n \phi^{1}_m \dx, \quad \foralls m \in \N^{1}, n \in \N^{2}.
\end{equation}
Again, for readability, we drop the superscript $\cdot^{1, 2}$ in the
following.  The composite local element tensors of $a$ can be computed
by selecting contributions from the elements of the star $S_K$, as
illustrated in in~\Cref{fig:local-tensors-mixed-dim}
and~\Cref{fig:contributions-local-tensor}. For this specific set of
$U^1$ and $U^2$, $\dim(\mathcal{N}_K^{1}) = 4$ and
$\dim(\mathcal{N}_K^{2}) = 2$.
\end{example}

In conclusion, a standard cell-wise finite element assembly algorithm can be
augmented by an additional inner loop over adjacent mesh entities
(stars) to allow for assembly of mixed dimensional cell integrals. The
algorithm is given in Algorithm~\ref{alg:mixeddim-assembly}.
\begin{algorithm}[H]
  \caption{Assembly of bilinear cell integrals over mixed dimensions}
  \label{alg:mixeddim-assembly}
  \begin{algorithmic}[1]
    \For{$K$ in $\mathcal{K}$ (with cell index $c^2$ relative to $\mathcal{T}^2 \equiv \mathcal{K}$)}
    \For{$\tilde K$ in $S_K$}
    \State{Compute the cell index $c^1$ of $\tilde K$ relative to $\mathcal{T}^1$}
    \State{Compute $A_{K, \tilde K}$}
    \For{$l_1 \gets 1, \dots, N^1_K$ and $l_2 \gets 1, \dots, N^2_K$}
    \State{Compute $m = \iota_{c^1}^1(l_1)$ and $n = \iota_{c^2}^2(l_2)$}
    \State{Zero previously computed rows $A^{\tilde K, K}$}
    \State{Add entry $(l_1, l_2)$ of $A^{\tilde K, K}$ to $A$ at entry $(m, n)$}
    \EndFor
    \EndFor
    \EndFor
  \end{algorithmic}
\end{algorithm}

Finally, the case where $U^1$ and $U^2$ have the same topological
dimensions $d^1 = d^2$, but the integration domain is of lower
dimension $d^{\mathcal{K}} < d^1 = d^2$ can be viewed as an extension
of the previous Case 2 where the star $S_K = \{\tilde K_1,\tilde
K_2\}$ is handled as $S_K = S_K^1 \cup S_K^2$ where $S_K^1 = \{\tilde
K_1\}$ (resp. $S_K^2 = \{\tilde K_2\}$) is the star relative to
$\mathcal{T}^1$ (resp. $\mathcal{T}^2$). We do not discuss this case
further here.

\begin{remark}[Assembly of facet and vertex integrals]
  The dedicated algorithms presented here for mixed dimensional
  assembly focus on cell integrals, assuming other integral types to
  be defined as cell integrals over a lower-dimensional mesh. Standard
  assembly algorithms may handle various other integral types,
  e.g.~facet or point assembly. We remark that these other types of
  integrals can be used as usual for diagonal blocks as the latter
  rely only on single domain assembly algorithms. 
\end{remark}

\begin{remark}[Mixed dimensional assembly with higher dimensional gaps]
  \label{remark:3d-1d}
  In general, and in particular for submeshes with codimension more
  than one, the star $S_K$ may contain an arbitrary number of cells in
  $\mathcal{T}$ sharing the lower dimensional cell $K$. We emphasize
  that the algorithms we have presented to form the composite local
  tensors $A^{K}$ can be applied to stars $S_K$ with an arbitrary
  number of elements. In particular, this design easily allows for
  extensions to higher dimensional gaps, e.g.~assembly of coupled 3D-1D
  variational forms.
\end{remark}

\section{Overview of new FEniCS user interface and pipeline}
\label{sec:interface-overview}

The abstractions and algorithms presented in this paper have been
implemented in the FEniCS finite element
framework~\cite{AlnaesBlechta2015a}. Both low level and high level
features are available in C++ and Python. Use of the high level
interface for solving mixed dimensional variational problems is
exemplified in \Cref{code:ref-poisson}, which solves the reference
Poisson problem introduced in \Cref{sec:introduction}
(\Cref{ex:reference-mixed-problem}). 

\begin{lstlisting}[mathescape=true, language=Python,
    caption={{[Python]} Illustration of mixed dimensional
      functionalities usage to solve
      the reference Poisson problem
      introduced as \Cref{ex:reference-mixed-problem}.
      The dedicated abstractions easily integrate
      the language, keeping the code syntax close
      to the mathematical formulation.},
    label=code:ref-poisson]
from dolfin import *

# Generate the meshes
mesh = UnitSquareMesh(n, n)
marker = MeshFunction("size_t",mesh,mesh.topology().dim()-1,0)
for f in facets(mesh):
    marker[f] = 0.5 - EPS < f.midpoint().x() < 0.5 + EPS
submesh = MeshView.create(marker,1)
    
# Initialize function spaces and basis functions
V = FunctionSpace(mesh, "CG", 1)
LM = FunctionSpace(submesh, "CG", 1)
W = MixedFunctionSpace(V,LM)
(u,l) = TrialFunctions(W)
(v,e) = TestFunctions(W)

# Dirichlet boundary condition(x = 0 or x = 1)
def boundary(x):
    return x[0] < DOLFIN_EPS or x[0] > 1.0 - DOLFIN_EPS
bc = DirichletBC(V, Constant(0.0), boundary)

# Variational formulation
dV = Measure("dx", domain=W.sub_space(0).mesh())
dL = Measure("dx", domain=W.sub_space(1).mesh())
a = inner(grad(u),grad(v))*dV + v*l*dL + u*e*dL
L = Constant(2)*v*dV + Constant(0.25)*e*dL

# Solve the problem
sol = Function(W)
solve(a == L, sol, bc)
\end{lstlisting}

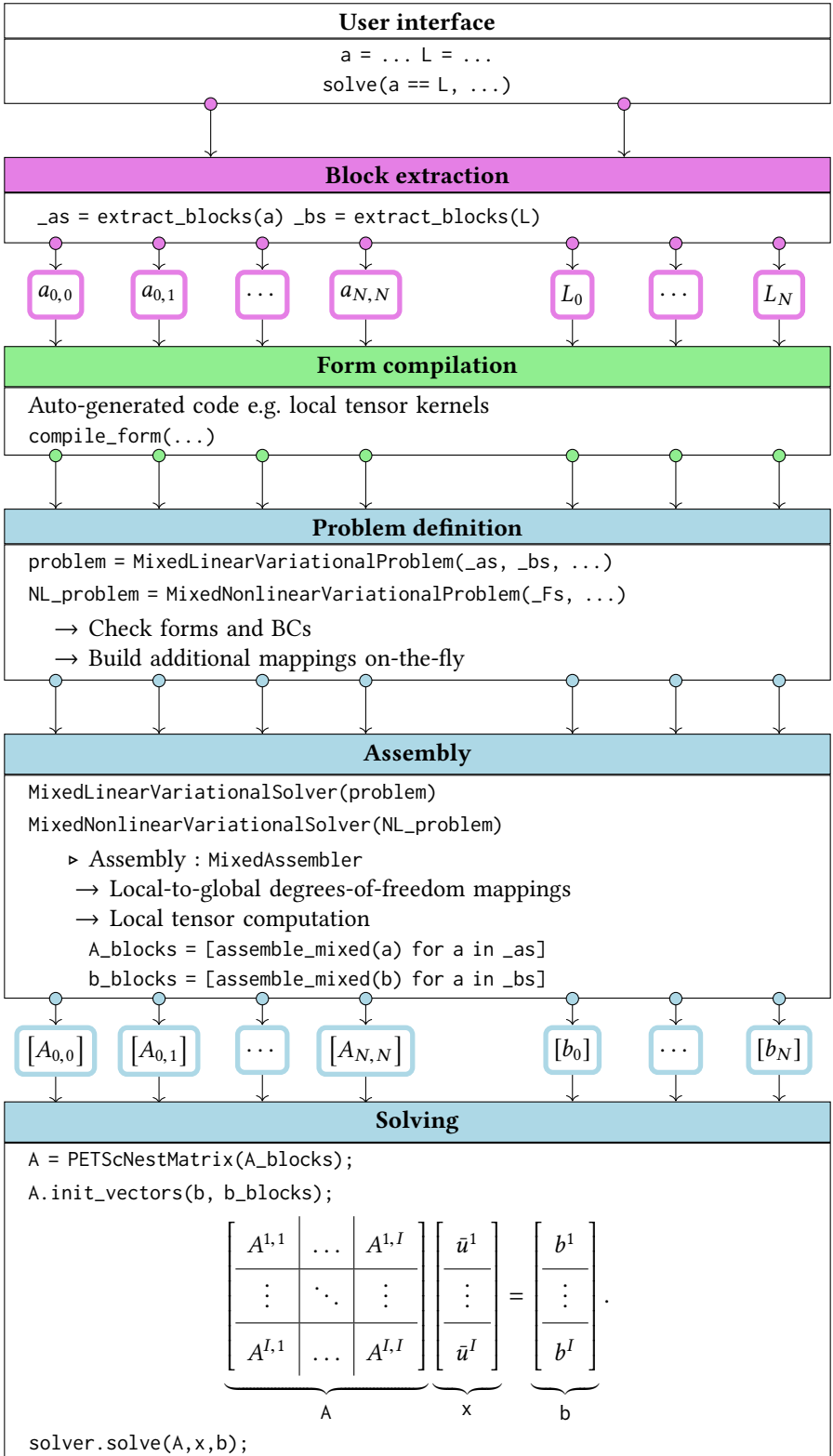
\begin{figure}
  \centering
  \begin{tikzpicture}
    \tikzstyle{title} = [draw=black, text=black, minimum height=10pt, minimum width=0.85\textwidth, text centered, font={\bf}]
    \tikzstyle{body} = [draw=black, text=black, anchor=north, minimum height=20pt, minimum width=0.85\textwidth, text width=0.8\textwidth]
    \tikzstyle{body_split} = [text=black, line width=2pt, rounded corners=1mm, anchor=north, minimum height=18pt, text centered]

    \def\ArrowShift{0.75}

    \node[title, fill=white](UserInterfaceTitle){User interface};
    \node[body, text width=0.25\textwidth, text centered](UserInterfaceBody) at ($(UserInterfaceTitle.south)$)
         {
           \code{a = ...} \code{L = ...}
           \code{solve(a == L, ...)}
         };
    \coordinate (A) at ($(UserInterfaceBody.south west) + 0.25*($(UserInterfaceBody.south east) - (UserInterfaceBody.south west)$)$);
    \coordinate (B) at ($(UserInterfaceBody.south east) - 0.25*($(UserInterfaceBody.south east) - (UserInterfaceBody.south west)$)$);
    \draw[->] (A) -- ($(A) + (0,-\ArrowShift)$);
    \draw[->] (B) -- ($(B) + (0,-\ArrowShift)$);
    \draw (0,0) [fill=UFL] (A) circle [radius=2.5pt];
    \draw (0,0) [fill=UFL] (B) circle [radius=2.5pt];

    \node[title, fill=UFL, anchor=north](BlockExtractionTitle) at ($(UserInterfaceBody.south) + (0,-\ArrowShift)$) {Block extraction};
    \node[body](BlockExtractionBody) at ($(BlockExtractionTitle.south)$)
         {
           ~~~~~\code{_as = extract_blocks(a)} ~~~~~~~~~~~~~~~~~~ \code{_bs = extract_blocks(L)} 
         };

    \coordinate (S) at ($0.5*($(BlockExtractionBody.south east) + (BlockExtractionBody.south west)$)$); 
    \coordinate (S1) at ($(BlockExtractionBody.south west) + 1/16*($(BlockExtractionBody.south east) - (BlockExtractionBody.south west)$)$);
    \coordinate (S2) at ($(S1) + 1/8*($(BlockExtractionBody.south east) - (BlockExtractionBody.south west)$)$);
    \coordinate (S3) at ($(S2) + 1/8*($(BlockExtractionBody.south east) - (BlockExtractionBody.south west)$)$);
    \coordinate (S4) at ($(S3) + 1/8*($(BlockExtractionBody.south east) - (BlockExtractionBody.south west)$)$);
    \coordinate (S5) at ($(S4) + 1/8*($(BlockExtractionBody.south east) - (BlockExtractionBody.south west)$)$);
    \coordinate (S6) at ($(S5) + 1/8*($(BlockExtractionBody.south east) - (BlockExtractionBody.south west)$)$);
    \coordinate (S7) at ($(S6) + 1/8*($(BlockExtractionBody.south east) - (BlockExtractionBody.south west)$)$);
    \coordinate (S8) at ($(S7) + 1/8*($(BlockExtractionBody.south east) - (BlockExtractionBody.south west)$)$);
    
    \draw[->] (S1) -- ($(S1) + (0,-0.5*\ArrowShift)$);
    \draw[->] (S2) -- ($(S2) + (0,-0.5*\ArrowShift)$);
    \draw[->] (S3) -- ($(S3) + (0,-0.5*\ArrowShift)$);
    \draw[->] (S4) -- ($(S4) + (0,-0.5*\ArrowShift)$);
    \draw[->] (S6) -- ($(S6) + (0,-0.5*\ArrowShift)$);
    \draw[->] (S7) -- ($(S7) + (0,-0.5*\ArrowShift)$);
    \draw[->] (S8) -- ($(S8) + (0,-0.5*\ArrowShift)$);

    \draw (0,0) [fill=UFL] (S1) circle [radius=2.5pt];
    \draw (0,0) [fill=UFL] (S2) circle [radius=2.5pt];
    \draw (0,0) [fill=UFL] (S3) circle [radius=2.5pt];
    \draw (0,0) [fill=UFL] (S4) circle [radius=2.5pt];
    \draw (0,0) [fill=UFL] (S6) circle [radius=2.5pt];
    \draw (0,0) [fill=UFL] (S7) circle [radius=2.5pt];
    \draw (0,0) [fill=UFL] (S8) circle [radius=2.5pt];

    \node[body_split,draw=white](Extracted) at ($(S) + (0,-0.5*\ArrowShift)$) {}; 
    \node[body_split,draw=UFL](Extracted1) at ($(S1) + (0,-0.5*\ArrowShift)$) {$a_{0,0}$};
    \node[body_split,draw=UFL](Extracted2) at ($(S2) + (0,-0.5*\ArrowShift)$) {$a_{0,1}$};
    \node[body_split,draw=UFL](Extracted3) at ($(S3) + (0,-0.5*\ArrowShift)$) {$\dots$};
    \node[body_split,draw=UFL](Extracted4) at ($(S4) + (0,-0.5*\ArrowShift)$) {$a_{N,N}$};
    \node[anchor=north] at ($(S5) + (0,-0.5*\ArrowShift)$) {};
    \node[body_split,draw=UFL](Extracted6) at ($(S6) + (0,-0.5*\ArrowShift)$) {$L_{0}$};
    \node[body_split,draw=UFL](Extracted7) at ($(S7) + (0,-0.5*\ArrowShift)$) {$\dots$};
    \node[body_split,draw=UFL](Extracted8) at ($(S8) + (0,-0.5*\ArrowShift)$) {$L_{N}$};

    \draw[->] ($(Extracted1.south)$) -- ($(Extracted1.south) + (0,-0.5*\ArrowShift)$);
    \draw[->] ($(Extracted2.south)$) -- ($(Extracted2.south) + (0,-0.5*\ArrowShift)$);
    \draw[->] ($(Extracted3.south)$) -- ($(Extracted3.south) + (0,-0.5*\ArrowShift)$);
    \draw[->] ($(Extracted4.south)$) -- ($(Extracted4.south) + (0,-0.5*\ArrowShift)$);
    \draw[->] ($(Extracted6.south)$) -- ($(Extracted6.south) + (0,-0.5*\ArrowShift)$);
    \draw[->] ($(Extracted7.south)$) -- ($(Extracted7.south) + (0,-0.5*\ArrowShift)$);
    \draw[->] ($(Extracted8.south)$) -- ($(Extracted8.south) + (0,-0.5*\ArrowShift)$);

    \node[title, fill=FFC, anchor=north](FormCompilationTitle) at ($(Extracted.south) + (0,-0.5*\ArrowShift)$) {Form compilation};
    \node[body](FormCompilationBody) at ($(FormCompilationTitle.south)$)
         {
           Auto-generated code e.g.~local tensor kernels\\
           \code{compile_form(...)}
         };
    \coordinate (S) at ($0.5*($(FormCompilationBody.south east) + (FormCompilationBody.south west)$)$); 
    \coordinate (S1) at ($(FormCompilationBody.south west) + 1/16*($(FormCompilationBody.south east) - (FormCompilationBody.south west)$)$);
    \coordinate (S2) at ($(S1) + 1/8*($(FormCompilationBody.south east) - (FormCompilationBody.south west)$)$);
    \coordinate (S3) at ($(S2) + 1/8*($(FormCompilationBody.south east) - (FormCompilationBody.south west)$)$);
    \coordinate (S4) at ($(S3) + 1/8*($(FormCompilationBody.south east) - (FormCompilationBody.south west)$)$);
    \coordinate (S5) at ($(S4) + 1/8*($(FormCompilationBody.south east) - (FormCompilationBody.south west)$)$);
    \coordinate (S6) at ($(S5) + 1/8*($(FormCompilationBody.south east) - (FormCompilationBody.south west)$)$);
    \coordinate (S7) at ($(S6) + 1/8*($(FormCompilationBody.south east) - (FormCompilationBody.south west)$)$);
    \coordinate (S8) at ($(S7) + 1/8*($(FormCompilationBody.south east) - (FormCompilationBody.south west)$)$);

    \draw[->] (S1) -- ($(S1) + (0,-\ArrowShift)$);
    \draw[->] (S2) -- ($(S2) + (0,-\ArrowShift)$);
    \draw[->] (S3) -- ($(S3) + (0,-\ArrowShift)$);
    \draw[->] (S4) -- ($(S4) + (0,-\ArrowShift)$);
    \draw[->] (S6) -- ($(S6) + (0,-\ArrowShift)$);
    \draw[->] (S7) -- ($(S7) + (0,-\ArrowShift)$);
    \draw[->] (S8) -- ($(S8) + (0,-\ArrowShift)$);

    \draw (0,0) [fill=FFC] (S1) circle [radius=2.5pt];
    \draw (0,0) [fill=FFC] (S2) circle [radius=2.5pt];
    \draw (0,0) [fill=FFC] (S3) circle [radius=2.5pt];
    \draw (0,0) [fill=FFC] (S4) circle [radius=2.5pt];
    \draw (0,0) [fill=FFC] (S6) circle [radius=2.5pt];
    \draw (0,0) [fill=FFC] (S7) circle [radius=2.5pt];
    \draw (0,0) [fill=FFC] (S8) circle [radius=2.5pt];
        
    \node[title, fill=DOLFIN, anchor=north](PbDefTitle) at ($(S) + (0,-\ArrowShift)$) {Problem definition};
    \node[body](PbDefBody) at ($(PbDefTitle.south)$)
         {
           \code{problem = MixedLinearVariationalProblem(_as, _bs, ...)}\\[0.05cm]
           \code{NL_problem = MixedNonlinearVariationalProblem(_Fs, ...)}
           \begin{itemize}
           \item[$\rightarrow$] Check forms and BCs
           \item[$\rightarrow$] Build additional mappings on-the-fly
           \end{itemize}
         };

    \coordinate (S) at ($0.5*($(PbDefBody.south east) + (PbDefBody.south west)$)$); 
    \coordinate (S1) at ($(PbDefBody.south west) + 1/16*($(PbDefBody.south east) - (PbDefBody.south west)$)$);
    \coordinate (S2) at ($(S1) + 1/8*($(PbDefBody.south east) - (PbDefBody.south west)$)$);
    \coordinate (S3) at ($(S2) + 1/8*($(PbDefBody.south east) - (PbDefBody.south west)$)$);
    \coordinate (S4) at ($(S3) + 1/8*($(PbDefBody.south east) - (PbDefBody.south west)$)$);
    \coordinate (S5) at ($(S4) + 1/8*($(PbDefBody.south east) - (PbDefBody.south west)$)$);
    \coordinate (S6) at ($(S5) + 1/8*($(PbDefBody.south east) - (PbDefBody.south west)$)$);
    \coordinate (S7) at ($(S6) + 1/8*($(PbDefBody.south east) - (PbDefBody.south west)$)$);
    \coordinate (S8) at ($(S7) + 1/8*($(PbDefBody.south east) - (PbDefBody.south west)$)$);

    \draw[->] (S1) -- ($(S1) + (0,-\ArrowShift)$);
    \draw[->] (S2) -- ($(S2) + (0,-\ArrowShift)$);
    \draw[->] (S3) -- ($(S3) + (0,-\ArrowShift)$);
    \draw[->] (S4) -- ($(S4) + (0,-\ArrowShift)$);
    \draw[->] (S6) -- ($(S6) + (0,-\ArrowShift)$);
    \draw[->] (S7) -- ($(S7) + (0,-\ArrowShift)$);
    \draw[->] (S8) -- ($(S8) + (0,-\ArrowShift)$);

    \draw (0,0) [fill=DOLFIN] (S1) circle [radius=2.5pt];
    \draw (0,0) [fill=DOLFIN] (S2) circle [radius=2.5pt];
    \draw (0,0) [fill=DOLFIN] (S3) circle [radius=2.5pt];
    \draw (0,0) [fill=DOLFIN] (S4) circle [radius=2.5pt];
    \draw (0,0) [fill=DOLFIN] (S6) circle [radius=2.5pt];
    \draw (0,0) [fill=DOLFIN] (S7) circle [radius=2.5pt];
    \draw (0,0) [fill=DOLFIN] (S8) circle [radius=2.5pt];

    \node[title, fill=DOLFIN, anchor=north](AssemblyTitle) at ($(S) + (0,-\ArrowShift)$) {Assembly};
    \node[body](AssemblyBody) at ($(AssemblyTitle.south)$)
         {
           \code{MixedLinearVariationalSolver(problem)}\\[0.05cm]
           \code{MixedNonlinearVariationalSolver(NL_problem)}
           \begin{itemize}
           \item[$\triangleright$] Assembly : \code{MixedAssembler}
             \begin{itemize}
             \item[$\rightarrow$] Local-to-global degrees-of-freedom mappings
             \item[$\rightarrow$] Local tensor computation
             \end{itemize}
           \item[]\code{A_blocks = [assemble_mixed(a)} \code{for a in _as]}
           \item[]\code{b_blocks = [assemble_mixed(b)} \code{for a in _bs]}
           \end{itemize}
         };

    \coordinate (S) at ($0.5*($(AssemblyBody.south east) + (AssemblyBody.south west)$)$); 
    \coordinate (S1) at ($(AssemblyBody.south west) + 1/16*($(AssemblyBody.south east) - (AssemblyBody.south west)$)$);
    \coordinate (S2) at ($(S1) + 1/8*($(AssemblyBody.south east) - (AssemblyBody.south west)$)$);
    \coordinate (S3) at ($(S2) + 1/8*($(AssemblyBody.south east) - (AssemblyBody.south west)$)$);
    \coordinate (S4) at ($(S3) + 1/8*($(AssemblyBody.south east) - (AssemblyBody.south west)$)$);
    \coordinate (S5) at ($(S4) + 1/8*($(AssemblyBody.south east) - (AssemblyBody.south west)$)$);
    \coordinate (S6) at ($(S5) + 1/8*($(AssemblyBody.south east) - (AssemblyBody.south west)$)$);
    \coordinate (S7) at ($(S6) + 1/8*($(AssemblyBody.south east) - (AssemblyBody.south west)$)$);
    \coordinate (S8) at ($(S7) + 1/8*($(AssemblyBody.south east) - (AssemblyBody.south west)$)$);

    \draw[->] (S1) -- ($(S1) + (0,-0.5*\ArrowShift)$);
    \draw[->] (S2) -- ($(S2) + (0,-0.5*\ArrowShift)$);
    \draw[->] (S3) -- ($(S3) + (0,-0.5*\ArrowShift)$);
    \draw[->] (S4) -- ($(S4) + (0,-0.5*\ArrowShift)$);
    \draw[->] (S6) -- ($(S6) + (0,-0.5*\ArrowShift)$);
    \draw[->] (S7) -- ($(S7) + (0,-0.5*\ArrowShift)$);
    \draw[->] (S8) -- ($(S8) + (0,-0.5*\ArrowShift)$);

    \draw (0,0) [fill=DOLFIN] (S1) circle [radius=2.5pt];
    \draw (0,0) [fill=DOLFIN] (S2) circle [radius=2.5pt];
    \draw (0,0) [fill=DOLFIN] (S3) circle [radius=2.5pt];
    \draw (0,0) [fill=DOLFIN] (S4) circle [radius=2.5pt];
    \draw (0,0) [fill=DOLFIN] (S6) circle [radius=2.5pt];
    \draw (0,0) [fill=DOLFIN] (S7) circle [radius=2.5pt];
    \draw (0,0) [fill=DOLFIN] (S8) circle [radius=2.5pt];

    \node[body_split,draw=white](Extracted) at ($(S) + (0,-0.5*\ArrowShift)$) {}; 
    \node[body_split,draw=DOLFIN](Extracted1) at ($(S1) + (0,-0.5*\ArrowShift)$) {$\left[ A_{0,0}\right]$};
    \node[body_split,draw=DOLFIN](Extracted2) at ($(S2) + (0,-0.5*\ArrowShift)$) {$\left[ A_{0,1}\right]$};
    \node[body_split,draw=DOLFIN](Extracted3) at ($(S3) + (0,-0.5*\ArrowShift)$) {$\dots$};
    \node[body_split,draw=DOLFIN](Extracted4) at ($(S4) + (0,-0.5*\ArrowShift)$) {$\left[ A_{N,N}\right]$};
    \node[anchor=north] at ($(S5) + (0,-0.5*\ArrowShift)$) {};
    \node[body_split,draw=DOLFIN](Extracted6) at ($(S6) + (0,-0.5*\ArrowShift)$) {$\left[ b_0\right]$};
    \node[body_split,draw=DOLFIN](Extracted7) at ($(S7) + (0,-0.5*\ArrowShift)$) {$\dots$};
    \node[body_split,draw=DOLFIN](Extracted8) at ($(S8) + (0,-0.5*\ArrowShift)$) {$\left[ b_N\right]$};

    \draw[->] ($(Extracted1.south)$) -- ($(Extracted1.south) + (0,-0.5*\ArrowShift)$);
    \draw[->] ($(Extracted2.south)$) -- ($(Extracted2.south) + (0,-0.5*\ArrowShift)$);
    \draw[->] ($(Extracted3.south)$) -- ($(Extracted3.south) + (0,-0.5*\ArrowShift)$);
    \draw[->] ($(Extracted4.south)$) -- ($(Extracted4.south) + (0,-0.5*\ArrowShift)$);
    \draw[->] ($(Extracted6.south)$) -- ($(Extracted6.south) + (0,-0.5*\ArrowShift)$);
    \draw[->] ($(Extracted7.south)$) -- ($(Extracted7.south) + (0,-0.5*\ArrowShift)$);
    \draw[->] ($(Extracted8.south)$) -- ($(Extracted8.south) + (0,-0.5*\ArrowShift)$); 

    \node[title, fill=DOLFIN, anchor=north](PbSolveTitle) at ($(Extracted.south) + (0,-0.5*\ArrowShift)$) {Solving};
    \node[body](PbSolveBody) at ($(PbSolveTitle.south)$)
         {
           \code{A = PETScNestMatrix(A_blocks);}\\[0.05cm]
           \code{A.init_vectors(b, b_blocks);}
           \begin{equation*}
             \renewcommand{\arraystretch}{1.75}
             \underbrace{\left[
                 \begin{array}{c|c|c}
                   A^{1,1} & \dots & A^{1,I} \\
                   \hline
                   \vdots & \ddots & \vdots \\
                   \hline
                   A^{I,1} & \dots & A^{I,I} \\
                 \end{array}
                 \right]}_{\text{\code{A}}}
             \underbrace{\left[
                 \begin{array}{ccc}
                   \bar{u}^1 \\
                   \hline
                   \vdots \\
                   \hline
                   \bar{u}^I \\
                 \end{array}
                 \right]}_{\text{\code{x}}}
             =
             \underbrace{\left[
                 \begin{array}{ccc}
                   b^1 \\
                   \hline
                   \vdots \\
                   \hline
                   b^I \\
                 \end{array}
                 \right]}_{\text{\code{b}}} .
           \end{equation*}
           \code{solver.solve(A,x,b);}
         };
       
  \end{tikzpicture}
  \caption{Overview of the FEniCS user interface for mixed dimensional problems.
    The block colors refer to the FEniCS component impacted :
    ~~\textcolor{UFL}{$\bullet$}:UFL ~~\textcolor{FFC}{$\bullet$}:FFC
    ~~\textcolor{DOLFIN}{$\bullet$}:DOLFIN}
  \label{fig:overview-framework}
\end{figure}

The \code{solve} function in the last line of \Cref{code:ref-poisson}
encapsulates the whole mixed domains problem solving process including
the extraction of block forms (see \Cref{sec:ufl}), the compilation
and code generation for each resulting subform, the block-by-block
assembly (see \Cref{sec:assembly}), and the solving of the recombined
block system using a given solver. The presented framework covers a
wide range of applications, and for the sake of flexibility,
intermediate lower-level functions are also available. For instance,
access to and manipulation of the separate matrix blocks can be used
for preconditioning and iterative solution purposes. The overall
structure of the mixed domain functionality is illustrated in
\Cref{fig:overview-framework}, while more implementation details are
presented in the next sections.

\subsection{Code generation of local tensors for mixed dimensional forms}
\label{subsec:ffc-mixed-dim}

The code for computing of the local tensors $A^K$
(\code{tabulate_tensor}) together with related quantities required for
the assembly is auto-generated by the form compiler FFC
\cite{LoggOlgaardEtAl2012a} given a variational form. Hence, each
variational form has its own kernel implementing the computation of
the corresponding local tensors $A^K$ depending on the finite element,
the integration domain and the form itself. In particular, the
measures \code{dx}, \code{ds} and \code{dS} discussed in Section
\ref{subsec:measures} to represent the different integral types (cell
integral, integral over exterior facets and over interior facets,
respectively) are mapped to different implementations of the
\code{tabulate_tensor} function with appropriate signatures.

The \code{tabulate_tensor} relative to cell integrals, shown in
\Cref{code:tabulate-tensor-cells} below, takes as argument the local
tensor \code{A} to be computed, together with information about the
cell geometry, the coordinates of its degrees of freedom and the form
coefficients if any.
\begin{lstlisting}[mathescape=true, language=C++,
    caption={{[C++]} Signature of the \code{tabulate_tensor} function
    dedicated to local tensor computation of cell integrals.},
    label=code:tabulate-tensor-cells]
void tabulate_tensor(double * A,
                     const double * const * w,
                     const double * coordinate_dofs,
                     int cell_orientation)
\end{lstlisting}
The computation of the local tensors for exterior facet integrals
requires the local index of the corresponding facet in the cell as an
additional argument, cf.~\Cref{code:tabulate-tensor-extf}.

\begin{lstlisting}[mathescape=true, language=C++,
    caption={{[C++]} Signature of the \code{tabulate_tensor} function
      dedicated to local tensor computation of exterior
      facet integrals.},
    label=code:tabulate-tensor-extf]
void tabulate_tensor(double * A,
                     const double * const * w,
                     const double * coordinate_dofs,
                     std::size_t facet,
                     int cell_orientation)
\end{lstlisting}

The algorithms presented in this paper focus on cell assembly,
assuming that measures in mixed-dimensional forms define integration
over cells of the lower dimensional mesh.  The cell assembly of mixed
forms with homogeneous dimension involves a single reference cell
whose local tensors $A^K$ are computed as usual using
\Cref{code:tabulate-tensor-cells}. On the other hand, the codimension
one local tensors $A^{\tilde{K}, K}$, relative to the cells
$\tilde{K}$ in the star $S_K$, can be assembled as exterior facet
integrals over the lower dimensional cell $K$. To accommodate for such
computations, the \code{tabulate_tensor} signature for cell integrals
has been revised, see~\Cref{code:tabulate-tensor-cells-new}. In
particular, we have added an optional input argument
\code{local_facet} to mimic the \code{facet} argument of
\Cref{code:tabulate-tensor-extf}.

\begin{lstlisting}[mathescape=true, language=C++,
    caption={{[C++]} Signature of the revised \code{tabulate_tensor} function
      dedicated to local tensor computation of cell integrals for handling
      codimension one local tensors.},
    label=code:tabulate-tensor-cells-new]
void tabulate_tensor(double * A,
                     const double * const * w,
                     const double * coordinate_dofs,
                     int cell_orientation,
                     std::size_t local_facet = 0)
\end{lstlisting}
  
\subsection{FEniCS interface to mixed domains assembly}

The FEniCS assembler implementation has been revised to tackle mixed
domain assembly of cell integrals as described in
\Cref{sec:assembly}. We present the revised assembly algorithms for
the case of homogeneous dimension (but mixed domains) and
heterogeneous dimension (mixed dimensional) in the respective sections
below.

The insertion of the local tensors $A^K$ and $A^{\tilde{K}, K}$ into
the block tensor requires the local-to-global degree of freedom
mappings discussed in \Cref{sec:dofs}. The cell indices $c^i$ relative
to $\mathcal{T}^i$ required for the mappings
$\iota^i_{c^i}$~\eqref{equ:local-to-global-dofs} are stored in a
double-indexed array \code{cell_index[i][j]}.  The first index
\code{i} represents the submesh $\mathcal{T}^i$ , \code{i=0}
(resp. \code{i=1}) corresponding to the test (resp. trial) function
space.  The second index \code{j} denotes the \code{j}-th contribution
$\tilde K$ of $S_K$ in the case of heterogeneous dimension.  These
cell indices are obtained from the parent-child and sibling mesh
mappings introduced in \Cref{sec:meshview}.

\subsubsection{Case 1 (homogeneous dimension) $d^{\mathcal{K}} = d^1 = d^2$}
Only one cell $K$ is involved in the computation of each local tensor
when the finite element spaces $U^1$ and $U^2$ are defined over the
same reference element.  The indices $c^i$ are the indices of this
integration cell $K$ in the corresponding submeshes $\mathcal{T}^i$
(see \Cref{fig:dof-local-to-global}). The computation of these indices
from the mesh mappings is shown in \Cref{code:cells-mappings-samedim}.

\begin{lstlisting}[mathescape=true, language=C++,
    caption={{[C++]} Each index $c^i$ (\code{cell_index[i][0]}) of the
      integration mesh cell $K$ (\code{cell}) relative to the basis
      function space mesh $\mathcal{T}^i$ is computed from the cell
      maps provided by the corresponding \code{MeshView} object
      \code{mapping}.}, label=code:cells-mappings-samedim]
  // Codimension 0 : $\mathcal{C}^i \rightarrow \mathcal{C}^j$
  cell_index[i][0] = mapping->cell_map()[cell->index()];
\end{lstlisting}

As described in \Cref{sec:assembly} the cell assembly algorithm
iterates over the cells $K$ of the integration domain $\mathcal{K}$
(\code{mesh}). The computation of the local tensor $A^K$ relative to
the integration mesh cell $K$ use the standard implementation of
\code{tabulate_tensor} kernel as given in Section
\ref{subsec:ffc-mixed-dim}. The indices $m$ and $n$ (e.g.~in the case of
bilinear forms) of the global degrees of freedom in the function
spaces $U^i$ for $i = 1, 2$ (\code{i = 0, 1}) are obtained from the
local-to-global mappings $\iota^i$ (\code{dofmaps[i].cell_dofs})
\eqref{equ:local-to-global-dofs}.

\begin{lstlisting}[mathescape=true, language=C++,
    caption={{[C++]} The homogeneous dimension cell assembly involves standard
      local tensor computation using \code{tabulate_tensor}.  The
      local-to-global mappings $\iota_{c^i}^i$ are defined from the
      local-to-global mappings \code{dofmaps[i].cell_dofs} to which we
      give the appropriate cell index \code{cell_index[i][0]} relative
      to $\mathcal{T}^i$.},
    label=code:samedim-assembly]
for (CellIterator cell(mesh); !cell.end(); ++cell)
{
  // Compute local element tensor $A^K$ (ufc.A)
  integral->tabulate_tensor(ufc.A.data(), ...);
  // Compute $m = \iota_{c^1}^1(l_1)$ and $n = \iota_{c^2}^2(l_2)$
  for(std::size_t i=0; i<form_rank; ++i)
  {
    auto dmap = dofmaps[i]->cell_dofs(cell_index[i][0]);
    dofs[i].set(dmap.size(), dmap.data());
  }
  // Add $A^{K}$ to $A$
  A.add_local(ufc.A.data(), dofs);
}
\end{lstlisting}

\subsubsection{Case 2 (codimension one) $d^{\mathcal{K}} = d^2 = d^1 - 1$}

The codimension one cell assembly involves the star $S_K$ of the
integration mesh cell $K$ (see \Cref{fig:dof-local-to-global}). The
computation of the indices (\code{cell_index[i][j]}) of the cells
$K^j$ in the star $S_K$ via the submesh mappings and the mesh
connectivity is shown in \Cref{code:cells-mappings-mixeddim}. The
local index of $K$ viewed as a facet relative to $K^j$
(\code{local_facets[j]}), required by the revised
\code{tabulate_tensor} implementation, can also be derived from the
mesh connectivity.
\begin{lstlisting}[mathescape=true, language=C++,
    caption={{[C++]} The index $c^i$ (\code{cell_index[i][j]}) of $K^j \in
      S_K$ relative to $\mathcal{T}^i$ is obtained from the set of
      cells adjacent to the corresponding facet \code{f} in the higher
      dimensional mesh. The local index of this facet in the cell
      $K^j$ (\code{local_facets[j]}) is required to compute the
      local tensors $A^{K^j, K}$.},
    label=code:cells-mappings-mixeddim]
  // Codimension 1 : $\mathcal{C}^i \rightarrow \mathcal{F}^j \rightarrow \{ \mathcal{C}^j, \mathcal{C}^j \}$
  Facet f(*(mapping->mesh()),mapping->cell_map()[cell->index()]);
  // Building $S_K$ as cells $\tilde K$ (dim D) that are adjacent to f
  const std::size_t D = mapping->mesh()->topology().dim();
  for(std::size_t j=0; j<f.num_entities(D);j++)
  {
    Cell mesh_cell(*(mapping->mesh()), f.entities(D)[j]);
    cell_index[i][j] = mesh_cell.index());
    // Local index of facet f in mesh_cell
    local_facets.push_back(mesh_cell.index(f));
  }
\end{lstlisting}

As detailed in \Cref{alg:mixeddim-assembly}, the assembly over mixed
dimensions involves an additional loop over the cells $\tilde K$ in
$S_K$.  The local tensors $A^{\tilde{K}, K}$ are computed from the
revised \code{tabulate_tensor} function taking the local index of the
corresponding facet as an additional argument
(see~Section~\ref{subsec:ffc-mixed-dim}). Again, the indices $m$ and
$n$ (e.g.~in the case of bilinear forms) of the global degrees of
freedom in the function spaces $U^i$ for $i = 1, 2$ (\code{i = 0, 1})
are obtained from the local-to-global mappings $\iota^i$
(\code{dofmaps[i].cell_dofs}) \eqref{equ:local-to-global-dofs}. As
specified in \Cref{sec:assembly}, entries of the local tensors
$A^{\tilde{K}, K}$ may have to be zeroed to avoid duplicates.
\begin{lstlisting}[mathescape=true, language=C++,
    caption={{[C++]} The codimension one assembly iterates over the cells $K$
      of the integration domain and the cells $\tilde K$ of its star $S_K$.
      The local tensors $A^{\tilde{K}, K}$ are computed from the revised \code{tabulate_tensor}
      taking the local index of the facet (\code{local_facets[j]}) as an additional argument.
      The local-to-global mappings $\iota_{c^i}^i$ are defined from the degrees of
      freedom mappings \code{dofmaps[i].cell_dofs} with the index \code{cell_index[i][j]}
      of the $j^{th}$ cell $\tilde K \in S_K$ relative to $\mathcal{T}^i$.},
    label=code:mixeddim-assembly]
for (CellIterator cell(mesh); !cell.end(); ++cell)
{
  // Iterating over cells $\tilde K$ in $S_K$
  for(std::size_t j=0; j<local_facets.size(); ++j)
  {
    // Compute $A^{\tilde K, K}$
    integral->tabulate_tensor(ufc.A.data(),...,local_facets[j]);
    // Compute $m = \iota_{c^1}^1(l_1)$ and $n = \iota_{c^2}^2(l_2)$
    for(std::size_t i=0; i<form_rank; ++i)
    {
      std::size_t jidx = (cell_index[i].size() > 1) ? j:0;
      auto dmap = dofmaps[i]->cell_dofs(cell_index[i][jidx]);
      zero(dmap, ...);
      dofs[i].set(dmap.size(), dmap.data());
    }
    // Add entry $(l_1, l_2)$ of $A^{\tilde K, K}$ to $A$
    A.add_local(ufc.A.data(), dofs);
  }
}
\end{lstlisting}

\subsection{Block linear algebra and representing assembled tensors}
\label{subsec:matnest}

Once assembled as described in \Cref{sec:assembly}, the block tensors
are recombined to form the block systems
\eqref{equ:block-shaped-system-linear} and
\eqref{equ:block-shaped-system-nonlinear}. The FEniCS/DOLFIN library
\cite{LoggWells2010a,LoggWellsEtAl2012a} uses the
software package PETSc \cite{petsc-web-page} for linear algebra
objects and algorithms.

In particular, the PETSc \code{MATNEST} structure is dedicated to
efficient representation and use of block tensors. This PETSc data
structure has been interfaced in FEniCS/DOLFIN as a
\code{PETScNestMatrix} class building a \code{MATNEST} object from a
list of matrices, cf.~\Cref{code:petsc-nest-mat}. The
\code{PETScNestMatrix} class is also equipped with a
\code{init_vectors} function for building vectors with the appropriate
block structure.
\begin{lstlisting}[mathescape=true, language=Python,
    caption={{[Python]} The assembled blocks are combined into a single block tensor
      of type \code{PETScNestMatrix} which interfaces the dedicated
      PETSc data structure \code{MATNEST}. The right hand side
      block vector \code{b} is built from \code{A} using the
      \code{init_vectors} function with a matching structure.},
    label={code:petsc-nest-mat}]
  A = PETScNestMatrix([A_00, ..., A_NN]);
  b = Vector()
  A.init_vectors(b, [b_0, ..., b_N]);
\end{lstlisting}

The wide selection of solvers and preconditioners available in
FEniCS/DOLFIN via PETSc can be applied to \code{PETScNestMatrix}
objects in the same manner as with standard single domain system.
However, direct
solvers are not directly compatible with this data structure.
Instead, \code{PETScNestMatrix} objects can easily be converted from
the \code{MATNEST} format to the more standard \code{MATAIJ} format
using the function \code{convert_to_aij} interfacing the corresponding
PETSc routine (see \Cref{code:solver-setup}).
\begin{lstlisting}[mathescape=true, language=Python,
    caption={{[Python]} The recombined block system $Ax=b$ can be solved using
      e.g.~iterative or direct solvers.},
    label=code:solver-setup]
# Solve block system using iterative solver
solver = PETScKrylovSolver()
solver.solve(A, x, b);

# Solve block system using direct solver
solver = PETScLUSolver()
A.convert_to_aij()
solver.solve(A, x, b);
\end{lstlisting}
Moreover, the \code{MATNEST} data structure is compatible with the
\code{PCFieldSplit} preconditioners offered by PETSc, allowing for the
application of specific preconditioners to each block matrix.

\section{Numerical results}
\label{sec:numerical-results}

The presented framework is applicable to a wide range of mixed
dimensional problems. In this section, we report on numerical results
for three selected cases. We start with the reference Poisson problem
introduced in \Cref{sec:introduction} as a demonstration of the
described features. Second, we study two Stokes problems with
non-standard boundary conditions as discussed
in~\cite{szopos:tel-01646867, bertoluzza:hal-01420651}. Finally, we
consider a mathematical model of ionic electrodiffusion in brain
tissue.
Our software to reproduce the presented results is
openly and freely available (see~\cite{DockerContainer,ZenodoExamples})

\subsection{Reference Poisson problem}
\label{subsec:ref-poisson-results}
\begin{figure}[H]
  \centering
  \begin{subfigure}[b]{0.45\textwidth}
    \centering
    \includegraphics[width=0.95\textwidth]{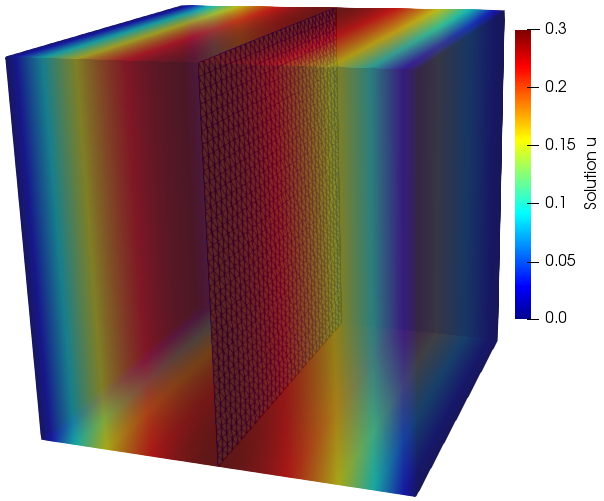}
    \caption{Solution $u_h$}
    \label{fig:ref-poisson-solution}
  \end{subfigure}
  \begin{subfigure}[b]{0.52\textwidth}
    \centering
    \begin{tikzpicture}
      \pgfplotsset{every axis legend/.append style = { at={(0.05,0.05)},anchor=south west}}
      \begin{loglogaxis} 
        [
          width=0.95\textwidth,
          ticklabel style = {font=\scriptsize},
          title style = {font=\scriptsize},
          legend style = {font=\scriptsize},
          legend cell align={left},
          xtick={2,4,8,16,32},
          xticklabels={2,4,8,16,32}
        ]

        \pgfplotstableread{poisson_3D_P1_np1_cvg.dat}\poissonA 
        \addplot[mark=*, color=blue] table[x=N,y=L2]{\poissonA};
        \addplot[mark=none, style=dashed, forget plot] table[y = {create col/linear regression={y=L2}}]{\poissonA};
        \pgfmathsetmacro{\slopeA}{abs(\pgfplotstableregressiona)}
        \addlegendentry{$\| r_u \|_{L_2}$: $\pgfmathprintnumber{\slopeA}$}
        \addplot[mark=o, color=blue] table[x=N,y=H1]{\poissonA};
        \addplot[mark=none, style=dashed, forget plot] table[y = {create col/linear regression={y=H1}}]{\poissonA};
        \pgfmathsetmacro{\slopeB}{abs(\pgfplotstableregressiona)}
        \addlegendentry{$\| r_u \|_{H_1}$: $\pgfmathprintnumber{\slopeB}$}        
      \end{loglogaxis}
    \end{tikzpicture}
    \caption{Convergence study}
    \label{fig:ref-poisson-cvg}
  \end{subfigure}
  \caption{Solution of the Poisson problem given in
    \Cref{ex:reference-mixed-problem} with $f=2$ and $c=0.25$ to
    obtain the manufactured solution $u(x,y) = x(1-x)$, and the
    corresponding convergence study plotting 
    $\| e_h \|_{L_2}$ and $\| e_h \|_{H_1}$ with order $k=1$
    depending on the mesh resolution.
    $\| r_u \|_{L_2}$ and $\| r_u \|_{H_1}$ are
    the resulting convergence rates.
  }
\end{figure}
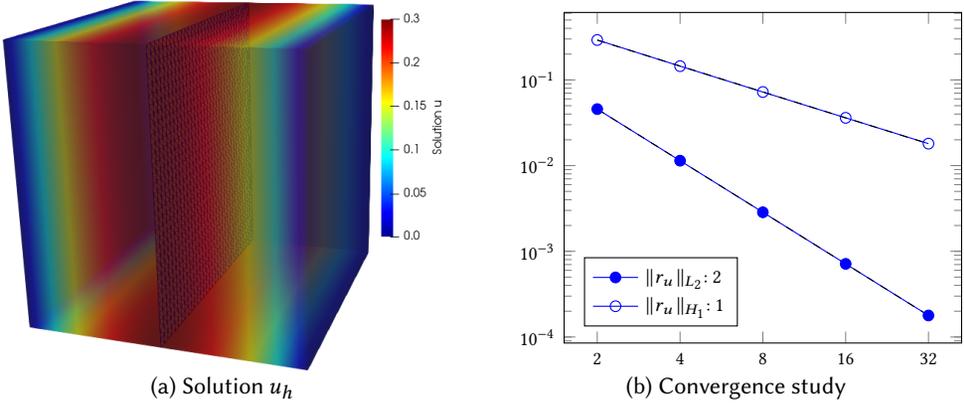

This numerical experiment presents a convergence study performed on
the 3D version of \Cref{ex:reference-mixed-problem} using the method
of manufactured solutions. The function $u(x,y) = x(1-x)$ is the exact
solution of \Cref{ex:reference-mixed-problem} with $f=2$ and
$c=0.25$. We consider a uniform tetrahedral mesh $\mathcal{T}^1$ of
the unit cube, and find the approximations $u_h$ using the space of
continuous piecewise linear polynomials $U^1$ relative to this mesh
and $c_h$ using continuous piecewise linear
polynomials defined relative to a mesh $\mathcal{T}^2$ of the
two-dimensional midplane. The obtained solution is shown in Figure
\ref{fig:ref-poisson-solution}.

We introduce the approximation error $e_h$ defined as $e_h = u - u_h$.
Given $k$ the polynomial order associated with $U^1(\mathcal{T}^1)$,
and $h$ the mesh size, there exists a constant $\alpha > 0$ such that
\begin{equation}
  \label{eq:cvg-properties}
  \| e_h \|_{L_2} \leqslant \alpha h^{k+1} ~~\text{and}~~ \| e_h \|_{H_1} \leqslant \alpha h^{k} .
\end{equation}
\Cref{fig:ref-poisson-cvg} plots the $L_2$ and $H_1$ norms of the
approximation error with respect to the mesh resolution, for the case
$k=1$. The slope of the corresponding lines -- indicated in legend --
shows that the expected orders of convergence
\eqref{eq:cvg-properties} are obtained.

\subsection{Fluid flow problems with non-standard boundary conditions}
\label{subsec:stokes-results}

\subsubsection{A Stokes-Brinkman problem with a manufactured solution}
\label{sec:stokes:1}

We consider the Stokes-Brinkman problem \eqref{eq:stokes-equation} solved on
the domain $\Omega = [0,1] \times [0,1]$.
\begin{equation}
\begin{split}
  \label{eq:stokes-equation}
  -\Delta u + u -\nabla p &= f ~\text{on}~ \Omega, \\
  \nabla \cdot u &= 0 ~\text{on}~ \Omega .
\end{split}
\end{equation}

We introduce the manufactured solutions $u$ and $p$
\eqref{eq:stokes-equation-exact} from which the right hand side $f$ in
\eqref{eq:stokes-equation} and the boundary conditions
\eqref{eq:stokes-equation-BC} are defined as
\begin{equation}
  \label{eq:stokes-equation-exact}
  u =
  \begin{pmatrix}
    \cos(\pi y)\sin(\pi x) \\
    - \cos(\pi x)\sin(\pi y)
  \end{pmatrix},
  \quad
  p = \pi \cos(\pi x)\cos(\pi y).
\end{equation}
We denote by $\Gamma_{t}$ and $\Gamma_{b}$ the top and bottom
boundaries, respectively, on which we impose homogeneous Neumann
conditions. Further, we set the velocity $g$ at the inlet
boundary $\Gamma_{\rm in}$ to be the manufactured solution
$u|_{\Gamma_{\rm in}}$, imposed through a Lagrange multiplier
$\lambda \in \Gamma_{\rm in}$.  In addition, a inhomogeneous
Neumann condition is imposed on $\Gamma_{\rm out}$, with traction $h$
defined from \eqref{eq:stokes-equation-exact}:
\begin{equation}
  \label{eq:stokes-equation-BC}
  \nabla u \cdot n + p n = h ~\text{on}~ \Gamma_{\rm out}.
\end{equation}

Combining, we obtain a mixed dimensional problem with the variational
formulation: find $(u,p,\lambda) \in U \times P
\times L$ such that
\begin{equation}
  \begin{split}
    \disp {\int_{\Omega} \nabla u \cdot \nabla v + u \cdot v
    +  p \, \nabla \cdot v + q \, \nabla \cdot u \dx }
    + \disp{ \int_{\Gamma_{\rm in}} \lambda \cdot v +  \eta \cdot u \ds } \\
    =
    \disp{ \int_{\Omega} f \cdot v \dx + \int_{\Gamma_{\rm out}} h \cdot v \ds } +
    \disp{ \int_{\Gamma_{\rm in}} \eta \cdot g \ds },
  \end{split}
\end{equation}
for all $(v,q,\eta) \in U \times P \times L$. We
define a uniform mesh of the unit square $\mathcal{T}^1$, and a
lower-dimensional mesh $\mathcal{T}^2$ of the boundary $\Gamma_{\rm in}$.
The velocity $u$ is approximated as $u_h \in
U$ through continuous piecewise vector fields of polynomial order $k+1
\geq 2$ while continuous piecewise polynomials of order $k \geq 1$ are
used for the pressure approximation $p_h \in P$. The Lagrange
multiplier $\lambda$ is approximated in the space $L =
L(\mathcal{T}^2)$ of continuous piecewise polynomials of order $k$
defined relative to $\mathcal{T}^2$.

Iterative solvers, such as e.g.~gmres with ilu preconditioning, easily
fail to converge for this problem due to the ill-conditioning of the
system. A fractional preconditioning strategy was introduced in
\cite{MiroKentPrecond2016}. In Figure \ref{fig:stokes-cvg}, we
demonstrate that we obtain the expected order of convergence
\eqref{eq:cvg-properties}, both using a direct solver and using said
preconditioning technique, for $k = 1, 2$.

\begin{figure}
  \centering
  \captionsetup[subfigure]{justification=centering}
  \begin{subfigure}[t]{0.48\textwidth}
    \centering
    \begin{tikzpicture}
      \pgfplotsset{every axis legend/.append style = { at={(0.5,1.03)},anchor=south }}
      \begin{loglogaxis} 
        [
          width=\textwidth,
          ticklabel style = {font=\scriptsize},
          title style = {font=\scriptsize},
          legend style = {font=\scriptsize},
          legend cell align={left},
          xtick={16,32,64,128},
          xticklabels={16,32,64,128},
          legend columns=2
        ]

        \pgfplotstableread{velocity_direct_P2P1P1.dat}\velocity
        \pgfplotstableread{pressure_direct_P2P1P1.dat}\pressure

        \addplot[mark=*, color=blue] table[x=N,y=L2]{\velocity};
        \addplot[mark=none, style=dashed, forget plot] table[y = {create col/linear regression={y=L2}}]{\velocity};
        \pgfmathsetmacro{\slopeA}{abs(\pgfplotstableregressiona)}
        \addlegendentry{$\|r_{u}\|_{L_2}$: $\pgfmathprintnumber{\slopeA}$}

        \addplot[mark=o, color=blue] table[x=N,y=H1]{\velocity};
        \addplot[mark=none, style=dashed, forget plot] table[y = {create col/linear regression={y=H1}}]{\velocity};
        \pgfmathsetmacro{\slopeB}{abs(\pgfplotstableregressiona)}
        \addlegendentry{$\|r_{u}\|_{H_1}$: $\pgfmathprintnumber{\slopeB}$}

        \addplot[mark=square*, color=red] table[x=N,y=L2]{\pressure};
        \addplot[mark=none, style=dashed, forget plot] table[y = {create col/linear regression={y=L2}}]{\pressure};
        \pgfmathsetmacro{\slopeC}{abs(\pgfplotstableregressiona)}
        \addlegendentry{$\|r_{p}\|_{L_2}$: $\pgfmathprintnumber{\slopeC}$}

        \addplot[mark=square, color=red] table[x=N,y=H1]{\pressure};
        \addplot[mark=none, style=dashed, forget plot] table[y = {create col/linear regression={y=H1}}]{\pressure};
        \pgfmathsetmacro{\slopeD}{abs(\pgfplotstableregressiona)}
        \addlegendentry{$\|r_{p}\|_{H_1}$: $\pgfmathprintnumber{\slopeD}$}
      \end{loglogaxis}
    \end{tikzpicture}
    \caption{$P_2^2 \times P_1 \times P_1^2$ - Direct solver}
    \label{fig:stokes-cvg-direct-P2P1P1}
  \end{subfigure}
  \begin{subfigure}[t]{0.48\textwidth}
    \centering
    \begin{tikzpicture}
      \pgfplotsset{every axis legend/.append style = { at={(0.5,1.03)},anchor=south }}
      \begin{loglogaxis} 
        [
          width=\textwidth,
          ticklabel style = {font=\scriptsize},
          title style = {font=\scriptsize},
          legend style = {font=\scriptsize},
          legend cell align={left},
          xtick={16,32,64,128},
          xticklabels={16,32,64,128},
          legend columns=2
        ]

        \pgfplotstableread{velocity_fracprec_P2P1P1.dat}\velocity
        \pgfplotstableread{pressure_fracprec_P2P1P1.dat}\pressure

        \addplot[mark=*, color=blue] table[x=N,y=L2]{\velocity};
        \addplot[mark=none, style=dashed, forget plot] table[y = {create col/linear regression={y=L2}}]{\velocity};
        \pgfmathsetmacro{\slopeA}{abs(\pgfplotstableregressiona)}
        \addlegendentry{$\|r_{u}\|_{L_2}$: $\pgfmathprintnumber{\slopeA}$}

        \addplot[mark=o, color=blue] table[x=N,y=H1]{\velocity};
        \addplot[mark=none, style=dashed, forget plot] table[y = {create col/linear regression={y=H1}}]{\velocity};
        \pgfmathsetmacro{\slopeB}{abs(\pgfplotstableregressiona)}
        \addlegendentry{$\|r_{u}\|_{H_1}$: $\pgfmathprintnumber{\slopeB}$}

        \addplot[mark=square*, color=red] table[x=N,y=L2]{\pressure};
        \addplot[mark=none, style=dashed, forget plot] table[y = {create col/linear regression={y=L2}}]{\pressure};
        \pgfmathsetmacro{\slopeC}{abs(\pgfplotstableregressiona)}
        \addlegendentry{$\|r_{p}\|_{L_2}$: $\pgfmathprintnumber{\slopeC}$}

        \addplot[mark=square, color=red] table[x=N,y=H1]{\pressure};
        \addplot[mark=none, style=dashed, forget plot] table[y = {create col/linear regression={y=H1}}]{\pressure};
        \pgfmathsetmacro{\slopeD}{abs(\pgfplotstableregressiona)}
        \addlegendentry{$\|r_{p}\|_{H_1}$: $\pgfmathprintnumber{\slopeD}$}
      \end{loglogaxis}
    \end{tikzpicture}
    \caption{$P_2^2 \times P_1 \times P_1^2$ - Preconditioned Minres}
    \label{fig:stokes-cvg-fracprec-P2P1P1}
  \end{subfigure}
  \begin{subfigure}[t]{0.48\textwidth}
    \centering
    \begin{tikzpicture}
      \pgfplotsset{every axis legend/.append style = { at={(0.5,1.03)},anchor=south }}
      \begin{loglogaxis} 
        [
          width=\textwidth,
          ticklabel style = {font=\scriptsize},
          title style = {font=\scriptsize},
          legend style = {font=\scriptsize},
          legend cell align={left},
          xtick={16,32,64,128},
          xticklabels={16,32,64,128},
          legend columns=2
        ]

        \pgfplotstableread{velocity_direct_P3P2P2.dat}\velocity
        \pgfplotstableread{pressure_direct_P3P2P2.dat}\pressure

        \addplot[mark=*, color=blue] table[x=N,y=L2]{\velocity};
        \addplot[mark=none, style=dashed, forget plot] table[y = {create col/linear regression={y=L2}}]{\velocity};
        \pgfmathsetmacro{\slopeA}{abs(\pgfplotstableregressiona)}
        \addlegendentry{$\|r_{u}\|_{L_2}$: $\pgfmathprintnumber{\slopeA}$}

        \addplot[mark=o, color=blue] table[x=N,y=H1]{\velocity};
        \addplot[mark=none, style=dashed, forget plot] table[y = {create col/linear regression={y=H1}}]{\velocity};
        \pgfmathsetmacro{\slopeB}{abs(\pgfplotstableregressiona)}
        \addlegendentry{$\|r_{u}\|_{H_1}$: $\pgfmathprintnumber{\slopeB}$}

        \addplot[mark=square*, color=red] table[x=N,y=L2]{\pressure};
        \addplot[mark=none, style=dashed, forget plot] table[y = {create col/linear regression={y=L2}}]{\pressure};
        \pgfmathsetmacro{\slopeC}{abs(\pgfplotstableregressiona)}
        \addlegendentry{$\|r_{p}\|_{L_2}$: $\pgfmathprintnumber{\slopeC}$}

        \addplot[mark=square, color=red] table[x=N,y=H1]{\pressure};
        \addplot[mark=none, style=dashed, forget plot] table[y = {create col/linear regression={y=H1}}]{\pressure};
        \pgfmathsetmacro{\slopeD}{abs(\pgfplotstableregressiona)}
        \addlegendentry{$\|r_{p}\|_{H_1}$: $\pgfmathprintnumber{\slopeD}$}
      \end{loglogaxis}
    \end{tikzpicture}
    \caption{$P_3^2 \times P_2 \times P_2^2$ - Direct solver}
    \label{fig:stokes-cvg-direct-P3P2P2}
  \end{subfigure}
  \begin{subfigure}[t]{0.48\textwidth}
    \centering
    \begin{tikzpicture}
      \pgfplotsset{every axis legend/.append style = { at={(0.5,1.03)},anchor=south }}
      \begin{loglogaxis} 
        [
          width=\textwidth,
          ticklabel style = {font=\scriptsize},
          title style = {font=\scriptsize},
          legend style = {font=\scriptsize},
          legend cell align={left},
          xtick={16,32,64,128},
          xticklabels={16,32,64,128},
          legend columns=2
        ]

        \pgfplotstableread{velocity_fracprec_P3P2P2.dat}\velocity
        \pgfplotstableread{pressure_fracprec_P3P2P2.dat}\pressure

        \addplot[mark=*, color=blue] table[x=N,y=L2]{\velocity};
        \addplot[mark=none, style=dashed, forget plot] table[y = {create col/linear regression={y=L2}}]{\velocity};
        \pgfmathsetmacro{\slopeA}{abs(\pgfplotstableregressiona)}
        \addlegendentry{$\|r_{u}\|_{L_2}$: $\pgfmathprintnumber{\slopeA}$}

        \addplot[mark=o, color=blue] table[x=N,y=H1]{\velocity};
        \addplot[mark=none, style=dashed, forget plot] table[y = {create col/linear regression={y=H1}}]{\velocity};
        \pgfmathsetmacro{\slopeB}{abs(\pgfplotstableregressiona)}
        \addlegendentry{$\|r_{u}\|_{H_1}$: $\pgfmathprintnumber{\slopeB}$}

        \addplot[mark=square*, color=red] table[x=N,y=L2]{\pressure};
        \addplot[mark=none, style=dashed, forget plot] table[y = {create col/linear regression={y=L2}}]{\pressure};
        \pgfmathsetmacro{\slopeC}{abs(\pgfplotstableregressiona)}
        \addlegendentry{$\|r_{p}\|_{L_2}$: $\pgfmathprintnumber{\slopeC}$}

        \addplot[mark=square, color=red] table[x=N,y=H1]{\pressure};
        \addplot[mark=none, style=dashed, forget plot] table[y = {create col/linear regression={y=H1}}]{\pressure};
        \pgfmathsetmacro{\slopeD}{abs(\pgfplotstableregressiona)}
        \addlegendentry{$\|r_{p}\|_{H_1}$: $\pgfmathprintnumber{\slopeD}$}
      \end{loglogaxis}
    \end{tikzpicture}
    \caption{$P_3^2 \times P_2 \times P_2^2$ - Preconditioned Minres}
    \label{fig:stokes-cvg-fracprec-P3P2P2}
  \end{subfigure}
  \caption{Convergence study for the Stokes-Brinkman problem
    \eqref{eq:stokes-equation}.
    The plots show the approximation error $\|e_h\|_Q$ \eqref{eq:cvg-properties}
    in the $Q$ norm depending on the mesh resolution.
    $\|r_{v}\|_Q$ defines the
    rate of convergence for the solution variable $v$ in the $Q$ norm.
    \subref{fig:stokes-cvg-direct-P2P1P1} and
    \subref{fig:stokes-cvg-fracprec-P2P1P1} show that the $P_2^2
    \times P_1 \times P_1^2$ approximation gives the expected
    convergence rates in the $L_2$ and $H_1$ norms, both using a
    direct solver and a preconditioned \cite{MiroKentPrecond2016}
    iterative solver. \subref{fig:stokes-cvg-direct-P3P2P2} and
    \subref{fig:stokes-cvg-fracprec-P3P2P2} show that the expected
    convergence rates are still obtained with higher order elements
    $P_3^2 \times P_2 \times P_2^2$.}
  \label{fig:stokes-cvg}
\end{figure}

\subsubsection{A Stokes problem with non-standard traction conditions}

The next test case presents the a variant of the Stokes problem (see
\Cref{fig:stokes-szopos-equation}) discussed in
\cite{szopos:tel-01646867, bertoluzza:hal-01420651}, involving
non-standard traction boundary conditions at the inlet and outlet
boundaries $\Gamma_{in}$ and $\Gamma_{\rm out}$, respectively. This
precise formulation involving the symmetric velocity gradient may be
useful in connection with e.g.~fluid-structure interaction
problems\cite{bertoluzza:hal-01420651}.
\begin{figure}[H]
  \centering
  \begin{minipage}[c]{0.42\textwidth}
    \begin{tikzpicture}[scale=0.25]
      \draw[black] (-10.0,-3.0) -- (10.0,-3.0);
      \draw[black] (10.0,3.0) -- (-10.0,3.0);
      \draw[thick] (-10.0,3.0) -- (-10.0,-3.0);
      \draw[thick] (10.0,-3.0) -- (10.0,3.0);
      \draw (0,0) node[] {\Large{$\Omega$}};
      \draw (-4,0) node[] {$\Gamma_{w}$};
      \draw[->] (-4,1) -- (-4,3);
      \draw[->] (-4,-1) -- (-4,-3);
      \draw (-9,0) node[right] {$\Gamma_{in}$};
      \draw (9,0) node[left] {$\Gamma_{\rm out}$};
      \draw[->] (-9,0) -- (-10,0);
      \draw[->] (9,0) -- (10,0);
    \end{tikzpicture}
  \end{minipage}
  \begin{minipage}[b]{0.52\textwidth}
    \begin{equation*}
      \renewcommand{\arraystretch}{1.5}
      \begin{array}{lll}
        -\mu \nabla ( \nabla u +\nabla u^T) + \nabla p = f ~\text{on}~ \Omega\\
        \nabla \cdot u = 0 ~\text{on}~ \Omega\\
        u \times n = 0 ~\text{on}~ \Gamma_{\rm in} \cup \Gamma_{\rm out} \\
      \end{array}
    \end{equation*}
  \end{minipage}
  \medskip
  \caption{The domain $\Omega$ (left) is the rectangle $[0,10] \times
    [0.5]$ where $\Gamma_{\rm in}$ and $\Gamma_{\rm out}$ represent
    the inlet and the outlet, and $\Gamma_w$ denotes the upper and
    lower boundaries (representing horizontal walls). The equations
    (right) are based on a standard formulation of the Stokes
    equations with additional traction conditions on $\Gamma_{\rm in}$
    and $\Gamma_{\rm out}$.}
  \label{fig:stokes-szopos-equation}
\end{figure}

The traction conditions are imposed using a Lagrange multiplier
$\lambda_{\rm in}$ (resp. $\lambda_{\rm out}$) relative to the inlet
$\Gamma_{\rm in}$ (resp. the outlet $\Gamma_{\rm out}$), acting on the
tangential component of the velocity $u \cdot t
\equiv u \times n$. The standard no-slip condition
$u=0$ is applied to the walls $\Gamma_w$, and a pressure
difference is imposed between the inlet where $p = p_{\rm in}$ and the
outlet where we impose $p = p_{\rm out}$.  We also apply homogeneous
Dirichlet boundary conditions for both Lagrange multipliers
$\lambda_i$ and $\lambda_{\rm out}$.

The variational formulation then reads: find $(u, p,
\lambda_{\rm in}, \lambda_{\rm out}) \in U \times P \times L_{\rm in}
\times L_{\rm out}$ such that
\begin{eqnarray}
  \disp{ \mu \int_{\Omega}} (\nabla u &+& \nabla u^T)
    \cdot (\nabla v + \nabla v^T) 
  - \disp{p(\nabla \cdot v)  }
  - \disp{q(\nabla \cdot u) \dx } \nonumber \\
  &+& \disp{ \int_{\Gamma_{\rm in}} \lambda_{\rm in} (v \times n) +
    \eta_{\rm in} (u \times n) \ds }
  + \disp{ \int_{\Gamma_{\rm out}} \lambda_{\rm out} (v \times n) +
     \eta_{\rm out} (u \times n) \ds } \label{equ:stokes-szopos-vf} \\
  &=&
  \disp{ \int_{\Omega} f \cdot v }
  - \disp{ \int_{\Gamma_{\rm in}} p_{\rm in} (v \cdot n)}
  - \disp{ \int_{\Gamma_{\rm out}} p_{\rm out} (v \cdot n)} \nonumber
\end{eqnarray}
for all $(v, q, \eta_{\rm in}, \eta_{\rm out}) \in U \times P \times L_{\rm in} \times L_{\rm out}$.

Again, we define a uniform mesh $\mathcal{T}^1$ of the domain $\Omega$
and lower-dimensional meshes $\mathcal{T}^2, \mathcal{T}^3$ for the
boundaries $\Gamma_{\rm in}$, $\Gamma_{\rm out}$, respectively. The
approximation of the velocity $u$ and the pressure $p$ of
\eqref{equ:stokes-szopos-vf} uses the standard Taylor-Hood elements,
\emph{i.e.} second order (resp. first order) Lagrange finite elements
for $u \in U$ (resp. for $p \in P$).  The Lagrange
multipliers are defined in continuous piecewise quadratic function
spaces $L_{\rm in} = L_{\rm in}(\mathcal{T}^2)$ and $L_{\rm out} =
L_{\rm out}(\mathcal{T}^3)$, ensuring the well posedness of the
problem as detailed in \cite{szopos:tel-01646867}.

\begin{figure}[H]
  \centering
  \begin{subfigure}[b]{0.49\textwidth}
    \centering
    \includegraphics[width=0.98\textwidth]{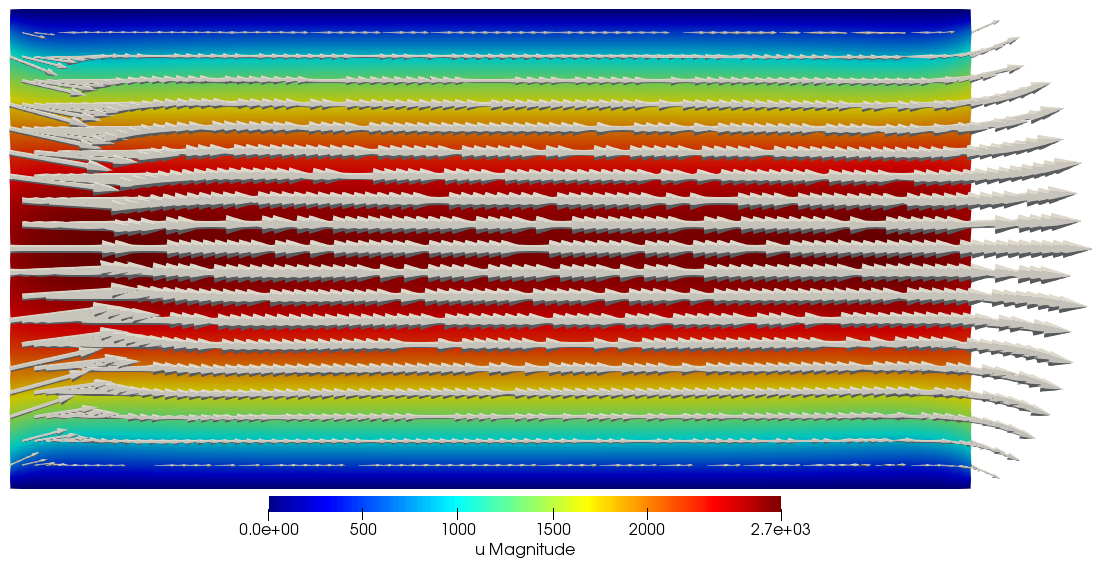}
    \caption{Without traction condition}
  \end{subfigure}
  \begin{subfigure}[b]{0.49\textwidth}
    \centering
    \includegraphics[width=0.98\textwidth]{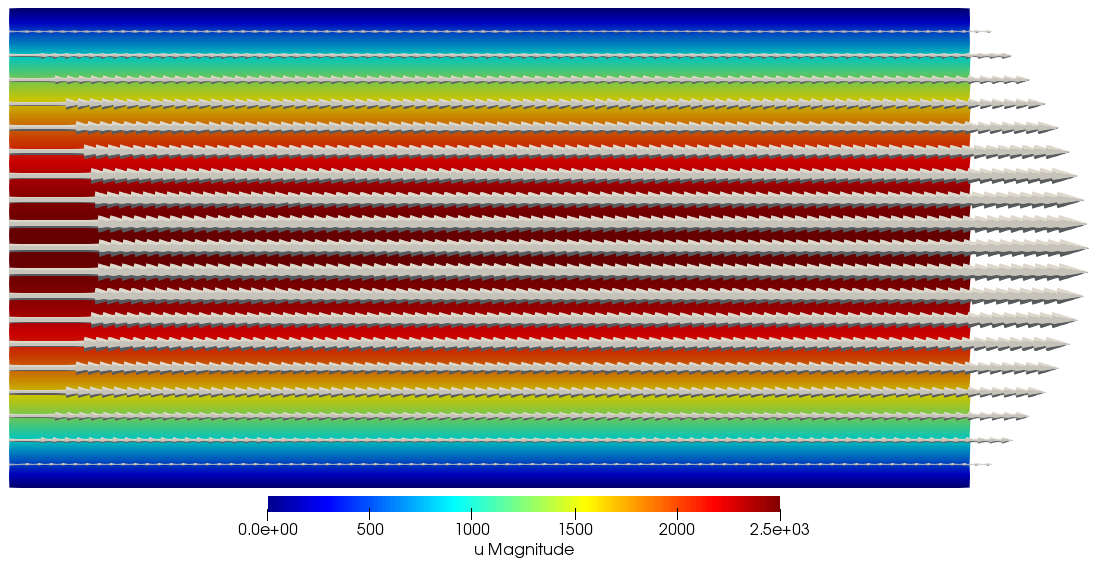}
    \caption{With traction condition}
  \end{subfigure}
  \caption{Comparison between the velocity field obtained from the
    Stokes problem with standard boundary conditions i.e.~without
    the Lagrange multipliers and the velocity field obtained
    with the traction condition on the inlet and the outlet. All other
    variables (mesh, material parameters etc.) were kept fixed.}
  \label{fig:stokes-szopos-velocity}
\end{figure}

Figure \ref{fig:stokes-szopos-velocity} compares the results obtained
with and without the traction condition, i.e.~with and
without the use of the Lagrange multipliers. Both solutions were
computed through the presented framework and preconditioned with the
fractional preconditioner \cite{MiroKentPrecond2016} introduced in
Section~\ref{sec:stokes:1}.

\subsection{Ionic electrodiffusion in cellular geometries}
\label{subsec:ion-dynamic-result}

Many cerebral pathological conditions
e.g.~spreading depression and epilepsy \cite{SpreadingDepression}
are associated with changes in ion
concentrations in the brain tissue. In this last example, we consider
a model of ionic electrodiffusion in intracellular and extracellular
domains, separated by a cell membrane. We represent the intracellular
and extracellular domains as separate two-dimensional subdomains with
the cell membrane as a topologically one-dimensional submesh. For more
details of the mathematical model and numerical method, we refer
to~\cite{EllingsrudEtAl2020}.

\begin{figure}[H]
  \centering
  \begin{minipage}[b]{0.45\textwidth}
    \begin{tikzpicture}[scale=1.0]
      \draw[black!50] (0.0,0.0) -- (4.0,0.0);
      \draw[black!50] (4.0,0.0) -- (4.0,4.0);
      \draw[black!50] (4.0,4.0) -- (0.0,4.0);
      \draw[black!50]  (0.0,4.0) -- (0.0,0.0);
      \draw[thick, dashed] (0.0,0.0) -- (4.0,0.0);
      \draw[thick, dashed] (4.0,0.0) -- (4.0,4.0);
      \draw[thick, dashed] (4.0,4.0) -- (0.0,4.0);
      \draw[thick, dashed] (0.0,4.0) -- (0.0,0.0);
      \draw[thick] (1.0,1.0) -- (3.0,1.0);
      \draw[thick] (3.0,1.0) -- (3.0,3.0);
      \draw[thick] (3.0,3.0) -- (1.0,3.0);
      \draw[thick] (1.0,3.0) -- (1.0,1.0);
      \draw[black] (0.75, 0.75) node[] {\large{$\Omega_e$}};
      \draw[black] (3.6, 3.6) node[] {\large{$\delta \Omega_e$}};
      \draw[black] (2.0, 2.0) node[] {\large{$\Omega^i$}};
      \draw[black] (1.0, 3.0) node[above right] {\large{$\Gamma$}};
    \end{tikzpicture}
  \end{minipage}
  \begin{minipage}[b]{0.45\textwidth}
    \centering
    \includegraphics[width=0.87\textwidth]{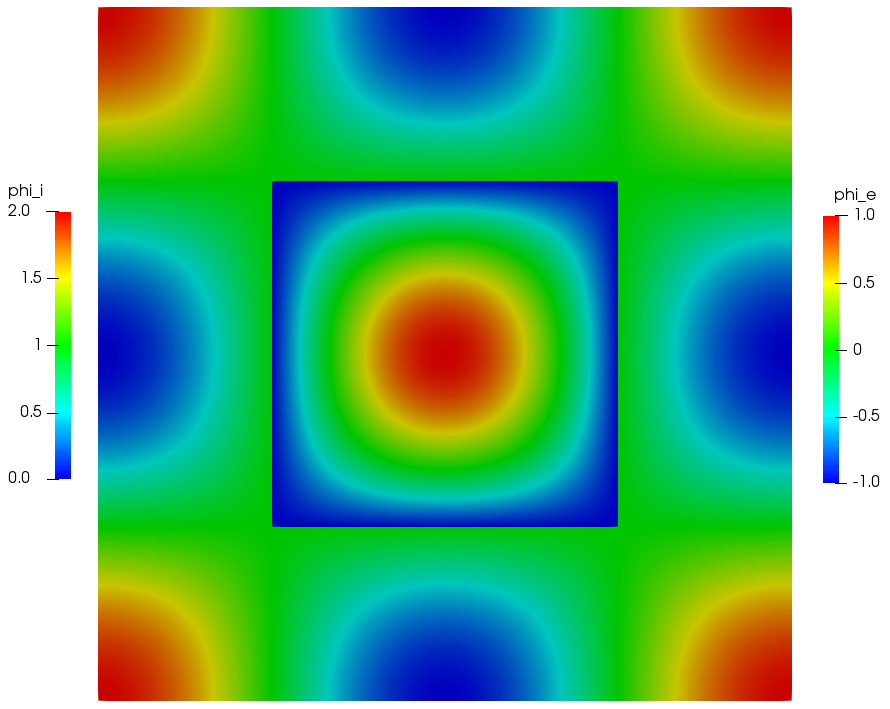}
  \end{minipage}
  \caption{Illustration of the domain composed of the extra cellular
    space $\Omega_e$, the intra cellular space $\Omega_i$ and the
    membrane $\Gamma$ as their interface (left).  First order
    approximation of the potentials $\phi_i \in \Omega_i$ and $\phi_e
    \in \Omega_e$ matching the manufactured solution
    \eqref{eq:ion:mms} (right).}
  \label{fig:knpemi-sol-phi}
\end{figure}
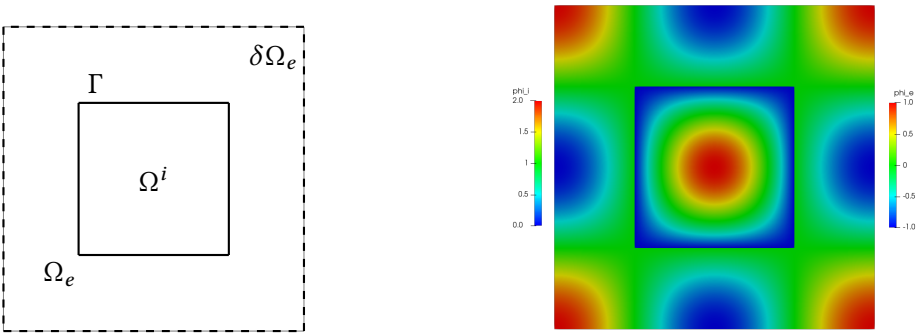

For this specific test case, illustrated in
Figure~\ref{fig:knpemi-sol-phi}, we consider a domain $\Omega \subset
\R^2$ consisting of $\Omega_i \subset \R^2$ and $\Omega_e \subset
\R^2$ representing the intracellular and extracellular spaces,
respectively, together with the cell membrane $\Gamma = \bar{\Omega}_i
\cap \bar{\Omega}_e$. The unknowns are the ion concentrations $[k]_i$
(resp. $[k]_e$) for each ion species $k \in K = \{\rm Na^{+}, K^{+},
Cl^{-}\}$, the electrical potential $\phi_i$ (resp. $\phi_e$) in the
intracellular space $\Omega_i$ (resp. extracellular space $\Omega_e$),
and the total ionic current density $I_M$ at the (lower dimensional)
cell membrane $\Gamma$.

The evolution and distribution of the ion concentration $[k]_r$, $r=\{i,e\}$
for $k \in K$ is described by the continuity equation
\begin{equation}
  \label{equ:knpemi-equation}
  \frac{\partial [k]_r}{\partial t} + \nabla \cdot J_r^k = f_r \text{ in } \Omega_r .
\end{equation}
The ion flux density $J_r^k$ ($r=\{i,e\}$) is expressed in
terms of the ion concentration gradients $\nabla [k]_r$ for $k \in K$
and the electrical potential gradients $\nabla \phi_r$ as
\begin{equation}
  \label{equ:knpemi-flux-density-J}
  J_r^k = -D_r^k \nabla [k]_r - \frac{D_r^k z^k}{\psi}[k]_r \nabla \phi_r 
\end{equation}
where $D_r^k$, $r=\{i,e\}$ is the effective diffusion coefficient and
$z^k$ the valence of the ion species $k \in K$.  $\psi = RTF^{-1}$
with $F$ the Faraday's constant, $T$ is the absolute temperature, and
$R$ the gas constant. Moreover, these ion flux densities are assumed
to satisfy the electroneutrality assumption:
\begin{equation}
  \label{equ:knpemi-flux-densities}
  F \sum \limits_{k \in K} z^k \nabla \cdot J_r^k = 0 \text{ in } \Omega_r ~~\forall r \in \{i,e\}
\end{equation}
Assuming that no charge can leave or enter the system yields the
exterior boundary condition
\begin{equation}
  \label{equ:knpemi-ext-bc}
  F \sum \limits_{k \in K} z^k J_e^k \cdot n_e = 0 \text{ on } \delta \Omega_e .
\end{equation}

The transmembrane potential is introduced as the difference between
the intracellular and extracellular potential $\phi_M = \phi_i -
\phi_e$ at the cell membrane, which satisfies
\begin{equation}
  \label{equ:knpemi-transmembrane-IM}
  \frac{\partial \phi_M}{\partial t} = \frac{1}{C_M}(I_M - I_{ch})
\end{equation}
where $C_M$ and $I_{\rm ch}$ are the given capacitance and ion species
specific channel current, respectively. This additional equation
\eqref{equ:knpemi-transmembrane-IM} giving the total ionic current
density $I_M \in \Gamma$ is coupled with \eqref{equ:knpemi-equation}
by the conservation equation of $I_M$ over the cell membrane as
\begin{equation}
  \label{equ:knpemi-IM-conservation}
  - F \sum \limits_{k \in K} z^k J_i^k \cdot n_i = F \sum \limits_{k \in K} z^k J_e^k \cdot n_e \equiv I_M .
\end{equation}

We introduce manufactured solutions $[k]_r$, $k \in K$ for the ion
concentrations and $\phi_r$ for the electric potential
($r=\{i,e\}$) satisfying
\eqref{equ:knpemi-equation}-\eqref{equ:knpemi-IM-conservation} on
$\Omega = \Omega_i \cup \Omega_e = [0,1] \times [0,1]$.
\begin{equation}
  \begin{split}
  \text{Na}_i &= 0.7 + 0.3\sin(2 \pi x) \sin(2 \pi y) (1 + e^{-t}) \\
  \text{Na}_e &= 1.0 + 0.6\sin(2 \pi x) \sin(2 \pi y) (1 + e^{-t}) \\
  \text{K}_i &= 0.3 + 0.3\sin(2 \pi x) \sin(2 \pi y) (1 + e^{-t}) \\
  \text{K}_e &= 1.0 + 0.2\sin(2 \pi x) \sin(2 \pi y) (1 + e^{-t}) \\
  \text{Cl}_i &= 1.0 + 0.6\sin(2 \pi x) \sin(2 \pi y) (1 + e^{-t}) \\
  \text{Cl}_e &= 2.0 + 0.8\sin(2 \pi x) \sin(2 \pi y) (1 + e^{-t}) \\
  \phi_i &= \cos(2 \pi x) \cos(2 \pi y) (1 + e^{-t}) \\
  \phi_e &= \cos(2 \pi x) \cos(2 \pi y)
  \end{split}
  \label{eq:ion:mms}
\end{equation}
As our test case here, we then consider the boundary conditions and
sources induced by inserting the manufactured
solution~\eqref{eq:ion:mms}
into~\eqref{equ:knpemi-equation}-\eqref{equ:knpemi-IM-conservation}.

We define a uniform mesh $\mathcal{T}^1$ of the unit square $\Omega$,
from which we build two two-dimensional meshes $\mathcal{T}^2$
and $\mathcal{T}^3$ relative to the intracellular domain
$\Omega_i$ and the extracellular domain $\Omega_e$, respectively.
We consider a lower-dimensional mesh $\mathcal{T}^4$ to define
the cell membrane $\Gamma$.
All the unknowns in our system i.e.~the ion concentrations
$[k]_r, r=\{i,e\}, k=\{\text{Na}^+, \text{K}^+, \text{Cl}^-\}$ and
the electrical potentials $\phi_i$ and $\phi_e$, are approximated
using piecewise continuous elements of order $l \geq 1$.

The corresponding solutions for the electrical potentials $\phi_i$ and
$\phi_e$ are shown in Figure \ref{fig:knpemi-sol-phi}. A convergence
study of the error of the approximation in the $L_2$ and $H_1$ norms
for the Sodium (Na$^{+}$) concentrations and electrical potentials
$\phi_i, \phi_e$ is shown in Figure~\ref{fig:knpemi-cvg-mms2}. The
convergence rates indicate that we obtain an
expected order of convergence \eqref{eq:cvg-properties} for both
polynomial orders $l=\{1, 2\}$. The analogous results are obtained for
the other ion concentrations.
\begin{figure}[H]
  \centering
  \captionsetup[subfigure]{justification=centering}
  \begin{subfigure}[t]{0.48\textwidth}
    \centering
    \begin{tikzpicture}
      \pgfplotsset{every axis legend/.append style = { at={(0.5,1.03)},anchor=south }}
      \begin{loglogaxis} 
        [
          width=\textwidth,
          ticklabel style = {font=\scriptsize},
          title style = {font=\scriptsize},
          legend style = {font=\scriptsize},
          legend cell align={left},
          xtick={8,16,32,64},
          xticklabels={8,16,32,64},
          legend columns=2
        ]

        \pgfplotstableread{Na_e_P1.dat}\NaExtOne
        \pgfplotstableread{Na_e_P2.dat}\NaExtTwo

        \addplot[color=blue, mark=*] table[x=N,y=L2]{\NaExtOne};
        \addplot[mark=none, style=dashed, forget plot] table[y = {create col/linear regression={y=L2}}]{\NaExtOne};
        \pgfmathsetmacro{\slopeA}{abs(\pgfplotstableregressiona)}
        \addlegendentry{$\|r_{\text{Na}_e^{+},1}\|_{L_2}$: $\pgfmathprintnumber{\slopeA}$}
        \addplot[color=blue, mark=o] table[x=N,y=H1]{\NaExtOne};
        \addplot[mark=none, style=dashed, forget plot] table[y = {create col/linear regression={y=H1}}]{\NaExtOne};
        \pgfmathsetmacro{\slopeB}{abs(\pgfplotstableregressiona)}
        \addlegendentry{$\|r_{\text{Na}_e^{+},1}\|_{H_1}$: $\pgfmathprintnumber{\slopeB}$}
        \addplot[color=red, mark=square*] table[x=N,y=L2]{\NaExtTwo};
        \addplot[mark=none, style=dashed, forget plot] table[y = {create col/linear regression={y=L2}}]{\NaExtTwo};
        \pgfmathsetmacro{\slopeC}{abs(\pgfplotstableregressiona)}
        \addlegendentry{$\|r_{\text{Na}_e^{+},2}\|_{L_2}$: $\pgfmathprintnumber{\slopeC}$}
        \addplot[color=red, mark=square] table[x=N,y=H1]{\NaExtTwo};
        \addplot[mark=none, style=dashed, forget plot] table[y = {create col/linear regression={y=H1}}]{\NaExtTwo};
        \pgfmathsetmacro{\slopeD}{abs(\pgfplotstableregressiona)}
        \addlegendentry{$\|r_{\text{Na}_e^{+},2}\|_{H_1}$: $\pgfmathprintnumber{\slopeD}$}
      \end{loglogaxis}
    \end{tikzpicture}
    \caption{Convergence study - Na$^+_e$}
    \label{fig:knpemi-cvg-nae-mms2}
  \end{subfigure}
  \begin{subfigure}[t]{0.48\textwidth}
    \centering
    \begin{tikzpicture}
      \pgfplotsset{every axis legend/.append style = { at={(0.5,1.03)},anchor=south }}
      \begin{loglogaxis} 
        [
          width=\textwidth,
          ticklabel style = {font=\scriptsize},
          title style = {font=\scriptsize},
          legend style = {font=\scriptsize},
          legend cell align={left},
          xtick={8,16,32,64},
          xticklabels={8,16,32,64},
          legend columns=2
        ]

        \pgfplotstableread{Na_i_P1.dat}\NaIntOne
        \pgfplotstableread{Na_i_P2.dat}\NaIntTwo

        \addplot[color=blue, mark=*] table[x=N,y=L2]{\NaIntOne};
        \addplot[mark=none, style=dashed, forget plot] table[y = {create col/linear regression={y=L2}}]{\NaIntOne};
        \pgfmathsetmacro{\slopeA}{abs(\pgfplotstableregressiona)}
        \addlegendentry{$\|r_{\text{Na}_i^{+},1}\|_{L_2}$: $\pgfmathprintnumber{\slopeA}$}
        \addplot[color=blue, mark=o] table[x=N,y=H1]{\NaIntOne};
        \addplot[mark=none, style=dashed, forget plot] table[y = {create col/linear regression={y=H1}}]{\NaIntOne};
        \pgfmathsetmacro{\slopeB}{abs(\pgfplotstableregressiona)}
        \addlegendentry{$\|r_{\text{Na}_i^{+},1}\|_{H_1}$: $\pgfmathprintnumber{\slopeB}$}
        \addplot[color=red, mark=square*] table[x=N,y=L2]{\NaIntTwo};
        \addplot[mark=none, style=dashed, forget plot] table[y = {create col/linear regression={y=L2}}]{\NaIntTwo};
        \pgfmathsetmacro{\slopeC}{abs(\pgfplotstableregressiona)}
        \addlegendentry{$\|r_{\text{Na}_i^{+},2}\|_{L_2}$: $\pgfmathprintnumber{\slopeC}$}
        \addplot[color=red, mark=square] table[x=N,y=H1]{\NaIntTwo};
        \addplot[mark=none, style=dashed, forget plot] table[y = {create col/linear regression={y=H1}}]{\NaIntTwo};
        \pgfmathsetmacro{\slopeD}{abs(\pgfplotstableregressiona)}
        \addlegendentry{$\|r_{\text{Na}_i^{+},2}\|_{H_1}$: $\pgfmathprintnumber{\slopeD}$}
      \end{loglogaxis}
    \end{tikzpicture}
    \caption{Convergence study - Na$^+_i$}
    \label{fig:knpemi-cvg-nai-mms2}
  \end{subfigure}
\end{figure}
\begin{figure}[H]
  \ContinuedFloat 
  \centering
  \captionsetup[subfigure]{justification=centering}
  \begin{subfigure}[t]{0.48\textwidth}
    \centering
    \begin{tikzpicture}
      \pgfplotsset{every axis legend/.append style = { at={(0.5,1.03)},anchor=south }}
      \begin{loglogaxis} 
        [
          width=\textwidth,
          ticklabel style = {font=\scriptsize},
          title style = {font=\scriptsize},
          legend style = {font=\scriptsize},
          legend cell align={left},
          xtick={8,16,32,64},
          xticklabels={8,16,32,64},
          legend columns=2
        ]

        \pgfplotstableread{Phi_e_P1.dat}\PhiExtOne
        \pgfplotstableread{Phi_e_P2.dat}\PhiExtTwo

        \addplot[color=blue, mark=*] table[x=N,y=L2]{\PhiExtOne};
        \addplot[mark=none, style=dashed, forget plot] table[y = {create col/linear regression={y=L2}}]{\PhiExtOne};
        \pgfmathsetmacro{\slopeA}{abs(\pgfplotstableregressiona)}
        \addlegendentry{$\|r_{\phi_e,1}\|_{L_2}$: $\pgfmathprintnumber{\slopeA}$}
        \addplot[color=blue, mark=o] table[x=N,y=H1]{\PhiExtOne};
        \addplot[mark=none, style=dashed, forget plot] table[y = {create col/linear regression={y=H1}}]{\PhiExtOne};
        \pgfmathsetmacro{\slopeB}{abs(\pgfplotstableregressiona)}
        \addlegendentry{$\|r_{\phi_e,1}\|_{H_1}$: $\pgfmathprintnumber{\slopeB}$}
        \addplot[color=red, mark=square*] table[x=N,y=L2]{\PhiExtTwo};
        \addplot[mark=none, style=dashed, forget plot] table[y = {create col/linear regression={y=L2}}]{\PhiExtTwo};
        \pgfmathsetmacro{\slopeC}{abs(\pgfplotstableregressiona)}
        \addlegendentry{$\|r_{\phi_e,2}\|_{L_2}$: $\pgfmathprintnumber{\slopeC}$}
        \addplot[color=red, mark=square] table[x=N,y=H1]{\PhiExtTwo};
        \addplot[mark=none, style=dashed, forget plot] table[y = {create col/linear regression={y=H1}}]{\PhiExtTwo};
        \pgfmathsetmacro{\slopeD}{abs(\pgfplotstableregressiona)}
        \addlegendentry{$\|r_{\phi_e,2}\|_{H_1}$: $\pgfmathprintnumber{\slopeD}$}
      \end{loglogaxis}
    \end{tikzpicture}
    \caption{Convergence study - $\phi_e$}
    \label{fig:knpemi-cvg-phie-mms2}
  \end{subfigure}
  \begin{subfigure}[t]{0.48\textwidth}
    \centering
    \begin{tikzpicture}
      \pgfplotsset{every axis legend/.append style = { at={(0.5,1.03)},anchor=south }}
      \begin{loglogaxis} 
        [
          width=\textwidth,
          ticklabel style = {font=\scriptsize},
          title style = {font=\scriptsize},
          legend style = {font=\scriptsize},
          legend cell align={left},
          xtick={8,16,32,64},
          xticklabels={8,16,32,64},
          legend columns=2
        ]

        \pgfplotstableread{Phi_i_P1.dat}\PhiIntOne
        \pgfplotstableread{Phi_i_P2.dat}\PhiIntTwo

        \addplot[color=blue, mark=*] table[x=N,y=L2]{\PhiIntOne};
        \addplot[mark=none, style=dashed, forget plot] table[y = {create col/linear regression={y=L2}}]{\PhiIntOne};
        \pgfmathsetmacro{\slopeA}{abs(\pgfplotstableregressiona)}
        \addlegendentry{$\|r_{\phi_i,1}\|_{L_2}$: $\pgfmathprintnumber{\slopeA}$}
        \addplot[color=blue, mark=o] table[x=N,y=H1]{\PhiIntOne};
        \addplot[mark=none, style=dashed, forget plot] table[y = {create col/linear regression={y=H1}}]{\PhiIntOne};
        \pgfmathsetmacro{\slopeB}{abs(\pgfplotstableregressiona)}
        \addlegendentry{$\|r_{\phi_i,1}\|_{H_1}$: $\pgfmathprintnumber{\slopeB}$}
        \addplot[color=red, mark=square*] table[x=N,y=L2]{\PhiIntTwo};
        \addplot[mark=none, style=dashed, forget plot] table[y = {create col/linear regression={y=L2}}]{\PhiIntTwo};
        \pgfmathsetmacro{\slopeC}{abs(\pgfplotstableregressiona)}
        \addlegendentry{$\|r_{\phi_i,2}\|_{L_2}$: $\pgfmathprintnumber{\slopeC}$}
        \addplot[color=red, mark=square] table[x=N,y=H1]{\PhiIntTwo};
        \addplot[mark=none, style=dashed, forget plot] table[y = {create col/linear regression={y=H1}}]{\PhiIntTwo};
        \pgfmathsetmacro{\slopeD}{abs(\pgfplotstableregressiona)}
        \addlegendentry{$\|r_{\phi_i,2}\|_{H_1}$: $\pgfmathprintnumber{\slopeD}$}
      \end{loglogaxis}
    \end{tikzpicture}
    \caption{Convergence study - $\phi_i$}
    \label{fig:knpemi-cvg-phii-mms2}
  \end{subfigure}
  \caption{Convergence study for the KNP-EMI model given
    by~\eqref{equ:knpemi-equation}-\eqref{equ:knpemi-IM-conservation}
    based on the manufactured solution \eqref{eq:ion:mms}.
    The plots show the approximation error $\|e_h\|_Q$ \eqref{eq:cvg-properties}
    in the $Q$ norm depending on the mesh resolution.
    $\|r_{v,l}\|_Q$ defines the rate of convergence
    for the solution variable $v$ in the $Q$ norm with
    approximation order $l$.
    The resulting convergence rates are given in legend.
    The expected convergence rates in $L_2$ and $H_1$
    norms are obtained for both
    first and second order approximations of sodium (Na$^{+}$) concentration
    \subref{fig:knpemi-cvg-nae-mms2}-\subref{fig:knpemi-cvg-nai-mms2}
    and for the electric potentials $\phi_e$
    \subref{fig:knpemi-cvg-phie-mms2} and $\phi_i$
    \subref{fig:knpemi-cvg-phii-mms2}.}
  \label{fig:knpemi-cvg-mms2}
\end{figure}
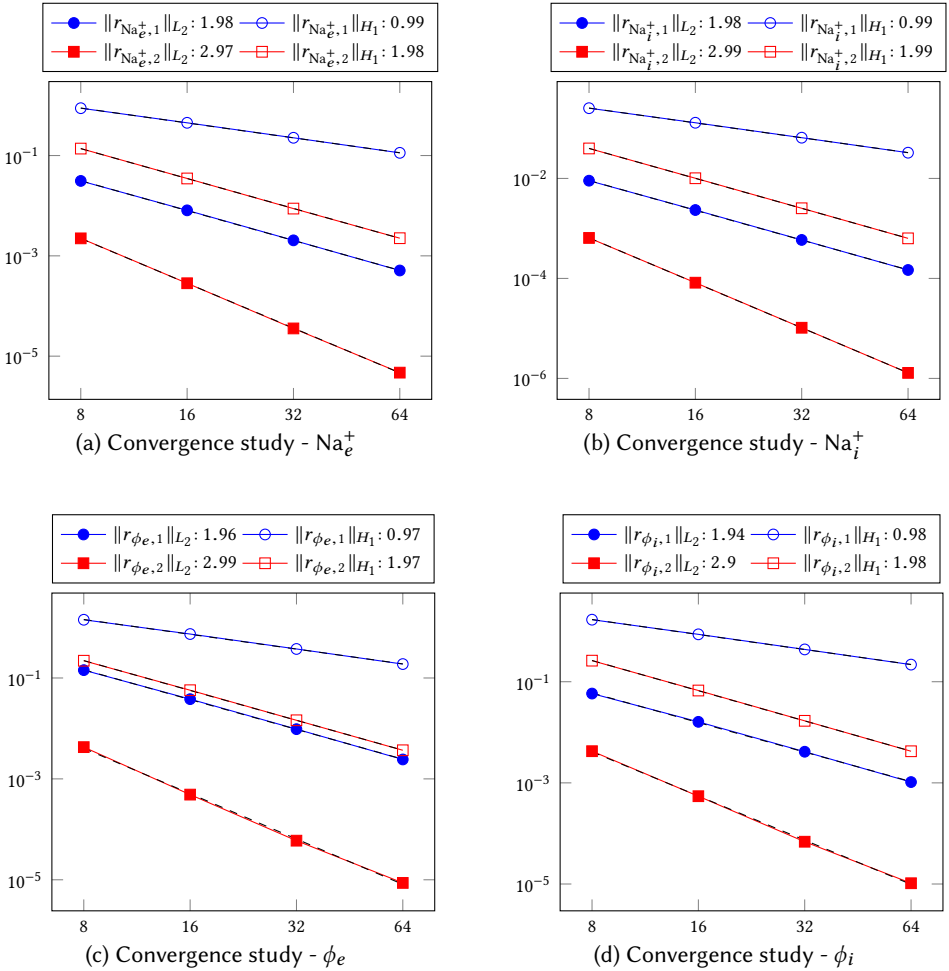

\section{Conclusions}
\label{sec:conclusions}

We have introduced a set of abstractions, algorithms and design
guidelines for the automated assembly and solution of mixed domain and
mixed dimensional finite element methods. Further, we have presented a
realization of these general concepts within the FEniCS finite element
software. These features thus allow for the solution of PDEs posed on
different domains, either of the same dimension or involving
codimension one subdomains. We have illustrated the features with a
number of numerical examples starting from a basic constrained Poisson
problem to a nontrivial model of ionic electrodiffusion. However, we
argue that this series of examples only begin to illustrate the
possibilities offered by the framework. Future work will focus on the
extension of the framework to coupled
problems with higher codimensions, non-conforming meshes, and optimal
mesh partitioning.

\begin{acks}
This project has received funding from the European Research Council
(ERC) under the European Union's Horizon 2020 research and innovation
programme under grant agreement 714892 (Waterscales).

We thank Martin Aln{\ae}s, Miroslav Kuchta and J{\o}rgen Dokken
(Simula Research Laboratory) for constructive discussion on topics
related to the manuscript.
\end{acks}

\bibliographystyle{ACM-Reference-Format}
\bibliography{references}


\end{document}